\documentstyle[eqsecnum,aps,prb,multicol,epsf]{revtex}
\begin{document}
\draft
\tighten

\title{Elasticity, Shape Fluctuations and Phase Transitions in the New
Tubule Phase of Anisotropic Tethered Membranes}
\author{Leo Radzihovsky} 
\address{Department of Physics, University of Colorado, Boulder, CO 80309}
\author{John Toner}
\address{Department of Physics, Institute for Theoretical Science, and
Material Science Institute, University of Oregon, Eugene, OR
97403-1274}

\date{\today}
\maketitle
\begin{abstract}

We study the shape, elasticity and fluctuations of the recently
predicted\cite{RT} and subsequently observed (in numerical
simulations)\cite{BFT} tubule phase of anisotropic membranes, as well
as the phase transitions into and out of it. This novel phase lies
between the previously predicted flat and crumpled phases, both in
temperature and in its physical properties: it is crumpled in one
direction, and extended in the other. Its shape and elastic properties
are characterized by a radius of gyration exponent $\nu$ and an
anisotropy exponent $z$. We derive scaling laws for the radius of
gyration $R_G(L_\perp,L_y)$ (i.e. the average thickness) of the tubule
about a spontaneously selected straight axis and for the tubule
undulations $h_{rms}(L_\perp,L_y)$ transverse to its average
extension. We show that for square membranes (with {\em intrinsic}
size $L_\perp=L_y=L$), $R_G\propto L^\nu$, and $h_{rms}\propto
L^{1-\eta_\kappa z/2}$, with $\eta_\kappa$ a bending rigidity
anomalous elasticity exponent related to $\nu$ and $z$.  For phantom
(i.e. {\em non}-self-avoiding) membranes, we predict $\nu=1/4$,
$z=1/2$ and $\eta_\kappa=0$, {\it exactly}, in excellent agreement
with simulations. For $D=2$ dimensional membranes embedded in the
space of dimension $d<11$, self-avoidance greatly swells the tubule
and suppresses its wild transverse undulations, changing its shape
exponents $\nu$, $z$, and $\eta_\kappa$. For a $D$-dimensional
membrane embedded in $d>d_*$ ($d_*(D=2)>7/2$), $\eta_\kappa=0$ and
$z=(D-1 + 2\nu)/3$, while for $d<d_*$, $\eta_\kappa>0$ and $z=(D-1 +
2\nu)/(3-\eta_\kappa)$.  ``Flory'' theory yields, in the physical case
of $D=2$ and $d=3$, $\nu=3/4$, while the recent $11-\epsilon$
expansion results\cite{BG} yields $\nu=0.52$. The actual value of
$\nu$ probably lies closer to the Flory estimate, between these two
limits. We give detailed scaling results for the shape of the tubule
of an arbitrary aspect ratio, i.e. for the tubule thickness, its
transverse undulations, and a variety of other correlation functions,
as well as for the anomalous elasticity of the tubules, in terms of
$\nu$ and $z$. Finally we present a scaling theory for the shape and
specific heat near the continuous transitions into and out of the
tubule phase and perform detailed renormalization group calculations
for the crumpled-to-tubule transition for phantom membranes.
\end{abstract}

\begin{multicols}{2}
\narrowtext

\section{Introduction}
\label{intro}
Tethered membranes\cite{Jerusalem} became a subject of great interest
when it was theoretically predicted\cite{NP} that, unlike polymers,
which are always orientationally disordered, membranes can exhibit two
distinct phases: crumpled and flat, with a ``crumpling'' transition
between them.  The flat phase is particularly novel and intriguing,
because it provides an example of a two dimensional system with a
continuous symmetry that nonetheless exhibits a long-ranged order
(specifically, long-ranged orientational order in the normal to the
membrane) in apparent violation of the Hohenberg-Mermin-Wagner
theorem\cite{MerminWagner}.  This ordering is made possible by
``anomalous elasticity''\cite{NP,AL,GD}: thermal fluctuations
infinitely enhance the bending rigidity $\kappa$ of the membrane at
long wavelengths, thereby stabilizing the orientational order against
these very fluctuations. This is perhaps the most dramatic
illustration yet found of the phenomenon of ``order from disorder''.

\begin{figure}[bth]
\centering
\setlength{\unitlength}{1mm}
\begin{picture}(150,40)(0,0)
\put(-17,-120){\begin{picture}(150,0)(0,0)
\includegraphics{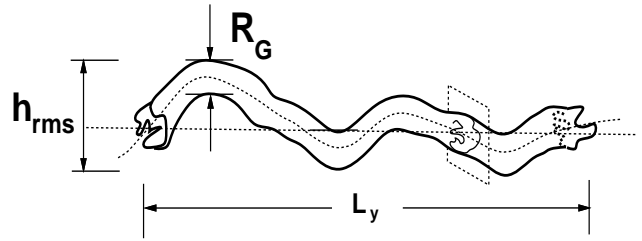}
\end{picture}}
\end{picture}
\caption{Schematic picture of the tubule phase of anisotropic
polymerized membrane, with the definition of its thickness $R_G$ and
roughness $h_{rms}$, our predictions for which are given in
Eqs.\ref{RG} and \ref{hrms}.}
\label{tubule_fig}
\end{figure}

Rich as these phenomena are, most past theoretical
work\cite{Jerusalem} has been restricted to {\it isotropic} membranes.
In a recent paper\cite{RT} we extended these considerations to {\it
intrinsically anisotropic} membranes (e.g., polymerized membranes with
in-plane tilt order\cite{LRunpublished}) and found, astonishingly,
that anisotropy, a seemingly innocuous generalization, actually leads
to a wealth of new phenomena.  Most dramatically, we found an entire
new phase of membranes, which we called the ``tubule'' phase,
ubiquitously intervenes between the high temperature crumpled and low
temperature ``flat'' phases.  The defining property of the tubule
phase is that it is crumpled in one of the two membrane directions,
but ``flat'' (i.e., extended) in the other.  Its average shape is a
long, thin cylinder of length $R_y = L_y\times O(1)$ and radius
$R_G(L_\perp)\ll L_\perp$, where $L_y$ and $L_\perp$ are the
dimensions the membrane would have in the extended and crumpled
directions respectively, were it to be flattened out.  It should be
clarified here that we use the term ``cylinder'' {\em extremely}
loosely; as illustrated in Fig.\ref{tubule_fig}, a cross section of the
membrane perpendicular to the tubule axis ($y$) will look as
disordered as a flexible polymer. These tubules, occurring as a low
temperature phase of anisotropic {\it polymerized} membranes, have
little in common (and therefore should not be confused) with
micro-tubules that are found in {\it liquid} phospholipid
membranes\cite{Rudolf}.

\begin{figure}[bth]
\centering
\setlength{\unitlength}{1mm}
\begin{picture}(150,60)(0,0)
\put(-20,-68){\begin{picture}(150,0)(0,0)
\includegraphics{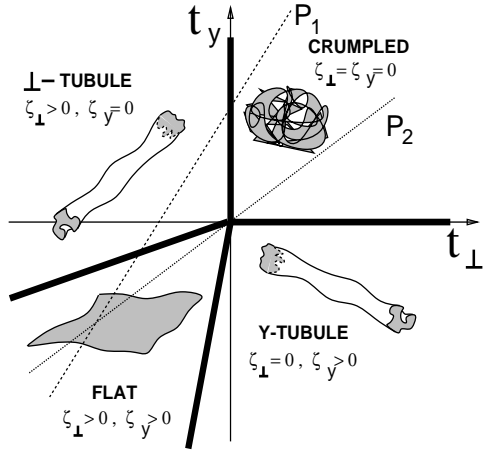}
\end{picture}}
\end{picture}
\caption{Phase diagram for anisotropic tethered
membranes showing the new tubule and previously studied flat and
crumpled phases.}
\label{phase_diagram1}
\end{figure}

Only in the special case of perfectly isotropic membranes
\cite{NotGeneric} is it possible for the membrane to undergo a direct
transition from the flat to the crumpled phase.  The theoretically
predicted\cite{RT} and recently observed\cite{BFT} phase diagram is
shown in Fig.\ref{phase_diagram1}.

The direct crumpling transition studied previously occurs in our more
generic model only for that special set of cuts through the phase
diagram (like P$_2$) that pass through the origin.  Generic paths
(like P$_1$) will experience {\it two} phase transitions,
crumpled-to-tubule, and tubule-to-flat, that are in new, heretofore
uninvestigated universality classes. 

This prediction was recently dramatically confirmed in Monte Carlo
simulations of phantom (i.e., non-self-avoiding) membranes by Bowick,
Falcioni and Thorleifsson (BFT).\cite{BFT}. They simulated membranes
with different bare bending moduli $\kappa_x$ and $\kappa_y$ in the
orthogonal $x$ and $y$ directions. As temperature (or one of the
bending rigidities e.g. $\kappa_x$) is varied, we predicted our model
would follow a generic path like P$_1$ in
Fig.\ref{phase_diagram1}. And, indeed, these simulations\cite{BFT}
observed two specific heat bumps, corresponding to two distinct
continuous transitions crumpled-to-tubule and tubule-to-flat (rounded
by finite membrane size), just as we predicted\cite{RT}. Furthermore,
the shape of the membrane in the phase between these two transitions
was exactly that of the tubule above (see Figure \ref{tubule_fig}),
and had, within numerical errors, precisely the scaling properties
and exponents that we predicted for phantom tubules\cite{RT}. Here we
present our detailed study of these transitions and the tubule phase,
in the presence of both thermal fluctuations and self-avoidance.

There are a number of possible experimental realizations of
anisotropic membranes. One is polymerized membranes with in-plane tilt
order\cite{LRunpublished}. {\it Fluid} membranes with such order have
already been found\cite{Sachmann,Bensimon}; it should be possible to
polymerize these without destroying the tilt order. Secondly,
membranes could be fabricated by cross-linking DNA molecules trapped
in a fluid membrane\cite{Sachmann,Bensimon}.  Performing the
cross-linking in an applied electric field would align the DNA and
"freeze in" the anisotropy induced by the electric field, which could
then be removed.

The tubule cross-sectional radius $R_G$, ( hereafter called the radius
of gyration), and its undulations $h_{rms}$ transverse to its average
axis of orientation, obey the scaling laws:
\begin{eqnarray}
R_G(L_\perp,L_y)&=& L_\perp^{\nu}S_R(L_y/L_\perp^z)\;,\label{RG}\\ 
h_{rms}(L_\perp,L_y)&=& L_y^{\zeta}S_h(L_y/L_\perp^z)\;,
\label{hrms}
\end{eqnarray}
where $\zeta=\nu/z$,
\begin{equation}
z={1\over3-\eta_\kappa}(1+2\nu)\;,
\label{z}
\end{equation}
we have specialized in Eq.\ref{z} to $D=2$ (with general expression
for a $D$-dimensional membrane given in the main text), the universal
exponents $\nu$ and $z$ are $<1$, $\eta_\kappa$ is the anomalous
elasticity exponent for the tubule bending rigidity $\kappa$ (as
defined by $\kappa\sim L_y^{\eta_\kappa}$, also see below), and for
convenience we chose to measure the intrinsic lengths $L_y$ and
$L_\perp$ in units of the ultraviolet cutoff, set approximately by the
the monomer (e.g.  phospholipid) size.
The scaling functions $S_{R,h}(x)$ have the limiting forms:
\begin{equation}
S_R(x)\propto
\left\{ \begin{array}{lr}
x^{\zeta-\nu_p/z}, & \mbox{for}\; x\rightarrow 0 \\
\mbox{constant}, & \mbox{for}\; x\rightarrow\infty
\end{array} \right.
\label{fR2d}
\end {equation}
\begin{equation}
\label{fh2d}
S_h(x)\propto
\left\{ \begin{array}{lr}
\mbox{constant}, & \mbox{for}\; x\rightarrow 0 \\
x^{{3\over 2} - \zeta}, & \mbox{for}\; x\rightarrow\infty
\end{array} \right.\;,
\end{equation}
where $\nu_p$ is the radius of gyration exponent of a coiled linear
polymer $\approx 3/5$. These scaling functions are {\it universal}
(i.e., independent of material parameters and temperature), up to an
overall non-universal multiplicative factor, which can, and will,
depend on material parameters and temperature.

The scaling forms, Eq.\ref{fR2d} and \ref{fh2d} imply that for a
"roughly square" membrane -- that is, one with $L_\perp\sim L_y\equiv
L$ -- in the limit $L\rightarrow\infty$
\begin{eqnarray}
R_G(L_\perp\sim L_y\equiv L)&\propto&L^{\nu}\;,\\
\label{Rsq}
h_{rms}(L_\perp\sim L_y\equiv L)&\propto&L^{1-\eta_\kappa z/2}\;,
\label{hsq}
\end{eqnarray}
where we have used the fact that for $ L_y \sim L_\perp$, the 
argument $x \equiv L_y/L_\perp^z$
of the scaling functions $S_{R,h}(x)$ goes to infinity as
$L\rightarrow\infty$, and used Eq.\ref{z} to simplify Eq.\ref{hsq}.

Detailed renormalization group calculations show that $\eta_\kappa$ is
strictly positive. Hence, $h_{rms}<<L$ for a roughly square membrane
as $L\rightarrow\infty$. Thus, the end-to-end orientational
fluctuations $\theta\sim h_{rms}/L\propto
L^{-\eta_\kappa z/2}\rightarrow 0$ as $L\rightarrow\infty$ for such a
roughly square membrane, proving that tubule order (which requires
orientational persistence in the extended direction) {\em is} stable
against undulations of the tubule embedded in $d=3$ dimensions.

On the other hand, in the limit $L_y>>L_\perp$, in which the tubule
looks more and more like a linear polymer (a ribbon of width $L_\perp$
and length $L_y$), we find
\begin{equation}
h_{rms}\propto{L_y^{3/2}\over L_\perp^{z(3/2-\zeta)}} ={L_y^{3/2}\over
L_\perp^{1/2+\eta_\kappa z/2}}\equiv L_y \left({L_y\over
L_P(L_\perp)}\right)^{1/2}\;,
\label{L_p}
\end{equation}
acting like a rigid polymer with a {\it polymer} bending rigidity 
\begin{equation}
\kappa_p(L_\perp)\propto L_\perp^{1+z\eta_{\kappa}}\;.
\label{kappa_polymer}
\end{equation}

It is well known\cite{Polymers}, of course, that a linear polymer does
{\em not} have long-ranged orientational order i.e., it has a finite
orientational persistence length $L_P$.  For length smaller than
$L_P(L_\perp)$ we recover the well-known\cite{Polymers} $L_y^{3/2}$
growth of transverse fluctuations.  By equating $h_{rms}$ from
Eq.\ref{L_p} with the length $L_y$ of the tubule, and defining (ribbon
width-dependent persistent length) $L_P(L_\perp)$ to be the value of
$L_y$ at which this equality occurs, we obtain an estimate for the
orientational persistence length $L_P$ of a long, skinny tubule:
\begin{equation}
L_P(L_\perp)\propto L_\perp^{1+\eta_\kappa z}\;.
\label{LP1}
\end{equation}

We see that only {\it very} long, skinny membranes ($L_y>>L_\perp$)
will be orientationally disordered; for any membrane with a reasonable
aspect ratio (i.e., $L_y\sim L_\perp$), $L_y$ is much less than
$L_P(L_\perp)$, and the orientational order of the tubule persists
throughout it. This proves that the tubule phase is stable in the
thermodynamic limit against thermal fluctuations.

Equation \ref{kappa_polymer} indicates that the effective polymer bend
modulus $\kappa_p(L_\perp)$ is ``anomalous'', by which we mean the
fact that $\kappa_p(L_\perp)$, grows as a power of $L_\perp$ greater
(by the ``anomalous dimension'' $\eta_\kappa z$) than $1$ (naively
expected based on dimensional analysis). This together with the
concomitant anomalous dimension of the persistent length
$L_P(L_\perp)$, Eq.\ref{LP1} embodies the phenomenon known as
``anomalous elasticity''.\cite{GP,NP,AL,GD} In addition to
fluctuating membranes, they have consequences for polymers whose
internal structure is that of a long ribbon of dimension
$L_\perp\times L_y$, with $L_y >> L_P(L_\perp)>>L_\perp$. Provided
that $L_\perp$ is large enough that the anomalous elasticity can
manifest itself, the radius of gyration ${\cal R}_G^p$ of this polymer
(which, since $L_y >> L_P$, will be coiled) will, in fact, grow more
rapidly with the {\it transverse} dimension $L_\perp$ of the polymer
than the conventional elastic theory would predict. Specifically, we
expect:
\begin{eqnarray}
{\cal R}_G^p&\approx& L_P(L_\perp)\left({L_y\over
L_P(L_\perp)}\right)^{\nu_p}\;,\nonumber\\ 
&\propto& L_y^{\nu_p}
L_\perp^{(1-\nu_p)(1+\eta_\kappa z)}\;,\label{RGcoil}
\end{eqnarray}
while conventional elastic theory would imply ${\cal R}_G^p\propto
L_\perp^{1-\nu_p}$. 

In addition to this anomalous elasticity in the effective polymer bend
modulus, the fluctuating tubule also displays anomalous elasticity for
stretching the tubule. In particular, experiments that attempt to
measure the stretching modulus $g_y$ of the tubule (defined more
precisely by the renormalized version of Eqs.\ref{uProp1} and
\ref{uProp2}) at wavevector $\bf q$ will produce results that depend
strongly on $\bf q$, {\it even in the limit} ${\bf q}\rightarrow
0$. In particular, this apparent wavevector-dependent stretching
modulus $g_y({\bf q})$ {\it vanishes} as $|{\bf q}|\rightarrow0$,
according to the scaling law
\begin{equation}
g_y({\bf q})=q_y^{\eta_u} S_g(q_y/q_\perp^z)\;,\label{gyAnsatz1}
\end{equation}
where $\eta_u>0$ is another universal exponent, and $S_g(x)$ another
universal scaling function. 

Similarly, the tubule bend modulus $\kappa$ (also defined more
precisely by the renormalized version of
Eqs.\ref{hProp1} and \ref{hProp2}) becomes strongly wavevector dependent
as ${\bf q}\rightarrow 0$, but it {\it diverges} in that limit:
\begin{equation}
\kappa({\bf q})=q_y^{-\eta_\kappa}
S_\kappa(q_y/q_\perp^z)\;,\label{kappaAnsatz1}
\end{equation}
with $\eta_\kappa\geq0$ yet another universal exponent, and
$S_\kappa(x)$ yet another universal scaling function.

\begin{figure}[bth]
\centering
\setlength{\unitlength}{1mm}
\begin{picture}(150,30)(0,0)
\put(-20,-75){\begin{picture}(150,30)(0,0)
\includegraphics{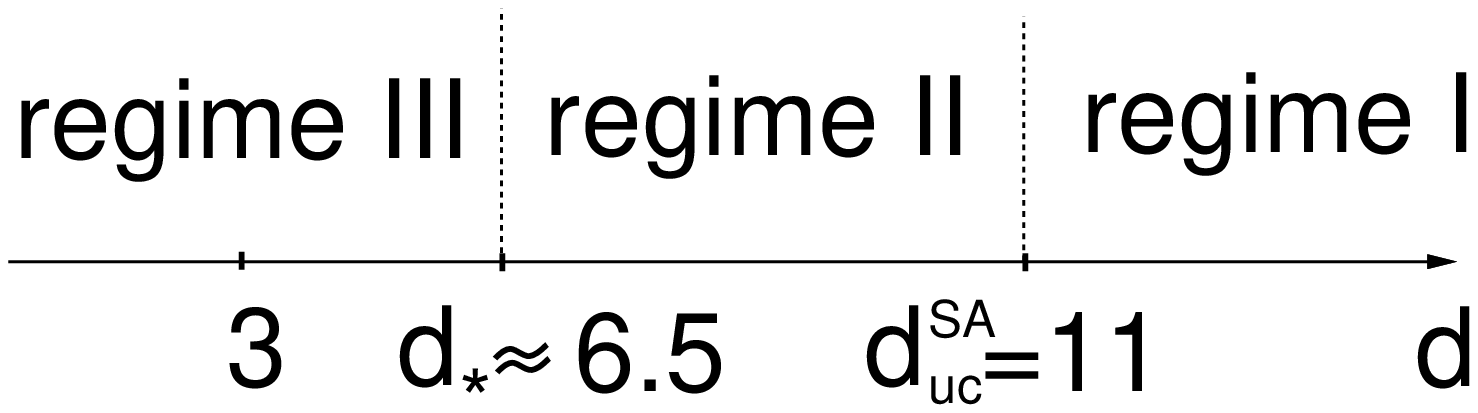}
\end{picture}}
\end{picture}
\caption{Illustration (in $D=2$) of the three regimes of embedding
dimension $d$ with qualitatively and quantitatively different tubule
shape scaling properties.  Our estimates of $d_*\approx 6.5$ place the
physical tubule ($d=3$) deep in regime III; the strict bound $d_*>7/2$
{\it guarantees} this.}
\label{regimes_d}
\end{figure}

The relations Eqs.\ref{RG}-\ref{hrms} summarize all of the scaling
properties in terms of the two universal exponents $\nu$ and $z$ (or
equivalently $\eta_\kappa$). Clearly, we would like to predict their
numerical values. There are three distinct cases to be considered, as
we decrease the embedding dimension $d$ in which the $D=2$-dimensional
membrane fluctuates, as illustrated in Fig.\ref{regimes_d} (the
generalization to arbitrary $D$ is given in the main text).

Regime I:\\
For a phantom membrane, or for a membrane with intrinsic dimension
$D=2$ embedded in a space of dimension $d\geq d_{uc}=11$,
self-avoidance effects can be asymptotically ignored in the tubule
phase, and we predict\cite{RT}
\begin{eqnarray}
\nu&=&{1\over4}\;,\label{nuI}\\
z&=&{1\over2}\;,\label{zI}\\
\eta_\kappa&=&0\;,\label{eta_kappaI}\\
\eta_u&=&1\;,\label{eta_uI}
\end{eqnarray}

Regime II:\\ For a self-avoiding membrane with $d_*<d<d_{uc}=11$ (with
$d_*> 7/2$), we have shown (as we describe in detail in
Sec.\ref{selfavoidance}) that the bending elasticity is {\it not}
anomalous, i.e., $\eta_\kappa=0$, as guaranteed by an {\em exact}
``tubule-gauge'' symmetry (see Sec.\ref{RGscaling_subsubsection}).
This, using Eq.\ref{z} immediately leads to the the exponent relation,
$z=(1+2\nu)/3$, which states that for $d>d_*$ all properties of a
self-avoiding tubule can be expressed in terms of a single radius of
gyration exponent $\nu$. In this range $d_*<d<d_{uc}=11$ of embedding
dimensionality, the exponents $\nu$ and $z$ can be computed in an
$\epsilon=11-d$-expansion. This has been done recently by Bowick and
Guitter (BG)\cite{BG} who have verified the validity of the Ward
identity $z=(1+2\nu)/3$ (for $D=2$) perturbatively, to all orders in
$\epsilon$. Furthermore, for {\em all} embedding dimensions $d>d_*$,
the absence of anomalous bend elasticity (i.e., $\eta_\kappa=0$)
renders the self-avoiding interaction ineffective in stabilizing wild
transverse tubule undulations and for a square membranes,
Eqs.\ref{hsq} and \ref{LP1} show that the $D=2$-dimensional tubule
phase is only {\em marginally} stable. For $D=2$, this
$d_*<d<d_{uc}=11$ regime has:

\begin{eqnarray}
{2\over5}&>&\nu>{1\over4}\;,\label{nuII}\\
z&=&{1\over3}(1+2\nu)\;,\label{zII}\\
\eta_\kappa&=&0\;,\label{eta_kappaII}\\
\eta_u&=&3-{1\over z}\;,\label{eta_uII}
\end{eqnarray}

Regime III:\\ Finally, as we describe in Sec.\ref{selfavoidance}, the
physics of the {\em physical} tubule (i.e., $D=2$-dimensional tubule
embedded in $d=3$ dimensions) is much richer than that for the
embedding dimensions $d>d_*$, where ``tubule-gauge'' symmetry imposes
strict {\it non}renormalization of the tubule bending rigidity
$\kappa$. For $d<d_*$, because of the presence of additional elastic
nonlinearities (which are irrelevant for $d$ near $d_{uc}=11$, but
become strongly relevant for physical dimensionality $d<d_*$), this
$\epsilon$-expansion about $d=d_{uc}=11$ gives no information about
the simultaneous role that the self-avoidance and elastic nonlinearity
play in the physical tubule ($D=2$, $d=3<d_*(D=2)$), where they are
{\em both} important. We find that, as the embedding dimension $d$ is
lowered below $d_*<d_{uc}=11$ ($d_*(D=2)>7/2$), the nonlinear
elasticity becomes relevant, destabilizing the fixed point studied in
Ref.~\onlinecite{BG}, and leading to the breakdown of the
$z=(1+2\nu)/3$ relation (with the amount of breakdown described by a
new anomalous elasticity exponent $\eta_\kappa$). Hence physical
tubules ($D=2$, $d=3$) are described by a new infra-red stable fixed
point, that is non-perturbative in $\epsilon=11-d$, which incorporates
the simultaneous effects of self-avoidance and nonlinear anomalous
elasticity. This new fixed point characterizes the $d<d_*$ regime
(appropriate to a physical tubule) with shape scaling exponents
\begin{eqnarray}
\nu&\geq&{2\over5}\;,\label{nuIIIa}\\
\nu&>&{1\over d-1}\;,\label{nuIIIb}\\
z&=&{1\over3-\eta_\kappa}(1+2\nu)\;,\label{zIII}\\
2\eta_\kappa+\eta_u&=&3-{1\over z}\;,\label{eta_kappaIII}\\
\eta_\kappa\;,\eta_u &>&0\;,\label{eta_uIII}
\end{eqnarray}

We cannot calculate exactly the critical embedding dimension $d_*(D)$
that separates regime II and regime III, but we {\it can} derive a
{\it rigorous} lower bound on it $d_*(2)>7/2$. Thus the physical
tubule, $D=2$, $d=3$ falls in regime III. Our best estimate of
$d_*(2)$ is that it lies between $5$ and $7$.

It should be emphasized that all of the exponents are {\it universal}
in a given embedding dimension $d$. Indeed, for $d_*<d<11$, where {\it
all} of the exponents are determined by the single unknown exponent
$\nu$, there are two different analytical approximations to $\nu$ that
agree to better than $1\%$ for $d>8$, and to better than $10\%$ for
$d$'s greater than the likely values of $d_*$. These analytical methods
are: Flory theory\cite{RT}, which predicts
\begin{equation}
\nu_F={3\over d+1}\;,
\label{nuFlory}
\end{equation}
and the leading order in $\epsilon=11-d$ expansion of Bowick and
Guitter\cite{BG}, which gives,
\begin{equation}
\nu_\epsilon={3\over 4-c\epsilon}-{1\over2}\;,
\label{nu_BG1}
\end{equation}
with 
\begin{equation}
c=0.13125\;,
\label{c_BG2}
\end{equation}
We suspect, based on the experience of comparing polymer exponents
obtained from Flory theory with those obtained from the
$\epsilon$-expansion, that, although BG's results are certainly more
accurate near $d=11$, when the BG and Flory results start to disagree
appreciably (i.e., below $d=7$), the Flory result is probably the more
accurate. Nonetheless, the {\it extremely} close agreement between
these two very different approaches in these high embedding dimensions
increases our faith in both of them.

In fact, as we describe in detail in
Sec.\ref{RGscaling_subsubsection}, for $D=2$-dimensional membrane,
$d_*$ is determined by the condition that $\nu(d)\rightarrow 2/5$ as
$d\rightarrow d_*^+$. Using the Flory result (Eq.\ref{nuFlory}), this
gives $d_*=13/2=6.5$; while using the BG result (Eq.\ref{nu_BG1})
gives $d_*=11-2/(3c)=5.92$.

All of the exponents jump discontinuously (as a function of $d$) at
$d_*$; figure \ref{exponent_fig} shows such a plot, schematically, for
$\nu(d)$ and $\eta_\kappa(d)$.

For a physical tubule, Flory theory, Eq.\ref{nuFlory}, implies
\begin{equation}
\nu_{F}(D=2,d=3)=3/4\;,
\label{nu_Flory_intro}
\end{equation}
in contrast to the BG result Eq.\ref{nu_BG1}, which implies
$\nu_{\epsilon}(D=2,d=3)=0.517$. What is the correct value of $\nu$ in $d=3$?
As discussed above, our experience with polymers suggests that Flory
theory is more reliable\cite{Flory_accurate} than the
$\epsilon$-expansion when both are pushed well below the upper
critical dimension. One might be concerned that this ceases to be true
for tubules, due to the discontinuous behavior of all of the exponents
at $d_*$, but we will present arguments later that suggest that this
is not the case, and that Flory theory is probably quite accurate in
the physical case of $d=3$.

It is widely\cite{Grest91,Plischke}, though not
universally,\cite{Baumgartner,CommentSA} believed that self-avoidance
destroys the crumpled phase. What is definitely known is that the
crumpled phase has only been seen in simulations of phantom membranes
and in more recent simulations by Baumgartner\cite{Baumgartner} of a
self-avoiding plaquette membrane model. It is therefore reasonable to
ask whether our tubule phase will suffer the same fate. We think not,
for the following reasons:

\begin{enumerate}
\item It is clear that self-avoidance, though a relevant perturbation
(in physical embedding dimension $d<d_{uc}=11$) has far less effect on
the tubule than the crumpled phase, since points on the membrane widely
separated in the y-direction never bump into each other in the tubule
phase, but do in the crumpled phase.
\item The {\em analytic} argument that self-avoidance destroys the
crumpled phase is based on the Gaussian variational (GV)
approximation\cite{Goulian91,Ledoussal92}, which predicts that the
radius of gyration exponent $\nu_{GV}^{crumpled}=4/d$, which implies
that $\nu\geq 1$ for $d\leq 4$, and hence that the membrane is
extended (i.e. flat) for those dimensions (which, of course, include
the physical case of $d=3$). We find that the same Gaussian
variational approximation leads to the same conclusion for the tubule
phase. Our result for $D=2$ is
\begin{equation}
\nu_{GV}^{tubule}={7\over 3d-5}\;,
\label{nuGV}
\end{equation}
and implies $\nu_{GV}^{tubule}\geq 1$ for $d\leq 4$, and hence an
instability of the self-avoiding tubule to an extended (i.e. flat)
membrane in physical dimensions.

We are not, however, overly concerned by this result, for a number of
reasons:
\begin{enumerate}
\item
The Gaussian variational approximation is known to be far from
trustworthy. For example, it predicts $\nu=2/d$ for linear polymers,
which not only is less accurate for {\em all} $d$ between $1$ and $4$
than the Flory result $\nu=3/(d+2)$, but also incorrectly predicts
that the lower critical dimension $d_{lc}$ below which linear polymers
are always extended is $d_{lc}=2$, whereas, in fact, it is known
exactly that $d_{lc}=1$, a result that is also predicted exactly by
the Flory theory. Thus, the Gaussian variational approximation is {\em
very} unreliable in predicting the lower critical dimension of a
crumpled object.
\item
There is a good reason to believe it is equally unreliable for our
problem as well. If we compare the Flory prediction for $\nu$ with the
$\epsilon$-expansion calculation of Ref.~\onlinecite{BG} (which is
asymptotically exact in $d\rightarrow 11$), in, e.g., $d=8$, we find
they differ less than $1/3$ of $1\%$: $\nu_{\epsilon}=0.332$\ \ 
\cite{BG}, $\nu_{Flory}=1/3$\ \ \cite{RT}; while the Gaussian
variational result $\nu_{GV}^{tubule}=7/19=0.3684$ is nearly $40$
times as far off $\nu_{\epsilon}$ as the Flory result. This strongly
suggests that both Flory theory and the $\epsilon$-expansion are more
reliable than the Gaussian variational approximation, and both of them
predict $\nu$ substantially $<1$ in $d=3$: $\nu_F=3/4$\ \ 
\cite{RT}, $\nu_{\epsilon}=0.517$\ \ \cite{BG}.
\item
Finally, on more general grounds, while the Gaussian variational
method can be quite useful, only some of its results can be trusted.
Certainly it is likely that the {\it trends} of, e.g., exponents with
dimensionality $d$ and $D$, are captured correctly by this theory. The
very existence of the crumpled phase relies on the precise value of
$\nu(d)$ (it disappears if the $d<d_{lc}$, with $d_{lc}$ defined by
$\nu(d_{lc})=1$). However, as with any approximate method, especially
with uncontrolled approximations such as the Gaussian method, there is
little credibility in the actual {\it values} of the
exponents. Furthermore, the Gaussian variational approximation is very
closely related to a large expansion in $1/d$ about the embedding
dimension $d\rightarrow\infty$ limit.\cite{magnets} It is therefore
intrinsically untrustworthy and ad hoc for small values of $d$ at
which one is assessing the stability of the tubule (or crumpled)
phase, which very delicately and sensitively depends on the precise
value of $\nu$ at small $d$.
\end{enumerate}
\end{enumerate}

In the remainder of this paper we present the details of our
calculations. In Sec.\ref{model} we introduce the
Landau-Ginzburg-Wilson free energy for our generalized model of
anisotropic polymerized membranes. In Sec.\ref{mft} we will first
solve this model in mean field theory.  From this solution we obtain
the phase diagram for anisotropic polymerized membranes, and identify
and characterize the new tubule phase as well as the previously
studied crumpled and flat phases.  In Sec.\ref{flat_and_crumpled} we
show that the scaling properties of the flat and crumpled phases are
unaffected by the anisotropy. In Secs.\ref{tubuleF} and
\ref{selfavoidance} we then consider the effects of both thermal
fluctuations and self-avoidance on the tubule phase. We treat this
problem using Flory theory, renormalization group and Gaussian
variational methods. We calculate the upper critical embedding and
intrinsic dimensions for both effects, and thereby show that both are
relevant for the physical case of two-dimensional membranes embedded
in three dimensions. We also show that, although there is {\em no}
anomalous elasticity for the bend modulus $\kappa$ along the tubule
near $d=d_{uc}=11$ (due to aforementioned ``tubule gauge'' symmetry),
such anomaly must set in for embedding dimensions $d < d_*$, with $d_*
> 7/2$. When this happens, the fixed point (perturbative around
$d=11$)\cite{BG} which describes a self-avoiding ({\it bend}
elastically {\em non}-anomalous) tubule, becomes unstable, and a new
fixed point controls the tubule phase. We derive new {\it exact}
relations Eqs.\ref{z_nu_new1} and \ref{z_nu_new2} between $\nu$ and
$z$, which involve anomalous elasticity exponent $\eta_\kappa$ (or
$\eta_u$, related to it) and are appropriate for a physical (with
anharmonic elasticity) tubule, described by this new fixed point.
We then use the Flory\cite{RT} and extrapolated
$\epsilon=11-d$--expansion\cite{BG} results for $\nu$ in this relation
to determine $z$ and all other tubule shape exponents in terms of two
constants that, unfortunately, we were not able to compute accurately.

In Section \ref{tubuleF} we also derive the scaling results
Eq.\ref{RG} and Eq.\ref{hrms} for $R_G$ and $h_{rms}$, and for the
anomalous elastic theory as well.

In Section \ref{transition_section} we use the renormalization group
to analyze the crumpled-to-tubule transition. We then construct a
scaling theory of the crumpled-to-tubule and tubule-to-flat
transitions, and compute within Flory theory the critical exponents
for these transitions.

In Section \ref{summary_section} we summarize, conclude, and make some
suggestions for further analytic, numerical, and experimental work.

\section{Model}
\label{model}
Our model for anisotropic membranes is a generalization of the
isotropic model considered in Ref.~\onlinecite{PKN}. As there, we
characterize the configuration of the membrane by giving the position
${\vec r}({\bf x})$ in the $d$-dimensional embedding space, of the
point in the membrane labeled by a $D$-dimensional internal
co-ordinate $\bf x$.  In the physical case, $d=3$ and $D=2$, of
course.  Throughout the remainder of this paper, we will distinguish
between $D$-dimensional "intrinsic" vectors and $d$-dimensional
"extrinsic" vectors by using boldface type for the former, and vector
arrows over the latter.

We now construct the Landau-Ginzburg-Wilson free energy $F$ for this
system, by expanding $F$ to leading order in powers of ${\vec r}({\bf
x})$ and its gradients with respect to internal space $\bf x$, keeping
only those terms consistent with the symmetries of the problem.  These
symmetries are global translation invariance ${\vec r}({\bf
x})\rightarrow{\vec r}({\bf x})+\vec{r}_o$, and global rotational
invariance ${\vec r}({\bf x})\rightarrow\vec{\vec{M}}\cdot{\vec
r}({\bf x})$, where $\vec{r}_o$ and $\vec{\vec{M}}$ are a constant
(i.e. $\bf x$-independent) vector and a constant rotation matrix,
respectively.  Global translational invariance requires that $F$ be
expanded only in powers of {\it gradients} with respect to $\bf x$.
We will furthermore take the membrane to be isotropic in the $D-1$
membrane directions (hereafter denoted by $\bf{x}_\perp$) orthogonal
to one special direction (which we call $y$). Since the physical case
is $D=2$, this specialization is innocuous.

The most general model consistent with all of these symmetries,
neglecting irrelevant terms, is,
\begin{eqnarray}
&&F[{\vec r}({\bf x})] = {1\over2}\int d^{D-1}x_\perp dy
\bigg[\kappa_\perp\left(\partial_\perp^2\vec{r}\right)^2 +
\kappa_y\left(\partial_y^2\vec{r}\right)^2 \nonumber\\
&+&
\kappa_{\perp y}\partial_y^2\vec{r}\cdot\partial_\perp^2\vec{r} +
t_\perp\left(\partial^\perp_\alpha\vec{r}\right)^2 +
t_y\left(\partial_y\vec{r}\right)^2 \nonumber\\
&+&
{u_{\perp\perp}\over2}\left(\partial^\perp_\alpha\vec{r}
\cdot\partial^\perp_{\beta}\vec{r}\right)^2 +
{u_{y y}\over2}\left(\partial_y\vec{r}\cdot\partial_y\vec{r}\right)^2 +
u_{\perp y}\left(\partial^\perp_\alpha\vec{r}
\cdot\partial_y\vec{r}\right)^2 \nonumber\\
&+&
{v_{\perp\perp}\over2}\left(\partial^\perp_\alpha\vec{r}
\cdot\partial^\perp_\alpha\vec{r}\right)^2 +
v_{\perp y}\left(\partial^\perp_\alpha\vec{r}\right)^2
\left(\partial_y\vec{r}\right)^2\bigg] \nonumber\\
&+&
{b\over2}\int d^{D}x \int d^{D}x'
\hspace{0.2cm}\delta^{(d)}\big({\vec r}({\bf x})-{\vec r}({\bf x'})\big)\;,
\label{Fc}
\end{eqnarray}
where the $\kappa$'s, $t$'s, $u$'s, $v$'s are elastic constants.  The
first three terms in $F$ (the $\kappa$ terms) represent the
anisotropic bending energy of the membrane.  The elastic constants
$t_\perp$ and $t_y$ are the most strongly temperature dependent
parameters in the model, changing sign from large, positive values at
high temperatures to negative values at low temperatures.  Their
positivity at high temperatures reflects the membrane's entropic
preference for crumpling.  To see this, note that this crumpled state
is one in which all the particles in the membrane attempt to cram
themselves into the same point $\vec{r}$; in this state, the gradients
with respect to the internal space $\partial^\perp_\alpha\vec{r}$ and
$\partial_y\vec{r}$ seek to minimize themselves, which is clearly
favorable when $t_\perp$, $t_y > 0$.  However, when either of these
becomes negative, it becomes favorable for the membrane to flatten
(i.e., extend) in the associated direction, as we shall show in a
moment.  The $u$ and $v$ quartic terms are higher order elastic
constants needed to stabilize the membrane when one or both of the
first order elastic constants $t_\perp$, $t_y$ become negative.
Stability requires that 
\begin{eqnarray}
u_{\perp\perp}'&>&0\;,\\
u_{y y}&>&0\;,\\
v_{\perp y}&>& -\sqrt{u_{\perp\perp}'u_{y y}}\;,
\end{eqnarray}
where 
\begin{equation}
u_{\perp\perp}'\equiv v_{\perp\perp}+u_{\perp\perp}/(D-1)\;.
\end{equation}
The final, $b$ term in Eq.\ref{Fc} represents the self-avoidance of
the membranes; i.e., its steric or excluded volume interaction.

Equation \ref{Fc} reduces to the model for isotropic membranes
considered in Ref.~\onlinecite{PKN} when $t_\perp=t_y$,
$\kappa_{\perp\perp}=\kappa_y$, $\kappa_{\perp y}=0$, $u_{y
y}=4(\tilde{v}+u)$, $u_{\perp\perp}=u_{\perp y}=4 u$, and
$v_{\perp\perp}=v_{\perp y}=4\tilde{v}$.

\section{Mean Field Theory}
\label{mft}

We begin our analysis of this model by obtaining its mean field
phase diagram, at first neglecting the self-avoidance interaction. Later,
we will consider both the effects of fluctuations
and self-avoidance.

In mean-field theory, we seek a configuration ${\vec r}({\bf x})$ that
minimizes the free energy Eq.\ref{Fc} (without the self-avoidance
term).  The curvature energies
$\kappa_\perp\left(\partial_\perp^2\vec{r}\right)^2$ and
$\kappa_y\left(\partial_y^2\vec{r}\right)^2$ are clearly minimized
when ${\vec r}({\bf x})$ is linear in $\bf x$.  We will therefore seek
minima of $F$ of the form
\begin{equation}
{\vec r}({\bf x})=\left(\zeta_\perp {{\bf x}_\perp}, \zeta_y y, 0, 0,
\ldots , 0\right)\;.
\label{ansatz}
\end{equation}
Obviously, uniform rotations ${\vec r}({\bf
x})\rightarrow\vec{\vec{M}}\cdot{\vec r}({\bf x})$, of any such
minimum, with $\vec{\vec{M}}$ a constant rotation matrix, will also be
minima.  A continuous degenerate set of minima is thereby obtained, as
usual for a system with a broken continuous symmetry. Uniform
translations of the entire membrane are also allowed, of course.

Inserting Eq.\ref{ansatz} into Eq.\ref{Fc}, and for now neglecting the
self-avoidance term, we obtain the mean-field free energy for
anisotropic membranes
\begin{eqnarray}
&&F={1\over2} L_\perp^{D-1} L_y\bigg[t_y\zeta_y^2+
t_\perp(D-1)\zeta_\perp^2 \nonumber\\
&+&
{1\over2}u'_{\perp\perp}(D-1)^2\zeta_\perp^4 +
{1\over2}u_{y y}\zeta_y^4 + v_{\perp y}(D-1)\zeta_\perp^2 \zeta_y^2\bigg]\;,
\label{Fzeta}
\end{eqnarray}
where  $L_\perp$ and  $L_y$  are the linear dimensions of the flattened
membrane in the $\perp$ and $y$ directions, respectively.

This mean field theory is precisely that studied long ago by Fisher et
al.\cite{NFAM} for a completely different (magnetic) problem.
Minimizing the free energy over $\zeta_\perp$ and $\zeta_y$ yields two
possible phase diagram topologies, depending on whether
$u_{\perp\perp}' u_{y y} > v_{\perp y}^2$ or $u_{\perp\perp}' u_{y y}
< v_{\perp y}^2$ .

For $u_{\perp\perp}' u_{y y} > v_{\perp y}^2$, we obtain the phase
diagram in Fig.\ref{phase_diagram1}.  Both $\zeta_\perp$ and $\zeta_y$
vanish for $t_\perp$, $t_y>0$. This is the crumpled phase: the entire
membrane, in mean-field theory, collapses into the origin,
$\zeta_\perp=\zeta_y=0$ i.e., ${\vec r}({\bf x})=0$ for all ${\bf x}$.

In the regime between the positive $t_\perp$-axis (i.e., the locus
$t_y = 0$ and $t_\perp > 0$) and the $t_y < 0$ part of the $t_y=(u_{y
y}/v_{\perp y})t_\perp$ line, lies our new y-tubule phase,
characterized by $\zeta_\perp=0$ and $\zeta_y=\sqrt{|t_y|/u_{y y}}>0$:
the membrane is extended in the y-direction but crumpled in all $D-1$
$\perp$-directions.

The $\perp$-tubule phase is the analogous phase with the $y$ and
$\perp$ directions reversed, $\zeta_y=0$ and
$\zeta_\perp=\sqrt{|t_\perp|/u_{\perp\perp}}>0$ (obviously a
symmetrical reversal for the physical case of $D=2$), and lies between
the $t_\perp < 0$ segment of the line $t_y=(v_{\perp
y}/u'_{\perp\perp}) t_\perp$ and the positive $t_y$ axis. Finally, the
flat phase, characterized by both 
\begin{eqnarray}
\zeta_\perp&=&\left[(|t_\perp| u_{y y}-|t_y| v_{\perp y})
/(u_{\perp\perp}' u_{y y} - v_{\perp y}^2)\right]^{1/2} > 0\;,
\label{zeta_perp}\\
\zeta_y&=&\left[(|t_y| u_{\perp\perp}-|t_\perp| v_{\perp y})
/(u_{\perp\perp}' u_{y y}-v_{\perp y}^2)\right]^{1/2} > 0\;,
\label{zeta_y}
\end{eqnarray}
lies between the $t_\perp < 0$ segment of the line $t_y=(u_{y
y}/v_{\perp y})t_\perp$ and the $t_y < 0$ segment of the line
$t_y=(v_{\perp y}/u'_{\perp\perp}) t_\perp$.

\begin{figure}[bth]
\centering
\setlength{\unitlength}{1mm}
\begin{picture}(150,60)(0,0)
\put(-20,-72){\begin{picture}(150,0)(0,0)
\includegraphics{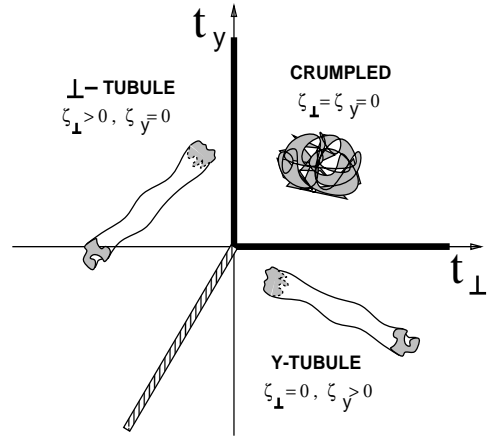}
\end{picture}}
\end{picture}
\caption{Phase diagram for tethered membranes showing our new tubule phase,
for the range of elastic parameters when the intermediate flat phase
disappears. A first-order phase transition separates $y$- and
$\perp$-tubule phases.}
\label{phase_diagram2}
\end{figure}

For $u_{\perp\perp}' u_{y y} < v_{\perp y}^2$, the flat phase
disappears, and is replaced by a direct first-order transition from
$\perp$-tubule to $y$-tubule along the locus $t_y=(v_{\perp
y}/u'_{\perp\perp}) t_\perp$ (see Fig.\ref{phase_diagram2}) .
The boundaries between the tubule and the crumpled phases remain the
positive $t_y$ and $t_\perp$ axes, as for $u_{\perp\perp}' u_{y y} >
v_{\perp y}^2$ case.

Note that a direct crumpling transition (i.e. a direct transition
between the crumpled and flat phase) is very non-generic in this
picture: only experimental loci that pass from $t_y$, $t_\perp>0$
through the origin (locus P$_2$ in Fig.\ref{phase_diagram1}) can
experience such a transition.  This transition is, in fact,
tetra-critical in this picture.

This does not, however, imply that direct crumpling transitions are
non-generic.  Many membranes will be perfectly isotropic, by virtue of
being formed under conditions of unbroken rotational symmetry (e.g.,
randomly polymerized membranes).  As discussed earlier, this set of
membranes, which is undoubtedly of finite measure, necessarily lies on
the special isotropic subspace of the full parameter space of the
model defined by Eq.\ref{Fc} specified by $t_\perp=t_y$,
$\kappa_{\perp\perp}=\kappa_y$, $\kappa_{\perp y}=0$, $u_{y
y}=4(\tilde{v}+u)$, $u_{\perp\perp}=u_{\perp y}=4 u$, and
$v_{\perp\perp}=v_{\perp y}=4\tilde{v}$.  The values of the quartic
couplings then satisfy $u_{\perp\perp}' u_{y y} > v_{\perp y}^2$ (for
$u$, $\tilde v > 0$), and hence the topology of the phase diagram is
Figure \ref{phase_diagram1}.  The boundaries of the flat phase for those
isotropic values of the quartic couplings become
$t_y=t_\perp\big(1+u/{\tilde v}/(D-1)\big)$ and $t_y=\left(1+u/{\tilde
v}\right)t_\perp$, respectively.  For $u$ and $v$ both positive (as
required by stability), the slopes of these lines are less than and
greater than $1$, respectively; the isotropic locus $t_y=t_\perp$
therefore lies between the two (i.e., in the flat phase), and hence,
that model {\it does} undergo a direct flat to crumpled transition.

Membranes with {\it any} intrinsic broken orientational symmetry
(e.g., in-plane tilt order \cite{LRunpublished}, which is quite
common\cite{Sachmann}), will generically have $t_y\neq t_\perp$.
Furthermore, they will not generically have both $t_\perp$ and $t_y$
vanish at the same temperature.  A generic locus through the phase
diagram in Fig.\ref{phase_diagram1} will be like locus P$_1$, and will
necessarily have one of the tubule phases intervening between the flat
and crumpled phases.  Our new tubule phase is not only generically
possible, but actually unavoidable, in membranes with any type or
amount of {\it intrinsic anisotropy}.

\section{Fluctuations and Self-avoidance in the Flat and Crumpled
Phases}
\label{flat_and_crumpled}
In this section, we show that both the flat and the crumpled phases of
anisotropic membranes are identical in their scaling properties, at
sufficiently long length scales, to the eponymous phases of isotropic
membranes.

Consider first the flat phase.  We can include fluctuations about the
mean-field solution by considering small deviations from the solution
in Eq.\ref{ansatz}
\begin{equation}
{\vec r}({\bf x})=\left(\zeta_\perp{\bf x_\perp}+{u_\perp}({\bf x}), \zeta_
y y +u_y({\bf x}), {\vec h}({\bf x})\right)\;,
\label{fluct}
\end{equation}
Inserting this into our initial free energy, Eq.\ref{Fc}, with
$t_\perp$ and $t_y$ both in the range in which the flat phase is
stable, we obtain the uniaxial elastic energy of
Ref.~\onlinecite{TonerAnisotropy}. As shown in that reference, fluctuation
effects in turn renormalize the anisotropic elastic energy into the
{\it isotropic} membrane elastic energy considered by
Refs.~\onlinecite{NP,AL,GD}.  In the flat phase, and at sufficiently long
scales, the anisotropic membranes therefore behave exactly like
isotropic membranes. This in particular implies that the flat phase of
anisotropic membranes is stable against thermal fluctuations. As in
isotropic membranes, this is due to the fact that these very thermal
fluctuations drive the bend modulus $\kappa$ to infinity at long
wavelengths\cite{NP,AL,GD}.

Specifically, $\kappa$ becomes wavevector dependent, and $\kappa(\bf
q)$ diverges like $q^{-\eta_\kappa}$ as $q\rightarrow 0$.  In the flat
phase the standard Lam\'e coefficients $\mu$ and $\lambda$\ \
\cite{LandauLifshitz} are also infinitely renormalized and become
wavevector dependent, vanishing in the $q\rightarrow0$ limit as
$\mu(q)\sim\lambda(q)\sim q^{\eta_u}$; the values of $\eta_\kappa$ and
$\eta_u$ in the flat phase differ from those in the tubule phase, as
does their physical interpretation. The flat phase is furthermore
novel in that it is characterized by a universal {\em negative}
Poisson ratio\cite{AL,LR} which for $D=2$ is defined as the long
wavelength limit $q\rightarrow0$ of
$\sigma=\lambda(q)/(2\mu(q)+\lambda(q))$. The transverse undulations
in the flat phase, i.e. the membrane roughness $h_{rms}$ scales with
the internal size of the membrane as $h_{rms}\sim L^\zeta$, with
$\zeta=(4-D-\eta_\kappa)/2$, exactly. Furthermore, an underlying
rotational invariance imposes an exact Ward identity between
$\eta_\kappa$ and $\eta_u$, $\eta_u+2\eta_\kappa=4-D$, leaving only a
single nontrivial independent exponent characterizing the properties
of the flat phase of even anisotropic membranes.  The best estimate
for $\eta_\kappa$ in the physical case of a two-dimensional membrane
($D=2$), embedded in a $d=3$-dimensional space comes from the
self-consistent screening approximation (SCSA) of Le Doussal and
Radzihovsky\cite{LR}, who find $\eta_\kappa=4/(1+\sqrt{15})\approx
0.82$.  The exponent relations above then predict $\eta_u=0.36$ and
$\zeta=0.59$. These exponents, together with the negative Poisson
ratio predictions of Le Doussal and Radzihovsky of $\sigma=-1/3$\ \
\cite{LR} have been recently spectacularly verified to high precision
in very large scale simulations (largest to date) by Falcioni, Bowick,
Guitter and Thorleifsson\cite{BowickPoisson}.

The root-mean-square (rms) thermal fluctuation $\langle\left|{\hat
n}({\bf x})-{\hat z}\right|^2\rangle\equiv \langle|{\delta\vec{
n}}({\bf x})|^2\rangle$ of the local membrane normal ${\hat
n}({\bf x})$ about its mean value (here taken to be $\hat z$) is
\begin{eqnarray}
\langle|\delta{\vec n}({\bf x})|^2\rangle
&=&
\langle|{\bbox{\nabla}}{\vec h}({\bf x})|^2\rangle\;,\nonumber\\
&=&
\int d^D q\;q^2\langle|{\vec h}
({\bf q})|^2\rangle\;,\nonumber\\
&\propto&\int {d^D q\over\kappa(q) q^2}
\propto\int {d^D q\over q^{2-\eta_\kappa}}\;,\nonumber\\
&\propto&L^{2-\eta_\kappa-D}\;,
\label{nfluct}
\end{eqnarray}
where we have imposed an infra-red cutoff $q > L^{-1}$, on the integral
over wavevectors, $L$ being the smaller of the intrinsic linear dimensions
$L_\perp$, $L_y$ of the flattened membrane.  These fluctuations are
finite as $L\rightarrow\infty$, when $2-\eta_\kappa-D<0$. In the physical case
$D=2$, this condition is always satisfied since $\eta_\kappa>0$.  Thus, membrane
orientational fluctuations remain bounded, and the flat phase is
stable against thermal fluctuations, for the physical case $D = 2$.
Indeed, the SCSA predicts that they remain bounded down to the lower
critical dimension $D = \sqrt2$.\ \ \cite{LR}

Note that this stability of the flat phase depends crucially on the
anomalous elasticity, i.e., the divergence of $\kappa(q)$ as
$q\rightarrow 0$.  In the absence of this effect, which would
correspond to $\eta_\kappa=0$, the integral over wavevector in
Eq.\ref{nfluct} would diverge logarithmically for $D=2$, describing
divergent orientational fluctuations leading to an instability of the
flat phase at any non-zero temperature.  Hence, the flat phase owes
its stability to the anomalous elasticity (i.e., the fact that
$\eta_\kappa>0$). In contrast, as we shall show in a moment, the tubule phase
is marginally {\it stable} against thermal fluctuations, even in the absence
of anomalous elastic effects.  Such effects are, nonetheless, actually
present for self-avoiding tubules, but they are not essential to the
stability of the phase.

Because of this persistent long-ranged orientational order (i.e.,
because the membrane is flat), widely intrinsically separated parts of
the membranes (i.e., points ${\bf x}$ and ${\bf x'}$, with $|{\bf
x}-{\bf x'}|$ large) do not bump into each other (i.e., never have
${\vec r}({\bf x})={\vec r}({\bf x'})$); hence, the self-avoidance
interaction in Eq.\ref{Fc} is irrelevant in the flat
phase.

That the crumpled phase of anisotropic membranes is identical to that
of isotropic membranes is even easier to see.  When both $t_\perp$ and
$t_y$ are positive, all of the other local terms in Eq.\ref{Fc}, i.e.,
the $\kappa$, $u$, and $v$-terms, are irrelevant at long wavelengths
(since they all involve more derivatives than the $t$-terms).  Once
these irrelevant terms are neglected, a simple change of variables
${\bf x}_\perp={\bf x}'\sqrt{t_\perp/t_y}$ makes the remaining energy
isotropic.  Thus, the entire crumpled phase is identical in its
scaling properties to that of isotropic membranes.

In particular, the membrane in this phase has a radius of gyration
$R_G(L)$ which scales with membrane linear dimension  $L$  like
$L^\nu$, with $\nu=(D+2)/(d+2)$ in Flory theory, and very similar
values predicted by $\epsilon$-expansion techniques\cite{KN_sa,KKN,AL_sa}.

\section{Fluctuations in phantom tubules}
\label{tubuleF}
In this section, we ignore self-avoidance (i.e., treat ``phantom''
membranes), and consider the effects of fluctuations on phantom
tubules.  We will show that these fluctuations do not destroy the
tubule phase, or change the topology of the phase diagram.  The
detailed properties of the tubule phase are, however, modified by the
fluctuations.

Let us consider the $y$-tubule phase (i.e., the tubule phase with the
tubule axis along the y-axis). To treat fluctuations, we perturb
around the mean-field solution ${\vec r}_o({\bf x})=\zeta_y\big(y,
\vec 0\big)$ by writing
\begin{equation}
{\vec r}({\bf x})=
\big(\zeta_y y + u({\bf x}),{\vec h}({\bf x})\big)\;,
\label{tubulecoordinate}
\end{equation}
where ${\vec h}({\bf x})$ is a $d-1$-component vector orthogonal to
the tubule's axis, which we take to be oriented along the $y$-axis.
The average extension factor $\zeta_y$ is near but not exactly equal
to its mean-field value, because fluctuations will change it.  Rather,
we will choose $\zeta_y$ so that all linear terms in ${\vec h}({\bf
x})$ and $u({\bf x})$ in the resultant elastic free energy for these
variables are exactly cancelled, in the long wavelength limit, by
their fluctuation renormalizations. This criterion guarantees that
${\vec h}({\bf x})$ and $u({\bf x})$ represent fluctuations around the
true ground state of $F$.  Precisely analogous choices have been used
in the study of bulk smectic A elasticity\cite{GP}, and the flat-phase
elasticity of isotropic membranes\cite{NP,AL,GD}.

Inserting the decomposition Eq.\ref{tubulecoordinate} into the free
energy Eq.\ref{Fc}, neglecting irrelevant terms, and, for the moment
ignoring the self-avoidance interaction, gives, after some algebra, the
elastic free energy $F_{tot}=F_{mft}+F_{el}$, where $F_{mft}$ is
simply the mean-field free energy for the tubule phase
\begin{equation}
F_{mft}={1\over2} L_\perp^{D-1} L_y\big[t_y\zeta_y^2 +
{1\over4}u_{y y}\zeta_y^4\big]\;,
\label{mftFtubule}
\end{equation}
and $F_{el}[u({\bf x}),{\vec h}({\bf x})]$ is the fluctuating elastic
free energy part
\begin{eqnarray}
F_{el}&=&{1\over2}\int d^{D-1}x_\perp dy
\bigg[\gamma\bigg(\partial_y u+{1\over2}(\partial_y{\vec h})^2 +
{1\over2}(\partial_y u)^2\bigg) \nonumber\\
&+&\kappa(\partial^2_y{\vec h})^2 + t(\partial^\perp_\alpha \vec h)^2
+ g_\perp(\partial^\perp_\alpha u)^2\nonumber\\
&+&
g_y\bigg(\partial_y u+{1\over2}(\partial_y{\vec h})^2 +
{1\over2}(\partial_y u)^2\bigg)^2\bigg]\;,
\label{elFtubule}
\end{eqnarray}
where $\kappa\equiv\kappa_y$, $t\equiv t_\perp +v_{\perp y}\zeta_y^2$,
$g_y\equiv u_{yy}\zeta_y^2/2$, $g_\perp\equiv t + u_{\perp
y}\zeta_y^2$, and $\gamma=t_y + u_{y y}\zeta_y^2$ are constant
coefficients.  Note first that the coefficient $\gamma$ of the linear
terms in $F_{el}$ is also the coefficient of the $(\partial_y{\vec
h})^2$ term. This is a consequence of the rotation invariance of the
original free energy Eq.\ref{Fc}, which leads to the existence of the
Goldstone mode $\partial_y\vec h$. The combination $E(u,{\vec
h})\equiv \partial_y u+{1\over2}(\partial_y{\vec h})^2 +
{1\over2}(\partial_y u)^2$ is the only combination of first
$y$-derivatives of $u$ and $\vec h$ that is invariant under global
rotations of the tubule.  It is analogous to the non-linear strain
tensor of conventional elasticity theory.\cite{LandauLifshitz}. On
these general symmetry grounds, therefore, the free energy can only
depend on $\partial_y u$ and $\partial_y{\vec h}$ through powers of
$E(u,{\vec h})$, and this property must be preserved upon
renormalization.  This has two important consequences: the first is
that, since, as discussed earlier, the coefficient of this linear term
will be chosen to vanish upon renormalization via a judicious choice
of the stretching factor $\zeta_y$, the coefficient of
$(\partial_y{\vec h})^2$ will likewise vanish.\cite{RTfuture}
This means that the $y$-direction becomes a ``soft'' direction for
fluctuations of ${\vec h}$ in the tubule phase.  We can trace this
softness back to the spontaneously broken rotational symmetry of the
tubule state. It is precisely analogous to the softness of height
fluctuations in the flat phase of isotropic membranes, manifested by
the absence of $(\partial_{x}{\vec h})^2$, $(\partial_{y}{\vec h})^2$
terms in the elastic free energy of the flat phase, analogous to
Eq.\ref{elFtubule} (when $\gamma$ is tuned to $0$).

The second important consequence is that the ratios of the
coefficients of the quadratic $(\partial_y u)^2$ and the anharmonic
$\partial_y u (\partial_y {\vec h})^2$ and $(\partial_y {\vec h})^4$
terms in $F_{el}$ must always be {\it exactly} $4:4:1$, since they
must appear together as a result of expanding $(\partial_y
u+{1\over2}(\partial_y{\vec h})^2 +{1\over2}(\partial_y u)^2)^2$.  We
will show in a few moments that, for this special value of these
ratios, the long-wavelength anomalous elastic behavior of the
``phantom'' tubule phase can be calculated {\it exactly}.

Recognizing that $\gamma$ vanishes after renormalization, we can now
calculate the propagators (i.e., the harmonic approximation to the
Fourier transformed correlation functions) by setting $\gamma = 0$ in
Eq.\ref{elFtubule}.   We thereby obtain
\begin{eqnarray}
\langle h_i({\bf q}) h_j(-{\bf q})\rangle
&=&
k_B T \delta^\perp_{i j} G_h({\bf q})\;,\label{hProp1}\\
\langle u({\bf q}) u(-{\bf q})\rangle &=& k_B T
G_u({\bf q})\;,\label{uProp1}
\end{eqnarray}
where
\begin{eqnarray}
G_h^{-1}({\bf q})&=&t q_\perp^2 + \kappa q_y^4\;,\label{hProp2}\\
G_u^{-1}({\bf q})&=&g_\perp q_\perp^2 + g_y q_y^2\;,\label{uProp2}
\end{eqnarray}
and $\delta_{ij}^\perp$ is a Kronecker delta when both indices $i$
and $j$ $\neq y$, and is zero if either $i$
or $j$ $=y$.

Inspection of the propagators $G_h$ and $G_u$ reveals that the $\vec
h$-fluctuations are much larger than the $u$-fluctuations for $|{\bf
q_\perp}|\approx q_y^2$, and that it is precisely this regime of
wavevectors that dominates the fluctuations.  Thus, in power counting
to determine the relevance or irrelevance of various operators, we
must count each power of $|{\bf q_\perp}|$ as {\it two} powers of
$q_y$. It is this power counting that leads to the identification of
the terms explicitly displayed in Eq.\ref{elFtubule} as the most
relevant ones.

Calculating the root-mean-squared (rms) real space positional fluctuations
$\langle|\vec{h}({\bf x})|^2\rangle$
in the harmonic approximation by integrating the propagators over all
wavevectors, we find
\begin{eqnarray}
\langle|\vec{h}({\bf x})|^2\rangle
&\propto&
\int_{q_\perp>L_\perp^{-1}}{d^{D-1} q_\perp d q_y\over(2\pi)^D}
{1\over t q_\perp^2 + \kappa q_y^4}\;,\nonumber\\
&\propto&
\int_{q_\perp>L_\perp^{-1}}{d^{D-1} q_\perp\over  q_\perp^{3/2}}
\propto L_\perp^{5/2-D}\;,
\label{hfluct}
\end{eqnarray}
where we have introduced an infra-red cutoff $|{\bf q_\perp}| >
L_\perp^{-1}$ in the last integral.  This expression clearly reveals
that for ``phantom'' tubules, the upper critical dimension $D_{uc}$
for this problem, below which transverse positional fluctuations
diverge is $D_{uc} = 5/2$; this in principle (but see discussion of
dominant zero modes in Sec.\ref{RGscaling_subsubsection}) allows a
quantitatively trustworthy $\epsilon=D_{uc}-2=1/2$ expansion for the
physical membrane of $D=2$. This should be contrasted with the result
$D_{uc} = 4$ for the analogous critical dimension in the flat
phase.\cite{AL,GD}

The lower critical dimension $D_{lc}$ below which the tubule is
necessarily crumpled in this problem is also lowered by the anisotropy.
Considering the fluctuations of the membrane normals $\bbox{\nabla}\vec h$
in the harmonic approximation, one sees immediately that the largest of
these is the fluctuation in the y-direction,
\begin{eqnarray}
\langle|\delta n_y({\bf x})|^2\rangle&=&
\langle|\partial_y\vec{h}({\bf x})|^2\rangle\;,\nonumber\\
&\propto&
\int_{q_\perp>L_\perp^{-1}}{d^{D-1} q_\perp d q_y\over(2\pi)^D}
{q_y^2\over t q_\perp^2 + \kappa q_y^4}\;,\nonumber\\
&\propto&
\int_{q_\perp>L_\perp^{-1}}{d^{D-1} q_\perp\over  q_\perp^{1/2}}\propto L_\perp^{3/2-D}\;,
\label{Dlc}
\end{eqnarray}
which clearly only diverges in the infra-red
$L_\perp\rightarrow\infty$ limit for $D\leq D_{lc}=3/2$ (but again see
discussion of dominant zero modes in
Sec.\ref{RGscaling_subsubsection}).

In the argot of the membrane field, the elasticity of phantom tubules
is anomalous.  In contrast to the flat phase, however, for phantom
tubules, the exponents characterizing the anomalous elasticity can be
calculated {\it exactly}.  To see this, we first note that the
$u$-fluctuations go like $1/q^2$ in {\it all} directions and hence are
negligible (in the relevant wavevector regime $|{\bf q_\perp}|\approx
q_y^2$) relative to the $\vec h$-fluctuations which scale like $1/q^4$
in this regime.  This justifies neglecting the ${1\over2}(\partial_y
u)^2$ piece of the invariant $E(u,{\vec h})$ operator. This also
emerges from a full renormalization group treatment\cite{RTfuture},
which shows that this term is strongly irrelevant. Once it is
neglected, the elastic free energy is quadratic in $u$, and these
phonon modes can therefore be integrated exactly out of the partition
function
\begin{equation}
Z=\int{\cal D} u {\cal D}\vec{h} e^{-\beta F_{el}[u,\vec{h}]}\;.
\label{Z}
\end{equation}
Once this is done, the only remaining anharmonic term in the effective
elastic free energy for $\vec h$ is, in Fourier space,
\begin{eqnarray}
F_{anh}[\vec{h}]={1\over4}\int_{{\bf k}_1,{\bf k}_3,{\bf k}_3}
\hspace{-3mm}&&\left(\vec{h}({\bf k}_1)\cdot\vec{h}({\bf k}_2)\right)
\left(\vec{h}({\bf k}_3)\cdot\vec{h}({\bf k}_4)\right)\times\nonumber\\
&& k_{y1} k_{y2} k_{y3}k_{y4} V_h({\bf q})\;,\label{Fanh}
\end{eqnarray}
where ${\bf q}={\vec k_1} + {\vec k_2}$ and ${\vec k_1} + {\vec k_2} +
{\vec k_3} + {\vec k_4}=0$.  The effective vertex $V_h({\bf q})$ above
reduces to 
\begin{equation}
V_h({\bf q})={g_y g_\perp q_\perp^2\over g_y q_y^2+g_\perp q_\perp^2}\;,
\label{vertexV_h}
\end{equation}
which is irrelevant near the Gaussian fixed point (but see
Sec.\ref{RGscaling_subsubsection}), as can be seen by the simple {\em
anisotropic} power counting described above.

The exact cancelation of the relevant terms in $F_{anh}[\vec{h}]$
above is a direct consequence of the $4:4:1$ ratios of the
coefficients of the quadratic $(\partial_y u)^2$ and the anharmonic
$\partial_y u (\partial_y {\vec h})^2$ and $(\partial_y {\vec h})^4$
terms in $F_{el}$ that was discussed earlier.  Given this
cancelation, $F_{anh}[\vec{h}]$ is now clearly less relevant than the
anharmonic vertices $\partial_y u (\partial_y {\vec h})^2$ and
$(\partial_y {\vec h})^4$ in the original free energy (before we
integrated out the phonons $u$). This is because the factor $V_h({\bf
q})\propto q_\perp^2/(g_y q_y^2+g_\perp q_\perp^2)$ vanishes like
$q_y^2$ in the relevant limit $|{\bf q_\perp}|\sim q_y^2$,
$q_y\rightarrow 0$ (the other factors in Eq.\ref{Fanh} are precisely
the Fourier transform of $(\partial_y\vec h)^4$, of course). This
lowers the upper critical dimension for anomalous elasticity {\it of
the $\vec h$ field} to $D_{uc} = 3/2$.  Thus, in the physical case $D
= 2$, there is no anomalous elasticity in $\vec h$ ; that is, the
elastic constants $t$ and $\kappa$ in Eq.\ref{hProp2} are
finite and non-zero as $q_y\rightarrow 0$.

\begin{figure}[bth]
\centering
\setlength{\unitlength}{1mm}
\begin{picture}(150,20)(0,0)
\put(-23,-97){\begin{picture}(150,0)(0,0)
\includegraphics{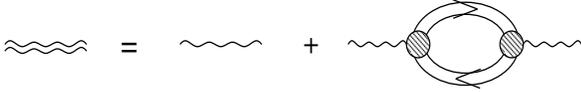}
\end{picture}}
\end{picture}
\caption{Feynman graph equation for the self-consistent
evaluation of $g_y({\bf q})$.}
\label{gySCloop}
\end{figure}

However, as asserted earlier, the {\it full} elasticity Eq.\ref{elFtubule},
{\it before} $u$ is integrated out, {\it is} anomalous, because $g_y$
is driven to zero as $q_y\rightarrow 0$.  Indeed, a self-consistent one-loop
perturbative calculation of $g_y({\bf q})$, obtained by evaluating the
Feynman graph in Fig.\ref{gySCloop}, gives
\begin{eqnarray}
&&g_y({\bf q})=g_y^o - \label{SCgy}\\
&&\int{k_B T g_y^2({\bf q}) p_y^2 (p_y-q_y)^2\;
d^{D-1} p_\perp d p_y/(2\pi)^D
\over\left(t p_\perp^2+\kappa({\bf p}) p_y^4\right)
\left(t |{\bf p_\perp}-{\bf q_\perp}|^2+
\kappa(|{\bf p}-{\bf q}|) (p_y-q_y)^4\right)}\;,\nonumber\\
\nonumber
\end{eqnarray}
where $g_y^o$ is the ``bare'' or unrenormalized value of $g_y$.
Our earlier argument shows that $\kappa({\bf p})$ can be replaced by
a constant in Eq.\ref{SCgy} as ${\bf p} \rightarrow 0$ , since the
${\vec h}$ elasticity is not anomalous.  The self-consistent equation
Eq.\ref{SCgy} can be solved by the ansatz,
\begin{equation}
g_y({\bf q})=q_y^{\eta_u} S_g(q_y/q_\perp^z)\;.\label{gyAnsatz}
\end{equation}
Simple power counting\cite{noteSC} then shows that we must choose
\begin{eqnarray}
z&=&{1\over2}\;,\label{z_phantom}\\
\eta_u&=& 5-2D\;.\label{etau_phantom}
\end{eqnarray}

It is straightforward to verify that these results hold to {\it all}
orders in perturbation theory; that is, at every order, the leading
dependence on ${\bf q}$ of the contribution to $g_y$ scales like
$q_y^{\eta_u} S_g(q_y/q_\perp^{1/2})$ with $\eta_u = 5-2D$.

It is straightforward to verify to {\it all} orders in perturbation
theory that there is no such renormalization of $g_\perp$. This is
because of the anisotropic scaling $q_\perp\sim q_y^2$, which implies
that all vertices proportional to powers of {\it perpendicular}
gradients of $\vec h$, i.e., powers of $\bbox{\nabla}_\perp\vec{h}$
are {\it irrelevant}. Since only such vertices can renormalize
$g_\perp |\bbox{\nabla}_\perp u|^2$, there are no {\it relevant}
renormalization of $g_\perp$. As a result, $g_\perp$ remains finite
and non-zero, or, in a word, non-anomalous, as $|{\bf
q}|\rightarrow0$.

Using the facts that $g_y({\bf q})$ is independent of ${\bf q}_\perp$
as $|{\bf q}_\perp|\equiv q_\perp\rightarrow0$ for fixed $q_y$, and,
likewise, to be independent of $q_y$ as $q_y\rightarrow0$ for fixed
$q_\perp$, we can obtain the limits of the scaling function
$S_g(x)$:
\begin{eqnarray}
S_g(x)&\propto&\left\{\begin{array}{lr}
\mbox{constant}, & x\rightarrow\infty\\
x^{-\eta_u}, & x\rightarrow 0\;.
\label{fg_limits}
\end{array} \right.
\end{eqnarray}
For {\em phantom} membranes with $D=2$, $\eta_u=1$ and $z=1/2$, so we
find:
\begin{eqnarray}
g_y({\bf q})&\propto&\left\{\begin{array}{lr}
q_y, & q_y>>\sqrt{q_\perp}\\
\sqrt{q_\perp}, &\;\;\; q_y<<\sqrt{q_\perp}\;.
\label{gy_limits}
\end{array} \right.
\end{eqnarray}

We will now use this result to compute the mean-squared real space
fluctuations
$\langle\left(u(L_\perp,y)-u(0,y)\right)^2\rangle\equiv\langle\Delta
u^2\rangle$ of $u({\bf x})$. These can be obtained via the
equipartition theorem and by summing all of the Fourier modes,
yielding:
\begin{equation}
\langle \Delta u^2\rangle\approx\int_{q_\perp>L_\perp^{-1},\;
q_y>L_y^{-1}}
{d q_\perp d q_y\over(2\pi)^2} {1-e^{i{\bf q_\perp}\cdot{\bf
L_\perp}}\over g_y({\bf q}) q_y^2 + g_\perp q_\perp^2}\;.
\label{uu1}
\end{equation}

Let us assume, and verify a posteriori, that the integral in this
expression is dominated by wavevectors with
$q_y<<\sqrt{q_\perp}$. Then, using Eq.\ref{gy_limits}, we see that
\begin{equation}
\langle \Delta u^2\rangle\approx\int_{q_\perp>L_\perp^{-1},\;
q_y>L_y^{-1}}
{d q_\perp d q_y\over(2\pi)^D} {1-e^{i{\bf q_\perp}\cdot{\bf
L_\perp}}\over c\sqrt{q_\perp} q_y^2 + g_\perp q_\perp^2}\;,
\label{uu2}
\end{equation}
where $c$ is a constant. Inspection of this integral reveals that it
is dominated by ${\bf q}$'s for which the two terms in the denominator
balance; this means $q_y\sim q_\perp^{3/4}<<\sqrt{q_\perp}$, the last
extreme inequality holding as $|{\bf q}|\rightarrow 0$. This verifies
our earlier a posteriori assumption that $q_y<<\sqrt{q_\perp}$ in the
dominant wavevector regime.

Now, changing variables in the integral $q_\perp\equiv
Q_\perp/L_\perp$, $q_y\equiv Q_y/L_\perp^{3/4}$, we find
\begin{equation}
\langle \Delta u^2\rangle=L_\perp^{1/4} S_u(L_y/L_\perp^{3/4})\;,
\label{uu_scaling_form}
\end{equation}
where
\begin{equation}
S_u(x)\equiv\int_{Q_\perp>1,\;
Q_y>x^{-1}} {d Q_\perp d
Q_y\over(2\pi)^2} {1-e^{i Q_\perp}\over c\sqrt{Q_\perp} Q_y^2 +
g_\perp Q_\perp^2}\;.
\label{uu3}
\end{equation}
We note that the scaling form for the $u$ phonon correlations is {\em
different} than that of the height field $\vec{h}$ as summarized in
e.g. Eqs.\ref{RG} and \ref{hrms}, and discussed in more detail below.

The limits of $S_u(x)$ scaling function can be obtained just as we did
for $S_g(x)$; we find, including ``zero modes'' (see below):
\begin{eqnarray}
S_u(x)&\propto&\left\{\begin{array}{lr}
\mbox{constant}, & x\rightarrow\infty\\
x^{-1}, & x\rightarrow 0\;.
\label{fu_limits}
\end{array} \right.
\end{eqnarray}

For roughly square membranes, $L_y\sim L_\perp=L$, so, as
$L\rightarrow\infty$, $L_y/L_\perp^{3/4}\rightarrow\infty$, and the
first limit of Eq.\ref{fu_limits} is the appropriate one. This gives
\begin{equation}
\langle \Delta u^2\rangle\propto L_\perp^{1/4}\;.
\label{uu4}
\end{equation}

The authors of Ref.~\onlinecite{BFT} measured a quantity that should
scale like $\langle \Delta u^2\rangle$ in their simulations of a
square anisotropic membrane.  They did this via their vividly named
``salami'' method: measuring the moment of inertia tensor of a
``salami'' slice, a set of $N$ points that all had the same internal
$y$ coordinate (for a $y$-tubule phase). It is straightforward to show
that the smallest eigenvalue of this tensor should scale like $N
\langle u({\bf x})^2\rangle$, since, as we shall see in a moment, the
mean squared displacements in the other directions are much bigger
than those in the $y$ direction. Therefore, from Eq.\ref{uu4} we
predict that the smallest eigenvalue of this salami slice moment of
inertia tensor scales like $N L^{1/4}$. BFT actually fit this
eigenvalue to $N\log L$, which might appear to disagree with our
prediction, until one recognizes that for $L$'s between $32$ and $100$
(where most of the data of Ref.~\onlinecite{BFT} is taken),
$L^{1/4}=(e/4)\log L$ to an accuracy of better than $1\%$. Thus, their
fit is certainly consistent with our predictions. To more strenuously
test our full scaling predictions Eq.\ref{uu_scaling_form} and
\ref{fu_limits}, one could simulate membranes with aspect ratios quite
different from $1$. In particular, we predict based on
Eq.\ref{uu_scaling_form} that increasing $L_y$ at fixed $L_\perp$ from
an initially square configuration would {\em not} increase this
smallest eigenvalue; nor would decreasing $L_y$ decrease it, until an
aspect ratio $L_y\sim L_\perp^{3/4}$ is reached, beyond which this
eigenvalue would increase like $L_y^{-1}$.

We now turn to the computation, for the phantom tubule, of the tubule
radius of gyration $R_G$ and roughness $h_{rms}$, defined by:
\begin{eqnarray}
R_G^2&\equiv&\langle|{\vec h}({\bf L}_\perp,y)- {\vec
h}(0_\perp,y)|^2\rangle\;,\label{R_def}\\
h_{rms}^2&\equiv&\langle|{\vec h}({\bf x_\perp},L_y)- {\vec h}({\bf
x_\perp},0)|^2\rangle\;,\label{h_def}
\end{eqnarray}
where ${\bf L}_\perp$ spans the intrinsic $\perp$ space of the
membrane.  Because $R_G$ is by definition the root-mean-square (rms)
distance between two points at the same $y$, it is roughly the radius
of a typical cross-section of the tubule perpendicular to the tubule
axis. Likewise, $h_{rms}$ measures fluctuations between points widely
separated {\it along} the tubule axis; hence, it gives the
polymer-like transverse "wandering" of the tubule. See
Fig.\ref{tubule_fig} for an illustration of $R_G$ and $h_{rms}$.

The reason we distinguish between these two quantities is that they
scale in different ways with the membrane dimensions $L_\perp$ and
$L_y$, in contrast to one's naive expectations. This happens because
there are large contributions to both quantities from "zero modes", by
which we mean Fourier modes with either ${\bf q_\perp}$ or $q_y=0$.
Those with ${\bf q_\perp}=0$ correspond to polymer-like undulations of
the entire tubule. Recognizing the existence of both types of modes,
we Fourier decompose ${\vec h}({\bf x})$ as follows:
\begin{eqnarray}
{\vec h}({\bf x})&=&{ 1 \over \sqrt{L_\perp^{D-1} L_y}}
\sum_{\bf q}
{\vec h}_B ({\bf q}) e^{i {\bf q}\cdot {\bf x}} + { 1 \over
\sqrt{L_y}}
\sum_{q_y}
{\vec h}_{0y}(q_y) e^{i q_y y}\nonumber\\
&+& {1 \over \sqrt{L_\perp^{D-1}}}
\sum_{{\bf q}_\perp} {\vec h}_{0\perp}({\bf q}_\perp) e^{i {\bf
q}_\perp\cdot {\bf x}_\perp}\;,\nonumber\\
\label{fodec}
\end{eqnarray}
where $B$, $0y$, and $0 \perp$ denote "bulk modes" (i.e., modes with
neither ${\bf q}_\perp$ nor $q_y$ = 0), and "zero modes" (i.e., modes
with either ${\bf q}_\perp$ or $q_y$ = 0), respectively. Note that we
have chosen different normalizations for the three types of modes.
For phantom membranes, we proceed by inserting this Fourier
decomposition into the harmonic, ${\vec h}$ dependent piece of the
elastic free energy $F_{el}$ (which is justified, since, as shown
above, the elasticity for ${\vec h}$ for a {\em phantom tubule} is not
anomalous), obtaining:
\begin{eqnarray}
F_0&=&{1 \over 2} \sum_{\bf q} (t q_\perp^2 + \kappa q_y^4) |{\vec
h}_B ({\bf q})|^2 + {1 \over 2}L_\perp^{D-1} \sum_{q_y} \kappa q_y^4 |{\vec
h}_{0y}(q_y)|^2\nonumber\\
&+& {1 \over 2}L_y \sum_{{\bf q}_\perp} t q_\perp^2 | {\vec
h}_{0\perp}({\bf q}_\perp)|^2)\;,\nonumber\\
\label{F0}
\end{eqnarray}
Note the explicit presence of the factors of $L_\perp^{D-1}$ and $L_y$
for the 0-modes. Applying equipartition to Eq.\ref{F0}, we can obtain the
mean squared fluctuations of the Fourier modes:
\begin{equation}
\langle|{\vec h}_B({\bf q})|^2\rangle\
= {k_B T (d-D) \over t q_\perp^2 + \kappa q_y^4}
\label{FluckB}
\end{equation}
\begin{equation}
\langle|{\vec h}_{0y}(q_y)
|^2\rangle\ = {k_B T (d-D) \over L_\perp^{D-1} \kappa q_y^4}
\label{Fluck0y}
\end{equation}
\begin{equation}
\langle|{\vec h}_{0\perp}({\bf q}_\perp)
|^2\rangle\ = {k_B T (d-D) \over L_y t q_\perp^2}
\label{Fluck0perp}
\end{equation}
Using these expressions inside Eqs.\ref{R_def} and \ref{h_def}, and
being careful about converting sums on ${\bf q}$ into integrals, we
get
\begin{eqnarray}
R_G^2 &=& 2(d-D)\bigg[{k_B T\over L_y}\int_{L_\perp^{-1}}{d^{D-1}
q_\perp\over(2\pi)^{D-1}} {1\over t q_\perp^2}
(1 - e^{ i {\bf q}_\perp\cdot{\bf L}_\perp})\nonumber\\ 
&+& k_B T\int_{L_\perp^{-1},L_y^{-1}}{d^{D-1} q_\perp d q_y\over(2\pi)^D} 
{(1-e^{i {\bf q}_\perp\cdot {\bf L}_\perp})
\over t q_\perp^2 + \kappa q_y^4}\bigg]\;,
\label{RgTubule}
\end{eqnarray}
where the subscripts $L_\perp^{-1}$ and $L_y^{-1}$ denote infra-red
cutoffs $|{\bf q}_\perp | > L_\perp^{-1} , q_y > L_y^{-1}$, with
$L_\perp \equiv | {\bf L}_\perp |$.

We observe here that $R_G$ in Eq.\ref{RgTubule} does not receive any
contribution from the ${\bf q}_\perp=0$ ''zero mode'' (i.e., in
addition to the bulk mode, $R_G$ receives a contribution only from the
$q_y=0$ ``zero mode'').

Scaling $L_\perp$ out of both integrals for $R_G$ by the change of
variables ${\bf Q_\perp} \equiv {\bf q}_\perp L_\perp$ and $Q_y\equiv
q_y\sqrt{ L_\perp}$, we obtain
\begin{equation}
R_G^2 = \big({C_1 L_\perp^{3-D} \over L_y} + L_\perp^{5/2-D}
I_R({  L_y \over \sqrt{  L_\perp}})\big)\;,\label{RG1}
\end{equation}
where
\begin{equation}
C_1 \equiv 2(d-D) k_B T\int_{1}{d^{D-1} q_\perp \over(2\pi)^{D-1}}
{(1-e^{i {\bf q}_\perp\cdot{\bf \hat L}_\perp})\over t q_\perp^2}\; .
\label{RG2}
\end{equation}
is a constant of $O(1)$, and
\begin{equation}
I_R(x) \equiv 2(d-D) k_B T\int_{1,x}{d^{D-1} q_\perp d
q_y\over(2\pi)^D} {(1-e^{i {\bf q}_\perp\cdot{\bf\hat L}_\perp})\over t
q_\perp^2 +
\kappa q_y^4} .
\label{RG3}
\end{equation}
with ${\bf\hat L}_\perp$ the unit vector along ${\bf L}_\perp$.
Defining the scaling function
\begin{equation}
S_R(x)\equiv \sqrt{{C_1\over x } + I_R(x)} ,
\label{RG4}
\end{equation}
we see that $R_G$ can be rewritten in the scaling form
\begin{equation}
R_G(L_\perp,L_y)= L_\perp^{\nu}S_R(L_y/L_\perp^z)
\label{RG5}
\end{equation}
with, for phantom membranes, 
\begin{eqnarray}
\nu &=& {5-2D \over 4}\;,\label{nu_phantom}\\
z &=& {1\over2}\;.\label{z_phantom2}
\end{eqnarray}
We will see later that the scaling form Eq.\ref{RG5} continues to
apply when self-avoidance is included, but with different values of
$\nu$ and $z$, and a different scaling function $S_R(x)$. For phantom
membranes, from our explicit expression for the scaling function
$S_R$, we see that it has the limiting forms:
\begin{equation}
S_R(x)\propto
\left\{ \begin{array}{lr}
1/\sqrt{x} & \mbox{for}\; x\rightarrow 0 \\
\mbox{constant}, & \mbox{for}\; x\rightarrow\infty
\end{array} \right.
\label{fR}
\end{equation}
In particular, the limiting form as $x \rightarrow \infty $ implies
that for the physically relevant case of a square membrane
$L_\perp\sim L_y\sim L\rightarrow\infty$, for which $L_y >>
L_\perp^z$, bulk modes dominate, and we obtain,
\begin{equation}
R_G\propto L_\perp^{\nu}\;.
\label{Rgdefine}
\end{equation}

The simulations of BFT\cite{BFT} measured $R_G$ for phantom tubules by
calculating the largest moment of inertia for for a set of membrane
points that all had the same value of the intrinsic coordinate
$y$. While we have used here a slightly different definition,
Eq.\ref{R_def}, the square root of this moment of inertia should
scale like our $R_G$. And, indeed, BFT found that it did scale like a
power of $L$, as in Eq.\ref{Rgdefine}, with $\nu=0.24\pm0.02$ in
excellent quantitative agreement with our predictions of $\nu=1/4$,
Eq.\ref{nu_phantom} evaluated in $D=2$. It would be of great interest to test
our full anisotropic scaling prediction of Eq.\ref{RG5} by varying the
aspect ratio of the membrane in such simulations. For instance, one
could fix $L_\perp$ and increase $L_y$; we predict that one should
observe {\em no} change in $R_G$. The same should hold if one {\em
decreased} $L_y$ at fixed $L_\perp$: $R_G$ should remain unchanged
until $L_y\sim\sqrt{L_\perp}$, at which point the tubule should begin
to get thinner (i.e. $R_G$ should decrease).

Equations \ref{RG4} and \ref{fR} also correctly recover the limit of
$L_y=\mbox{constant} << L_\perp^z\rightarrow\infty$, where the $q_y=0$
``zero modes'' dominate, the tubule simply becomes a phantom, coiled
up, $D-1$-dimensional polymeric network of size $L_\perp$ embedded in
$d-1$ dimensions, with the radius of gyration $R_G(L_\perp)\sim
L_\perp^{(3-D)/2}$. In the physical dimensions ($D=2$ and $d=3$) in
particular this gives a coiled up ideal polymer of length $L_\perp$
with $R_G\sim L_\perp^{1/2}$, as expected.

We now turn our attention to the calculation of the tubule roughness
$h_{rms}$.  As we will see, here the ${\bf q}_\perp=0$ zero mode will
play an essential role and will dominate the transverse undulations
for ``very long'' tubules, which (because of anisotropic scaling) in
particular includes tubules made from square membranes. Using the
definition of $h_{rms}$, Eq.\ref{h_def}, we have
\begin{eqnarray}
h_{rms}^2 &=& 2(d-D)\bigg[{k_B T\over L_\perp^{D-1}}\int_{L_y^{-1}}
{d q_y\over(2\pi)}
{1\over\kappa(q_y) q_y^4} (1-e^{i q_y L_y})
\;\nonumber\\
& + &k_B T\int_{L_\perp^{-1},L_y^{-1}}{d^{D-1} q_\perp d q_y\over(2\pi)^D}
{(1-e^{i q_y L_y})\over t q_\perp^2 +
\kappa q_y^4}\bigg]\;.\label{hTubule}
\end{eqnarray}
Here we observe that $h_{rms}$ in Eq.\ref{hTubule} does not receive
any contribution from the $q_y=0$ ''zero mode'' (i.e., in addition to
the bulk mode, $h_{rms}$ receives a contribution only from the ${\bf
q}_\perp=0$ ``zero mode''). This is to be contrasted with the behavior
of $R_G$ that we noted following Eq.\ref{RgTubule}, and is responsible
for the differences in scaling properties of $R_G$ and $h_{rms}$,
notes above.

Now, for perverse and twisted reasons of our own, we choose to scale
$L_y$, rather than $L_\perp$, out of the integrals in this expression,
via the change of variables $Q_y \equiv q_y L_y , {\bf Q}_\perp \equiv
{\bf q_\perp} L_y^2$, which leads to
\begin{equation}
h_{rms}^2 = ({C_2 L_y^3 \over L_\perp^{D-1}} + L_y^{5-2D} I_h({ 
L_y \over \sqrt{  L_\perp}}))
\label{h1}
\end{equation}
where
\begin{equation}
C_2 \equiv 2(d-D) k_B T\int_{1}{d Q_y \over 2\pi} {(1-e^{i Q_y})\over
\kappa Q_y^4} .
\label{h2}
\end{equation}
is yet another constant of $O(1)$, and
\begin{equation}
I_h(x) \equiv 2(d-D) k_B T\int_{x^{2},1}{d^{D-1} Q_\perp d
Q_y\over(2\pi)^D} {(1-e^{i Q_y})\over t Q_\perp^2 + \kappa Q_y^4}\;.
\label{h3}
\end{equation}
Defining the scaling function
\begin{equation}
S_h(x) \equiv \sqrt{{C_2 x^{2(D-1)} } + I_h(x)} ,
\label{h4}
\end{equation}
we see that $h_{rms}$ can be rewritten in the scaling form
\begin{equation}
h_{rms}(L_\perp,L_y)= L_y^{\zeta}S_h(L_y/L_\perp^z)
\label{h5}
\end{equation}
with, for phantom membranes, 
\begin{eqnarray}
\zeta &=& {5-2D \over 2}\;,\label{zeta}\\
z &=&{1\over2}\;.\label{z_phantom3}
\end{eqnarray}
Again, this scaling law Eq.\ref{h5} continues to apply when
self-avoidance is included, but with different values of $\zeta$ and
$z$.

Equations \ref{RG5} and \ref{h5} give information about the tubule
roughness for arbitrarily large size $L_\perp$ and $L_y$, and
arbitrary aspect ratio.  For the physically relevant case of a square
membrane $L_\perp\sim L_y\sim L\rightarrow\infty$, for which $ L_y >>
L_\perp^z$, we obtain,
\begin{eqnarray}
h_{rms}&\propto&{L_y^{\zeta+(D-1)/2z}\over L_\perp^{(D-1)/2}}\;,\\
&\propto&L^{\zeta+(D-1)(1-z)/2z}\;,
\label{hrmsL}
\end{eqnarray}

Equations \ref{zeta}, \ref{z_phantom3} then give, for a $D=2$ phantom
tubule, $\zeta=1/2$, $z=1/2$
\begin{equation}
h_{rms}\sim {L_y^{3/2}\over L_\perp^{1/2}}\;,
\label{hrmsL1}
\end{equation}
and therefore predicts for a square membrane
\begin{equation}
h_{rms}\sim L\;.
\label{hrmsL2}
\end{equation}

This prediction for square phantom membranes has also been
spectacularly quantitatively confirmed in simulations by
BFT\cite{BFT}. Their ingenious procedure for determining $h_{rms}$ is
rather involved, and the interested reader is referred to their paper
for a clear and complete discussion of it. The bottom line, however,
is that they find $h_{rms}\sim L^\gamma$, (our $\gamma$ is $\zeta$ in
their notation) with $\gamma=0.895\pm 0.06$, in excellent agreement
with our prediction $\gamma=1$ from Eq.\ref{hrmsL2} above.  As with
$R_G$, it would be interesting to test the full scaling law
Eq.\ref{h5} by simulating non-square membranes, and testing for the
independent scaling of $h_{rms}$ with $L_y$ and $L_\perp$. Note that,
unlike $R_G$, according to Eq.\ref{hrmsL1}, $h_{rms}$ will show
immediate growth (reduction) when one increases (decreases) $L_y$ at
fixed $L_\perp$.

Because, unlike the flat phase, no $\log(L/a)$ correction arises, the
($D=2$) phantom tubule is just marginally {\it stable}, but with wild
transverse undulations which scale linearly with its length. As we
will see in Sec.\ref{selfavoidance}, these wild fluctuations will be
suppressed when the effects of self-avoidance are included.

The above discussion also reveals that our earlier conclusions about
the lower critical dimension $D_{lc}$ for the existence of the tubule
are strongly dependent on how $L_\perp$ and $L_y$ go to infinity
relative to each other; i.e., on the membrane aspect ratio. The
earlier conclusion that $D_{lc}=3/2$ only strictly applies when the
bulk modes dominate the physics, which is the case for a very squat
membrane, with $ L_y\approx L_\perp^z$, in which case $L_y<<L_\perp$.
For the physically more relevant case of a square {\it phantom}
membrane, from the discussion above, we find that $D_{lc}=2^-$, where
the $^-$ superscript means that there are no logarithmic corrections
at $D=2$ and therefore strictly speaking the $D=2$ tubule {\it is}
marginally stable.

Equations \ref{RG5} and \ref{h5} also correctly recover the limit of
$L_\perp^z=\mbox{constant}<<   L_y\rightarrow\infty$,
where the tubule simply becomes a polymer of thickness $R_G(L_\perp)$
given in Eq.\ref{R_def} of length $L_y$ embedded in $d-1$ dimensions.
As already discussed in the Introduction for a more general case of a
self-avoiding tubule, these equations then correctly recover this
polymer limit giving
\begin{equation}
h_{rms}\approx L_P (L_y/L_P)^{3/2}\;,
\end{equation}
with $L_\perp$-dependent persistent length 
\begin{equation}
L_P(L_\perp)\propto L_\perp^{D-1}\;.
\label{LP2}
\end{equation}
which agrees with Eq.\ref{LP1} of Sec.\ref{intro}, for $D=2$ when one
remembers that, for the phantom membranes, $\eta_\kappa=0$.  So, as
expected for a phantom tubule, if $L_\perp$ does not grow fast enough
(e.g. remains constant), while $L_y\rightarrow\infty$, the tubule
behaves as a linear polymer and crumples along its axis and the
distinction between the crumpled and tubule phases disappears.  

To summarize: the radius of gyration $R_G$ and the tubule roughness
$h_{rms}$ scale differently with membrane size $L$ for a square
membrane because the former is dominated by bulk modes, while the
latter is dominated by ${\bf q_\perp}=0$ ``zero modes''.

\section{Self-avoidance in the tubule phase}
\label{selfavoidance}
We now look at the effects of self-avoidance on the tubule phase, and
begin by calculating the upper critical embedding dimension $d_{uc}$
below which the self-avoidance becomes relevant in the tubule phase. A
model of a self-avoiding membrane in the the tubule phase is described
by a free energy functional which is a combination of the elastic free
energy $F_{el}$ from Eq.\ref{elFtubule} and the self-avoiding
interaction $F_{SA}$ from Eq.\ref{Fc} specialized to the tubule
extended in $y$-direction using Eq.\ref{tubulecoordinate} for
$\vec{r}({\bf x})$
\begin{eqnarray}
F_{SA}&=&{b\over2}\int\hspace{-0.1cm}dy dy'\hspace{0.05cm} 
d^{D-1}x_\perp d^{D-1}x'_\perp
\hspace{0.05cm}\delta^{(d-1)}\big({\vec h}({\bf x_\perp},y)- 
{\vec h}({\bf x'_\perp},y)\big)\nonumber\\
&\times&\delta\big(\zeta_y y+u({\bf x_\perp},y)-
\zeta_y y'-u({\bf x'_\perp},y')\big)\;.
\label{SAt1}
\end{eqnarray}
If the in-plane fluctuations $u$ scale sub-linearly with $y$ (which we
will self-consistently verify a posteriori that they do), at long
length scales one can ignore the phonons inside the self-avoiding
interaction above.  This can be confirmed more formally by an explicit
renormalization group analysis\cite{RTfuture}. We then obtain a
self-avoiding interaction that is local in $y$, with corrections that
are irrelevant in the renormalization group sense and therefore
subdominant at long length scales. The appropriate free energy that
describes a self-avoiding tubule is then given by
\begin{eqnarray}
F&=&{1\over2}\int d^{D-1}x_\perp dy
\bigg[\gamma\bigg(\partial_y u+{1\over2}(\partial_y{\vec h})^2 +
{1\over2}(\partial_y u)^2\bigg) \nonumber\\
&+&\kappa(\partial^2_y{\vec h})^2 + t(\partial^\perp_\alpha \vec h)^2
+ g_\perp(\partial^\perp_\alpha u)^2\nonumber\\
&+&
g_y\bigg(\partial_y u+{1\over2}(\partial_y{\vec h})^2 +
{1\over2}(\partial_y u)^2\bigg)^2\bigg]\;\nonumber\\
&+&{v}\int\hspace{-0.1cm}dy\hspace{0.05cm} 
d^{D-1}x_\perp d^{D-1}x'_\perp
\hspace{0.05cm}\delta^{(d-1)}\big({\vec h}({\bf x_\perp},y)-
{\vec h}({\bf x'_\perp},y)\big)\;,\nonumber\\
\label{F_Tubule_full}
\end{eqnarray}
where $v=b/2\zeta_y$. 

It is important for simulators to note that, although the
self-avoiding interaction is effectively local in {\it intrinsic}
coordinate $y$, this does {\it not} mean that the effects of
self-avoidance can be included in simulations that have each particle
on the membrane avoid only those labeled by the same {\it intrinsic}
$y$ coordinate. Such a simulation, rather, models the very different
(unphysical) self-avoiding interaction
\begin{eqnarray}
F_{SA}^{wrong}&=&{v}\int\hspace{-0.1cm}dy\hspace{0.05cm}d^{D-1}x_\perp
d^{D-1}x'_\perp
\delta(u({\bf x_\perp},y)-u({\bf x'_\perp},y)\nonumber\\
&&\times\hspace{0.1cm}\delta^{(d-1)}\big({\vec h}({\bf x_\perp},y)-
{\vec h}({\bf x'_\perp},y)\big)\;
\label{SAwrong},
\end{eqnarray}
which accounts for interaction only of particles that have the same
intrinsic coordinate $y$ {\it and} the same extrinsic coordinate.  For
large membranes, this unphysical interaction is smaller than the true
self-avoiding interaction in Eq.\ref{F_Tubule_full} by a factor that
scales like the inverse of the rms fluctuations of $u$: $\langle
u^2\rangle^{-1/2}$, as can be seen trivially from the scaling of the
$\delta$-function of $u$ in Eq.\ref{SAwrong}. Since these fluctuations
of $u$ diverge as $L_\perp\rightarrow\infty$ like $u_{rms}\sim
L_\perp^{\zeta_u}$, with $\zeta_u>0$ (e.g., $\zeta_u=1/8$, for $d\geq
11$ and $D=2$), the {\it wrong} self-avoiding interaction in
Eq.\ref{SAwrong} drastically {\it underestimates} the true
self-avoiding interaction by a factor that diverges in the
thermodynamic limit. Although it is tempting to do so in simulations,
one must be careful not to implement the unphysical self-avoiding
interaction in Eq.\ref{SAwrong}. Since it might be difficult to
implement the approximate (but asymptotically exact) self-avoiding
interaction of Eq.\ref{F_Tubule_full} in simulations, it is easiest to
simulate the unapproximated interaction in Eq.\ref{SAt1}.

In the next three subsections we analyze the properties of a
self-avoiding tubule described by this nonlinear elastic free energy,
using Flory theory\cite{RT}, the renormalization group, and the
Gaussian variational method\cite{RT}.

\subsection{Flory theory}
\label{Flory}

The effects of self-avoidance in the tubule phase can be estimated by
generalizing standard Flory arguments from polymer
physics\cite{Polymers} to the extended tubule geometry. The total
self-avoidance energy scales as
\begin{equation}
E_{SA}\propto V\rho^2\;,
\end{equation}
where 
\begin{equation}
V\propto R_G^{d-1} L_y\;
\end{equation}
is the volume in the embedding space occupied by the tubule and
$\rho=M/V$ is the embedding space density of the tubule. Using the
fact that the tubule mass $M$ scales like $L_\perp^{D-1} L_y$, we see
that
\begin{equation}
E_{SA}\propto {L_y L_\perp^{2(D-1)}\over R_G^{d-1}}\;,\label{Esa}
\end{equation}

Using the radius of gyration $R_G\propto L_\perp^\nu$, and
considering, as required by the anisotropic scaling, a membrane with
$L_\perp\propto L_y^2$, we find that $E_{SA}\propto
L_y^{\lambda_{SA}}$ around the phantom fixed point, with
\begin{equation}
\lambda_{SA}=1+4(D-1)-2(d-1)\nu\;,\label{lambda}
\end{equation}

Self-avoidance is relevant when $\lambda_{SA} > 0$, which, from the
above equation, happens for $\nu=\nu_{ph}=(5-2D)/4$ (as per
Eq.\ref{nu_phantom}) when the embedding dimension
\begin{equation}
d<d_{uc}^{SA}={6D-1\over5-2D}\;.\label{duc}
\end{equation}
For $D=2$-dimensional membranes, $d_{uc}^{SA} = 11$.  Thus, self-
avoidance is strongly relevant for the tubule phase in $d=3$, in
contrast to the flat phase.

We can estimate the effect of the self-avoidance interactions on
$R_G(L_\perp)$ in Flory theory, by balancing the estimate Eq.\ref{Esa}
for the self-avoidance energy with a similar estimate for the elastic
energy:
\begin{equation}
E_{elastic}=t\left({R_G\over L_\perp}\right)^2 L_\perp^{D-1} L_y\;.
\label{Eel}
\end{equation}
Equating $E_{elastic}$ with $E_{SA}$, we obtain a Flory estimate for
the radius of gyration $R_G$:
\begin{equation}
R_G(L_\perp)\propto L_\perp^{\nu_{F}}\;,\; {\nu_{F}}={D+1\over
d+1}\;,\label{SAnuDd}
\end{equation}
which should be contrasted with the Flory estimate of
$\nu_{F}^c=(D+2)/(d+2)$ for the {\em crumpled} phase. The similarity
of the expressions is not surprising, since for the tubule phase the
y-dimension decouples in both the intrinsic and the embedding spaces
and is not affected by the self-avoidance.  For the physical case
$D=2$, $d=3$ Eq.\ref{SAnuDd} gives
\begin{equation}
R_G\propto L_\perp^{3/4}\;,\label{SAnu}
\end{equation}
a result that is known to be {\it exact} for the radius of gyration of
a $D=1$-polymer embedded in $d=2$-dimensions.\cite{2dPolymer} Since
the cross-section of the $D=2$-tubule, crudely speaking, traces out a
crumpled polymer embedded in two dimensions (see
Fig.\ref{tubule_fig}), it is intriguing to conjecture that ${\nu}=3/4$
is also the {\it exact} result for the scaling of the thickness of the
tubule. Unfortunately, we have no strong arguments supporting this
appealing conjecture.

For a square membrane, $L_y\sim L_\perp$, it is straightforward to
argue, as we did previously, that the $q_y=0$ zero modes do not
contribute to $R_G$, and $L_\perp$ is the relevant cutoff. Hence
Eq.\ref{SAnuDd} gives the correct radius of gyration. More generally,
we expect
\begin{equation}
R_G(L_\perp,L_y)\propto L_\perp^{\nu} S_R\left({  L_y\over
L_\perp^z}\right)
\;,\label{nuSAscale}
\end{equation}
where $S_R(x)$ is the scaling function given in Eq.\ref{fh2d} and $z$
is the anisotropy exponent given in Eq.\ref{z}.

\subsection{Renormalization group and scaling relations}
\label{RGscaling_subsubsection}
In this subsection, we present a renormalization group analysis of the
physical {\em self-avoiding} membrane, which will also require a
simultaneous treatment of the nonlinear elasticity that was already
present in a phantom membrane, as discussed in Sec.\ref{tubuleF}.

The correct model, which incorporates the effects of both the
self-avoiding interaction and the anharmonic elasticity, is defined by
the free energy Eq.\ref{F_Tubule_full}.
\begin{eqnarray}
F&=&{1\over2}\int d^{D-1}x_\perp dy
\bigg[\kappa(\partial^2_y{\vec h})^2 + t(\partial^\perp_\alpha \vec h)^2
+ g_\perp(\partial^\perp_\alpha u)^2\nonumber\\
&+&g_y\bigg(\partial_y u+{1\over2}(\partial_y{\vec h})^2\bigg)^2\bigg]\nonumber\\
&+&{v}\int\hspace{-0.1cm}dy\hspace{0.05cm} 
d^{D-1}x_\perp d^{D-1}x'_\perp
\hspace{0.05cm}\delta^{(d-1)}\big({\vec h}({\bf x_\perp},y)-
{\vec h}({\bf x'_\perp},y)\big)\;,\nonumber\\
\label{SAtubuleRG}
\end{eqnarray}
where we have set $\gamma=0$ and dropped the subdominant phonon
anharmonicity.

It is convenient for the purposes of this section to choose the units
of length such that $t=\kappa=1$ throughout, and choose the
renormalization group rescalings to keep them fixed at $1$ even after
the diagrammatic corrections are taken into account (i.e., beyond the
tree-level). We follow the standard renormalization group
procedure\cite{Wilson}:

(i) Integrate out fluctuations of the Fourier modes $u({\bf q})$ and
$\vec{h}({\bf q})$ of the fields $u({\bf x})$ and $\vec{h}({\bf x})$
with wavevectors in the high wavevector shell $\Lambda
e^{-l}<q_\perp<\Lambda$, $-\infty<q_y<\infty$, where the ultraviolet
cutoff $\Lambda$ is of order an inverse microscopic length, and $l$ is
a parameter known as the ``renormalization group time''. This
integration can, of course only be accomplished perturbatively in the
nonlinear couplings $v$ and $g_y$.

(ii) Anisotropically rescale lengths (${\bf x}_\perp$, $y$) and
fields (${\vec h}({\bf x}),u({\bf x})$), so as to restore the
ultraviolet cutoff to $\Lambda$:
\begin{mathletters}
\begin{eqnarray}
{\bf x}_\perp&=&e^{l}{\bf x}_\perp'\;,\\
y&=&e^{z l} y' \;,\\
{\vec h}({\bf x})&=&e^{\nu l}{\vec h}'({\bf x}')\;,\\
u({\bf x})&=&e^{(2\nu-z) l}u'({\bf x}')\;,
\label{rescalings}
\end{eqnarray}
\end{mathletters}
where we have chosen the convenient (but not necessary) rescaling of
the phonon field $u$ so as to preserve the form of the
rotation-invariant operator $(\partial_y u+{1\over2}(\partial_y{\vec
h})^2)^2$.

(iii) Define the effective length-scale dependent coupling constants
so as to bring the resulting long wavelength effective free energy
into the same form as Eq.\ref{SAtubuleRG}.

As discussed above, we will choose the arbitrary rescaling exponents
$\nu$ and $z$ so as to keep the renormalized $\kappa(l)$ and $t(l)$
equal to one. This choice of $\nu$ and $z$ can be shown by standard
renormalization group arguments to be the $\nu$ and $z$ that appear in
the scaling function Eqs.\ref{RG} and \ref{hrms}, as we will demonstrate
later in this subsection.

The result of the three steps of the above renormalization group
transformation (i.e., mode integration, rescaling, and coupling
redefinition) can be summarized in differential recursion relations
for the flowing coupling constants:
\begin{eqnarray}
{d t\over d l}&=&[2\nu+z+D-3-f_t(v)]t\;,
\label{t_rr}\\
{d \kappa\over d l}&=&[2\nu-3z+D-1+f_\kappa(g_y,g_\perp)]\kappa\;,
\label{kappa_rr}\\
{d g_y\over d l}&=&[4\nu-3z+D-1-f_g(g_y)]g_y\;,
\label{gy_rr}\\
{d g_\perp\over d l}&=&[4\nu-z+D-3]g_\perp\;,
\label{gp_rr}\\
{d v\over d l}&=&[2D-2+z-(d-1)\nu-f_v(v)]v\;,
\label{v_rr}
\end{eqnarray}
where the various $f$-functions represent the graphical (i.e.,
perturbative) corrections.  Since the self-avoiding interaction only
involves $\vec{h}$, and the parameters in the $\vec{h}$ propagator
($t$ and $\kappa$) are going to be held fixed at $1$, the graphical
corrections coming from self-avoiding interaction alone depend only on
the strength $v$ of the self-avoiding interaction. Therefore, to {\em
all} orders in $v$, and leading order in $g_y$, $f_t(v)$ and $f_v(v)$
are only functions of $v$ and $f_\kappa(g_y,g_\perp)$ and $f_g(g_y)$
are only functions of $g_y$ and $g_\perp$.

It is important to note that $g_\perp$ suffers no graphical
corrections, i.e., Eq.\ref{gp_rr} is {\em exact}. This is enforced by
an exact symmetry
\begin{equation}
u({\bf x}_\perp,y)\rightarrow u({\bf x}_\perp,y)+\chi({\bf
x}_\perp)\;,
\label{u_symmetry}
\end{equation}
where $\chi({\bf x}_\perp)$ is an arbitrary function of ${\bf
x}_\perp$, under which the nonlinearities in $F$ are invariant.

We further note that there is an additional tubule ``gauge''-like
symmetry for $g_y=0$
\begin{equation}
\vec{h}({\bf x}_\perp,y)\rightarrow \vec{h}({\bf x}_\perp,y)+\vec{\phi}(y)\;,
\label{h_symmetry}
\end{equation}
under which the only remaining nonlinearity, the self-avoiding
interaction, being local in $y$, is invariant. This ``tubule gauge''
symmetry demands that $f_\kappa(g_y=0,g_\perp)=0$, which implies that
if $g_y=0$, there is no divergent renormalization of $\kappa$, {\em
exactly}, i.e., the self-avoiding interaction {\em alone} cannot
renormalize $\kappa$. This {\em non}-renormalization of $\kappa$ by
the self-avoiding interaction, in a truncated (unphysical) membrane
model with $g_y=0$, has been recently verified to all orders in a
perturbative renormalization group calculation\cite{BG}.

To see that the $\nu$ and $z$ obtained as fixed point solutions of
Eqs.\ref{t_rr}-\ref{v_rr} have the same physical significance as the
$\nu$ and $z$ defined in the scaling expressions Eqs.\ref{RG} and
\ref{hrms} for the radius of gyration $R_G$ and tubule wigglyness
$h_{rms}$, we use the renormalization group transformation to relate
these quantities in the unrenormalized system to those in the
renormalized one. This gives, for instance, for the radius of gyration
\begin{eqnarray}
&&\hspace{-.3in}R_G(L_\perp,L_y;t(0),\kappa(0),\ldots)=\nonumber\\
&=&\langle|{\vec h}({\bf L}_\perp,y)-
{\vec h}({\bf 0}_\perp,y)|^2\rangle^{1/2}
\bigg{|}_{L_y,t(0),\kappa(0),\ldots}\nonumber\\ 
&=&e^{\nu l}\langle|{\vec h}(e^{-l}{\bf L}_\perp,y)- {\vec h}({\bf
0}_\perp,y)|^2\rangle^{1/2} 
\bigg{|}_{e^{-z l}L_y,t(l),\kappa(l),\ldots}\nonumber\\ 
&=&e^{\nu l} R_G(e^{-l}L_\perp,e^{-z l} L_y;t(l),\kappa(l),\ldots)\;,
\nonumber\\
\label{matching1}
\end{eqnarray}
where $t(l),\kappa(l),\ldots$ stand for all flowing coupling constants
whose evolution with $l$ is determined by the recursion relations
Eqs.\ref{t_rr}-\ref{v_rr}. Choosing $l=l_*=\log L_\perp$ this becomes:
\begin{equation}
R_G(L_\perp,L_y;t,\kappa,\ldots)=
L_\perp^\nu R_G(1,L_y/L_\perp^z;t(l_*),\kappa(l_*),\ldots)\;.
\label{matching2}
\end{equation}
This relation holds for {\it any} choice of the (after all, arbitrary)
rescaling exponents $\nu$ and $z$. However, {\it if} we make the
special choice such that Eqs.\ref{t_rr}-\ref{v_rr} lead to fixed
points (see Eqs.\ref{t_fixed}-\ref{v_fixed}), $t(l_*)$,
$\kappa(l_*),\ldots$ in Eq.\ref{matching2} go to {\it constants},
independent of $l_*$ (and hence $L_\perp$), as $L_\perp$ and hence
$l_*$, go to infinity. Thus, in this limit, we obtain from
Eq.\ref{matching2}
\begin{equation}
R_G(L_\perp,L_y;t,\kappa,\ldots)=
L_\perp^\nu R_G(1,L_y/L_\perp^z;t_*,\kappa_*,\ldots)\;,
\label{matching3}
\end{equation}
where $t_*, \kappa_*, \ldots$ are the fixed point values of coupling
constants. This result clearly agrees with the scaling forms for
$R_G$, Eq.\ref{RG} (with analogous derivation for $h_{rms}$) if we
define $S_R(x)\equiv R_G(1,x;t_*,\kappa_*,g_y^*,v^*)$.

The recursion relations Eqs.\ref{t_rr}-\ref{v_rr} reproduce all of
our phantom membrane results, as well as the upper critical embedding
dimension $d_{uc}^{SA}$ for self-avoidance predicted by Flory theory,
Eq.\ref{lambda}, {\em and} the upper critical {\it intrinsic}
dimension $D_{uc}=5/2$ for anomalous elasticity for phantom
membranes. To see this, consider first the phantom membrane; i.e.,
$v=0$. In this case, $f_t(v)=0$, and to keep $t(l)$ fixed we see from
the recursion relation Eq.\ref{t_rr} for $t(l)$ that we must choose
\begin{equation}
2\nu+z+D-3=0\;.\label{phantom1}
\end{equation}
Assuming for the moment that $f_\kappa(g_y,g_\perp)\rightarrow 0$ as
$l\rightarrow\infty$, which, as we shall see in a moment, it does for
phantom membranes for $D>3/2$, we see from the recursion relation
Eq.\ref{kappa_rr} for $\kappa(l)$ that we must choose
\begin{equation}
2\nu-3z+D-1=0\;.\label{phantom2}
\end{equation}
Solving Eqs.\ref{phantom1} and \ref{phantom2} for $z$ and $\nu$ yields
the phantom membrane results $z=1/2$, $\nu=(5-2D)/4$, as obtained in
Eqs.\ref{nu_phantom} and \ref{z_phantom2}.

To extract the upper-critical embedding dimension $d_{uc}^{SA}$ for
self-avoidance from the renormalization group recursion relations, we
construct from them a flow equation for a dimensionless coupling
constant 
\begin{equation}
\tilde{v}=v t^a \kappa^b\;,
\end{equation}
where $a$ and $b$ will be chosen to eliminate the arbitrary rescaling
exponents $\nu$ and $z$ from the recursion relation for
$\tilde{v}$. This requirement lead to the choice
\begin{eqnarray}
a&=&(3d-5)/8\;,\label{a}\\
b&=&(d+1)/8\;,\label{b}
\end{eqnarray}
which implies:
\begin{eqnarray}
{d \tilde{v}\over d l}&=&\left(\left[6D-1-(5-2D)d\right]/4-
f_v+{d+1\over8}f_\kappa\right.\nonumber\\
&&\left.-{3d-5\over8}f_t\right)\tilde{v}\;,
\label{vtilde_rr}
\end{eqnarray}
Of course, an identical flow equation is obtained for $v(l)$ if one
instead requires that $t(l)$ and $\kappa(l)$ are fixed, i.e.,
independent of $l$, thereby determining $\nu$ and $z$ and using them
inside Eq.\ref{v_rr}.

It is easy to see that the sign of the terms in the square bracket
determines the relevance of the self-avoiding interaction, which
becomes relevant when 
\begin{equation}
6D-1-(5-2D)d>0\;,
\end{equation}
i.e., for $d<d_{uc}^{SA}=(6D-1)/(5-2D)$, consistent with the analysis
of the Flory theory, Eq.\ref{lambda}. 

Likewise, the renormalization group flow equations contain information
about the upper-critical {\it intrinsic} dimension for the anomalous
elasticity, $D_{uc}$, below which tubule elasticity becomes
anomalous. This can be seen (analogously to the discussion of the
relevance of self-avoidance coupling $v$) by using
Eqs.\ref{t_rr}-\ref{gy_rr} to construct the renormalization group flow
equation for the dimensionless coupling constant
\begin{equation}
\tilde{g}_y={g_y\over t^{3/4} \kappa^{5/4}}\;,
\end{equation}
chosen such that its flow
\begin{equation}
{d \tilde{g}_y\over d l}=\left({5\over2}-D-f_g - {5\over4} f_\kappa + 
{3\over4} f_t\right)\tilde{g}_y\;,
\label{gtilde_rr}
\end{equation}
is independent of the arbitrary rescaling exponents $z$ and
$\nu$. Again the same recursion relation can be obtained by instead
using the values of $z$ and $\nu$ required to keep $t(l)$ and
$\kappa(l)$ fixed inside the flow equation for $g_y(l)$,
Eq.\ref{gy_rr}.  It is then obvious that anharmonic elasticity becomes
relevant for $D<D_{uc}=5/2$, where anomalous elasticity of the tubule
is induced. As we will see below, in a phantom tubule or a tubule
embedded in $d>d*$, this anomalous elasticity manifests itself {\it
only} in phonon ($u$) fluctuations, i.e., softens $g_y$, but does {\it
not} renormalize the bending rigidity $\kappa$.  In {\it physical}
tubules, however, which are self-avoiding and are embedded in
$d=3<d_*\approx6.5$, the elasticity is fully anomalous, both with
respect to the phonon $u$ fluctuations (i.e. $g_y$ vanishes as
$q\rightarrow0$) and the height $\vec{h}$ undulations (i.e. $\kappa$
diverges as $q\rightarrow0$).

To further analyze the renormalization of $\kappa$ in a self-avoiding
membrane, it is convenient to integrate out the phonon field $u$ as we
did in Sec.\ref{tubuleF} for the phantom tubule, obtaining
\begin{eqnarray}
F&=&{1\over2}\int d^{D-1}x_\perp dy \big[\kappa(\partial^2_y{\vec
h})^2 + t(\partial^\perp_\alpha \vec h)^2\big]\nonumber\\
&+&F_{anh}[\vec{h}]+F_{SA}[\vec{h}]
\label{SAtubuleRG2}
\end{eqnarray}
where, $F_{anh}$ is the non-local interaction, Eq.\ref{Fanh}, mediated
by integrated out phonons, with a kernel
\begin{equation}
V_h({\bf q})={g_y g_\perp q_\perp^2\over g_y q_y^2+g_\perp q_\perp^2}\;,
\label{vertexV_h2}
\end{equation}
and $F_{SA}$ is the self-avoiding interaction. 

The long wavelength properties of the tubule phase will very much
depend on the behavior of the denominator in the kernel $V_h$ at long
length scales. If $g_y({\bf q}) q_y^2 >> g_\perp({\bf q}) q_\perp^2$
(as we saw for a phantom tubule) then at long scales $V_h({\bf
q})\approx g_\perp q_\perp^2/q_y^2$, which behaves like $\sim q_y^2$
in the relevant limit of $q_\perp\sim q_y^2$. In this case, simple
power counting around the {\it Gaussian} fixed point then shows that
this elastic nonlinearity only becomes relevant for $D<D_{uc}=3/2$,
i.e. is irrelevant for a physical $D=2$-dimensional tubule, as we
argued in Sec.\ref{tubuleF}.

On the other hand, if the scaling is such that $g_\perp({\bf q})
q_\perp^2$ dominates over $g_y({\bf q}) q_y^2$, then $V_h({\bf
q})\approx g_y$, i.e. a constant at long length scales. Simple
power-counting then shows that this coupling is relevant for
$D<D_{uc}=5/2$ and the bending rigidity modulus of a $D=2$-dimensional
tubule {\it is} anomalous in this case.

As we saw in our analysis of a {\it phantom} tubule, for which one is
perturbing around a {\it Gaussian} fixed point described by
$q_\perp\sim q_y^2 << q_y$ (in the long wavelength limit), the
anharmonic nonlinearity is irrelevant for $D>3/2$ and $\kappa$ is {\it
not} anomalous.  We now need to extend this analysis to a physical
tubule, i.e., to include the effects of self-avoidance.

The analysis of the behavior of $V_h({\bf q})$ (which determines the
relevance of anharmonic elasticity) at long scales, around an
arbitrary fixed point, is more conveniently done using the language of
the renormalization group through the recursion relations
Eqs.\ref{t_rr} and \ref{v_rr}. At the globally stable fixed point, in
the presence of both the nonlinear elasticity and the self-avoiding
interaction, we can keep $t=\kappa=1$ and $g_y$ and $v$ fixed at fixed
point values, by requiring
\begin{eqnarray}
2\nu+z+D-3-f_t(v^*)&=&0\;,
\label{t_fixed}\\
2\nu-3z+D-1+f_\kappa(g_y^*,g_\perp^*)&=&0\;,
\label{kappa_fixed}\\
4\nu-3z+D-1-f_g(g_y^*)&=&0\;,
\label{gy_fixed}\\
2(D-1) + z - \nu(d-1)-f_v(v^*)&=&0\;.
\label{v_fixed}
\end{eqnarray}
In light of the above discussion, the anharmonic vertex for $\vec{h}$
in this renormalization group picture becomes relevant when
$g_\perp(l\rightarrow\infty)$ renormalizes to infinity, while it is
{\it irrelevant} when $g_\perp(l\rightarrow\infty)$ flows to
zero. Thus, the relevance of $V_h$ is decided by the sign of the
renormalization group flow eigenvalue of $g_\perp(l)$ in
Eq.\ref{gp_rr}
\begin{equation}
\lambda_{g_\perp}=4\nu-z+D-3\;,
\label{lambda_gp}
\end{equation}
which is {\it exactly} determined by the values of $\nu$ and $z$,
since $g_\perp$ suffers no graphical renormalization.

As we have discussed in previous sections, for a phantom tubule
$\nu=(5-2D)/4$ and $z=1/2$. For $d<d_{uc}^{SA}=(6D-1)/(5-2D)$ ($=11$
for $D=2$), these values are modified by the self-avoiding
interaction, but only by order $\epsilon\equiv
d-d_{uc}^{SA}$, i.e.
\begin{eqnarray}
\nu&=&(5-2D)/4+O(\epsilon)\;,\\
z&=&1/2+O(\epsilon)\;.
\end{eqnarray}
Hence a $D=2$-dimensional tubule, embedded in $d$ dimensions close to
$d_{uc}^{SA}=11$, $\lambda_{g_\perp}=-1/2$ and $g_\perp(l)$ flows
according to
\begin{equation}
{d g_\perp\over d l}=[-{1\over 2} + O(\epsilon)]g_\perp\;,
\label{gp_rr11}
\end{equation}
i.e. $g_\perp$ is {\em irrelevant} near $d=11$ (for $\epsilon\ll 1$),
$V_h({\bf q})\sim g_\perp q_\perp^2/q_y^2\sim q_y^{2-O(\epsilon)}$ is
irrelevant for a physical $D=2$-dimensional tubule , and, hence,
$f_\kappa$ in Eq.\ref{kappa_rr} vanishes as $l\rightarrow\infty$. So
$\kappa$ is unrenormalized near $d=11$, for $D=2$. That is, as we
described above, the anharmonic elasticity is irrelevant to the {\it
bend} elasticity for embedding dimensions near $d_{uc}^{SA}$, and in
this case the full model of a self-avoiding tubule with nonlinear
elasticity reduces to the {\em linear} elastic truncated model
introduced by us\cite{RT} and recently further analyzed in
Ref.~\onlinecite{BG}.

In this simpler (but unphysical) case, one is justified in ignoring
the nonlinear elasticity. One is then able to analyze (perturbatively
in $\epsilon=d_{uc}^{SA}-d$) the effects of the self-avoiding
interaction alone, by computing the functions $f_t(v)$ and $f_v(v)$
appearing in Eqs.\ref{t_rr} and \ref{v_rr}.\cite{BG} Since, as we
discussed above, the ``tubule gauge'' symmetry guarantees that in this
case the self-avoiding interaction alone cannot renormalize $\kappa$,
$f_\kappa=0$. Thus, for $d$ near $d_{uc}^{SA}$, Eq.\ref{kappa_fixed},
leads to $\eta_\kappa=0$ and an {\it exact} exponent relation (leaving
only a single independent tubule shape exponent):
\begin{equation}
z={1\over 3}(2\nu+D-1)\;,
\label{z_nu_phantom}
\end{equation}
which is exact for a finite {\it range} $d_*<d<d_{uc}^{SA}$ of
embedding dimensions, and for phantom tubules in any embedding
dimension. This result has been independently obtained in
Ref.\onlinecite{BG}.

However, this simple scenario, and, in particular, the scaling
relation Eq.\ref{z_nu_phantom}, is {\it guaranteed} to break down as
$d$ is reduced. The reason for this is that, as $d$ decreases, $\nu$
increases, and eventually becomes so large that the eigenvalue
$\lambda_{g_\perp}$ of $g_\perp$ changes sign and becomes positive. As
discussed earlier, once this happens, the nonlinear vertex
Eq.\ref{vertexV_h2} becomes relevant, and $\kappa$ acquires a
divergent renormalization, i.e., $f_\kappa\neq 0$, and bend tubule
elasticity becomes anomalous. We will now show that the critical
dimension $d_*$ below which this happens for $D=2$ is {\it guaranteed}
to be $>7/2$, and hence, obviously, $>3$.

To show this, we use the exponent relation Eq.\ref{z_nu_phantom},
which is valid for $d>d_*$, inside the expression for the eigenvalue
$\lambda_{g_\perp}$, Eq.\ref{lambda_gp}, obtaining
\begin{equation}
\lambda_{g_\perp}={1\over 3}(10\nu + 2D-8)\;.
\label{lambda_gp2}
\end{equation}
We then take advantage of a rigorous lower bound on $\nu$
\begin{equation}
\nu>{D-1\over d-1}\;,
\label{nu_bound}
\end{equation}
imposed by the condition that the monomer density $\rho\propto
L_\perp^{D-1}/R_G^{d-1}\propto L_\perp^{D-1-\nu(d-1)}$ remain finite
in the thermodynamic $L_\perp\rightarrow\infty$ limit. Using this
bound inside Eq.\ref{lambda_gp2} we obtain
\begin{equation}
\lambda_{g_\perp}\geq {1\over 3}\left(10{D-1\over d-1} + 2D - 8\right)\;,
\label{lambda_gp3}
\end{equation}
from which it follows that $\lambda_{g_\perp}$ {\it must} become
positive for $d<d_*^{lb}(D)$ with
\begin{eqnarray}
d_*^{lb}(D)&=&{4D-1\over 4-D}\;,\label{d_starD}\\
d_*^{lb}(2)&=&7/2\;,\label{d_star2}
\end{eqnarray}
as asserted above.

\begin{figure}[bth]
\centering
\setlength{\unitlength}{1mm}
\begin{picture}(150,70)(0,0)
\put(-20,-70){\begin{picture}(150,70)(0,0)
\includegraphics{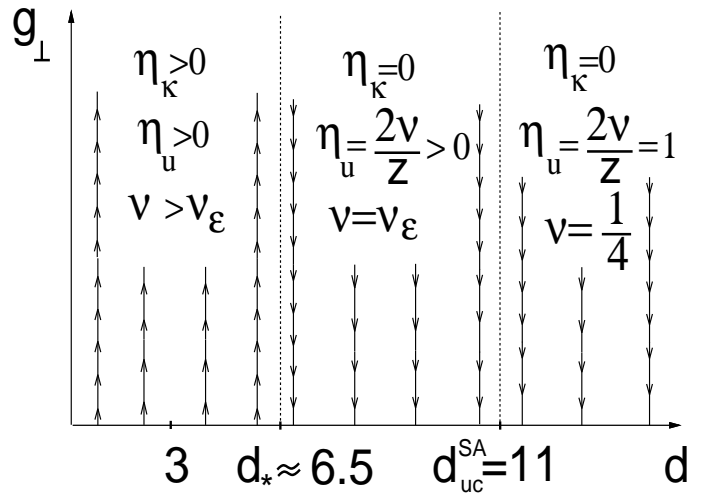}
\end{picture}}
\end{picture}
\caption{Schematic illustration (specialized to $D=2$) of change in
relevance of $g_\perp(l)$ which occurs at $d_*$. For embedding
dimensions below $d_*$ (which includes the physical case of $d=3$),
$g_\perp(l)$ becomes relevant, leading to anomalous bending elasticity
with $\kappa({\bf q})\sim q_y^{-\eta_\kappa}$, which diverges at long
length scales. Other consequences of this qualitative and quantitative
change for $d<d_*$ are discussed in the text.}
\label{gperp_d}
\end{figure}

In fact, $d_*(2)$ is probably quite a bit bigger than its $7/2$ lower
bound, as two estimates of it indicate. If, for example, we take the
Flory tubule exponent $\nu=(D+1)/(d+1)$ in Eq.\ref{lambda_gp2}, we
obtain:
\begin{eqnarray}
d_*^{F}&=&{6D+1\over 4-D}\;,\label{d_starDF}\\
d_*^{F}(2)&=&13/2\;,\label{d_star2F}
\end{eqnarray}
while if we use the $\epsilon=11-d$--expansion result for $\nu$ of
Bowick and Guitter, ($D=2$)\cite{BG}
\begin{equation}
\nu_\epsilon={3\over4+\delta}-{1\over2}\;,\label{nu_eps}
\end{equation}
with 
\begin{equation}
\delta=-1.05 \left({\epsilon\over 8}\right)\;,\label{delta_eps}
\end{equation}
we obtain
\begin{equation}
d_*^\epsilon=5.92\;.\label{d_star_eps}
\end{equation}
So, based on the above estimates, we expect that in a
$D=2$-dimensional tubule, embedded in $d<d_*\approx 6$, the fixed
point of the truncated tubule model introduced by us\cite{RT} and
studied in Ref.~\onlinecite{BG}, is {\em unstable} to anharmonic
elasticity $F_{anh}$. This means that $\kappa$ diverges at long length
scales, and the scaling relation Eq.\ref{z_nu_phantom} between $z$ and
$\nu$ breaks down. Thus, for the physical embedding dimension $d=3$,
the tubule bend elasticity is certainly anomalous, in the sense that
$\kappa$ diverges, and probably quite strongly. We have summarized the
above discussion in Fig.\ref{gperp_d}, schematically illustrating how
the renormalization group flow of $g_\perp$, and therefore the
anomalous $\kappa$ elasticity, change (at $d_*$) as a function of
embedding dimension $d$.

Once $d<d_*$, the new nontrivial relations Eqs.\ref{kappa_fixed} and
\ref{gy_fixed} hold, with functions $f_\kappa(g_y, g_\perp)$ and
$f_g(g_y)$ evaluated at the fixed point values $g_y^*$ and
$g_\perp^*$. 

Using the sort of renormalization group correlation function matching
calculations described earlier, Eqs.\ref{matching1}-\ref{matching3},
it is straightforward to show that the correlation functions of the
tubule, including anomalous elastic effects, are correctly given by
the harmonic results, Eqs.\ref{hProp2} and \ref{uProp2}, {\it except}
that the elastic {\it constants} $g_y$ and $\kappa$ must be replaced
by wavevector dependent quantities that vanish and diverge,
respectively as ${\bf q}\rightarrow 0$:
\begin{eqnarray}
g_y({\bf q})&=&q_y^{\eta_u} S_g(q_y/q_\perp^z)\;,\label{gyScaleForm}\\
\kappa({\bf q})&=&q_y^{-\eta_\kappa}
S_\kappa(q_y/q_\perp^z)\;,\label{kappaScaleForm}
\end{eqnarray}
with 
\begin{eqnarray}
z\eta_\kappa&=&f_\kappa(g_y^*, g_\perp^*)\;,
\label{eta_kappa}\\
z\eta_u&=&f_g(g_y^*)\;.
\label{eta_u}
\end{eqnarray}

Our earlier conclusion that the relevance of $V_h$ is determined by
the sign of $\lambda_{g_\perp}$ (Eq.\ref{lambda_gp}) can be reproduced
by simply noting that $g_y({\bf q}) q_y^2$ scales like
$q_y^{\eta_u+2}$, and in the long wavelength limit is therefore
subdominant to $g_\perp q_\perp^2\sim q_y^{2/z}$ when
\begin{equation}
z\eta_u > 2 - 2z\;,
\label{z_etau}
\end{equation}
which, upon using Eq.\ref{gy_fixed} and the definition of
$\eta_u=f_g(g_y^*)/z$, is identical to the condition that
$\lambda_{g_\perp}>0$.

The scaling functions have the asymptotic forms
\begin{eqnarray}
S_g(x\rightarrow 0)&\rightarrow& x^{-\eta_u}\;,
\label{S_g}\\
S_\kappa(x\rightarrow 0)&\rightarrow& x^{\eta_\kappa}\;.
\label{S_kappa}
\end{eqnarray}

Combining the expressions Eqs.\ref{eta_kappa} and \ref{eta_u} for
$\eta_\kappa$ and $\eta_u$ with the RG fixed point conditions
Eqs.\ref{kappa_fixed} and \ref{gy_fixed} shows that, at this new
globally stable fixed point, {\em two} exact relations hold between
{\em four} independent exponents $z$, $\nu$, $\eta_\kappa$, and
$\eta_u$ (instead of a single relation Eq.\ref{z_nu_phantom} between
two exponents)
\begin{eqnarray}
z&=&{1\over 3-\eta_\kappa}(2\nu+D-1)\;,
\label{z_nu_new1}\\
z&=&{1\over 3+\eta_u}(4\nu+D-1)\;.
\label{z_nu_new2}
\end{eqnarray}
That is, in contrast to the behavior for $d>d_*$, for $d<d_*$ there
are {\it two} independent exponents characterizing the tubule phase,
{\em not one}. We furthermore note that these exponent relations
automatically contain the rotational symmetry Ward identity. This can
be easily seen by eliminating $\nu$ from Eqs.\ref{z_nu_new1} and
\ref{z_nu_new2}, obtaining
\begin{equation}
2\eta_\kappa+\eta_u=3-(D-1)/z\;.
\label{rot_symm}
\end{equation}
Ultimately, the origin of this relation is the requirement that
graphical corrections do not change the form of the rotationally
invariant operator $(\partial_y u+{1\over2}(\partial_y{\vec h})^2)$.

\begin{figure}[bth]
\centering
\setlength{\unitlength}{1mm}
\begin{picture}(150,70)(0,0)
\put(-20,-70){\begin{picture}(150,70)(0,0)
\includegraphics{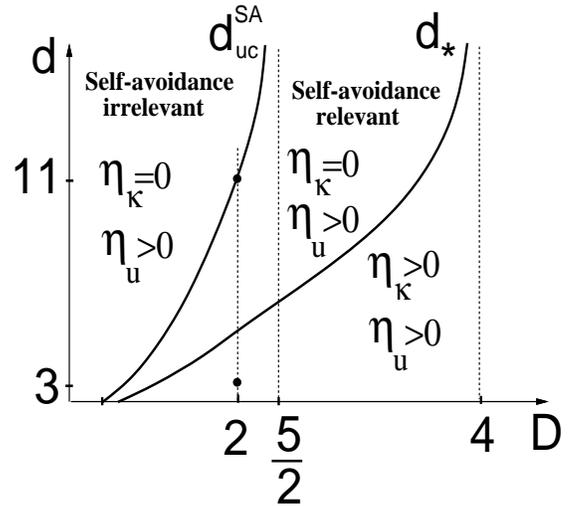}
\end{picture}}
\end{picture}
\caption{Schematic of the tubule ``phase'' diagram in the embedding
$d$ vs intrinsic $D$ dimensions. Self-avoiding interaction becomes
relevant for $d<d_{uc}^{SA}(D)=(6D-1)/(5-2D)$, ($=11$, for
$D=2$). Below the $d_*(D)$ curve (for which the lower bound is
$d_*^{lb}(D)=(4D-1)/(4-D)$) the anharmonic elasticity becomes
relevant, leading to anomalous elasticity with a divergent bending
rigidity.}
\label{d_vs_D}
\end{figure}

Just as the divergence of $\kappa$ is controlled by $f_\kappa(g_y^*,
g_\perp^*)$, the softening of $g_y({\bf q})\sim q_y^{\eta_u}$ is
determined by the $\eta_u=z f_g(g_y^*)$. Because $f_g(0)=0$, this {\em
physical} $g_y({\bf q})$ remains non-zero and finite as
$q\rightarrow0$, {\em only if} the running coupling $g_y(l)$ in the
renormalization group recursion equation Eq.\ref{gy_rr} {\em does} go
to zero (because then the graphical piece $f_g(g^*)$
vanishes). Examining the flow equation for $g_y(l)$, Eq.\ref{gy_rr},
for $g_y(l)$ to vanish, we must have
\begin{equation}
4\nu - 3z + D-1 < 0\;.
\label{condition_gy}
\end{equation}
However, using the lower bound on $\nu$, Eq.\ref{nu_bound} in the
physical case of $D=2$ and $d=3$, we find $\nu>1/2$. Hence, as long as
$z<1$, the condition Eq.\ref{condition_gy} is {\em not} satisfied, and
therefore $g_y({\bf q}\rightarrow 0)\rightarrow 0$, that is,
$\eta_u>0$. We summarize the above discussion in Fig.\ref{d_vs_D}.
We now show that the above general analysis of tubule anomalous
elasticity in the presence of self avoidance, obtained using the
renormalization group, can be reproduced via a heuristic, but
beautiful physical argument similar to that used by Landau and
Lifshitz\cite{LandauLifshitz} to derive shell theory.  For a tubule of
diameter $R_G$, the non-zero shear $g_y$ elasticity leads to an
effective $R_G$-dependent bending rigidity modulus which will be
$L_\perp$ and $L_y$-dependent if the tubule diameter depends on
$L_\perp$ and $L_y$. This can be seen as follows (see
Fig.\ref{bend_tubule}):
\begin{figure}[bth]
\centering
\setlength{\unitlength}{1mm}
\begin{picture}(150,57)(0,0)
\put(-17,-95){\begin{picture}(150,57)(0,0)
\includegraphics{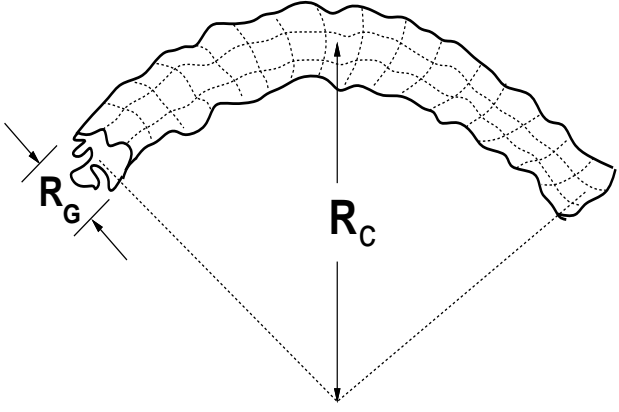}
\end{picture}}
\end{picture}
\caption{Illustration of the physical mechanism for the enhancement of
the bending rigidity $\kappa$ by the shear $g_y$ elasticity. To bend a
polymerized tubule of thickness $R_G$ into an arc of radius $R_c$
requires $R_G/R_c$ fraction of bond stretching and therefore costs
elastic shear energy, which when interpreted as bending energy leads
to a length-scale dependent renormalization of the bending rigidity
$\kappa$ and to the Ward identity Eq.\ref{expRelation}, as described
in more detail in the text.}
\label{bend_tubule}
\end{figure}
If we bend the tubule with some radius of curvature $R_c\gg R_G$,
simple geometry tells us that this will induce a strain
$\varepsilon\sim\partial_y u$ {\it along} the tubule axis of order
$\varepsilon\sim R_G/R_c$, since the outer edge of the tubule must be
stretched by this factor, and the inner edge compressed by it, in
order to accomplish the required bend. This strain induces an
additional elastic energy density (i.e., additional to those coming
from the bare $\kappa$), namely those coming from the $u$ elastic
energy. This goes like
$g_y(L_y,L_\perp)\varepsilon^2=g_y(L_y,L_\perp)(R_G(L_y)/R_c)^2$.
Interpreting this additional energy as an effective bending energy
density $\kappa_y(L_\perp , L_y)/R_c^2$, leads to the {\it effective}
bending modulus $\kappa_y(L_\perp ,L_y)$,
\begin{equation}
\kappa_y(L_\perp,L_y)\sim
g_y(L_\perp,L_y)R_G(L_\perp,L_y)^2\;,\label{kappaInduced}
\end{equation}
Inserting the scaling forms
$\kappa_y(L_\perp,L_y)=L_y^{\eta_\kappa}S_\kappa( L_y/ L_\perp^z)$,
$g_y(L_\perp,L_y)=L_y^{-\eta_u}S_g(L_y/ L_\perp^z)$ and
$R_G(L_\perp,L_y)=L_\perp^{\nu}S_R(L_y/ L_\perp^z)$ into above
expression, we obtain a relation between the scaling exponents
\begin{equation}
2\nu=z(\eta_\kappa+\eta_u)\;.\label{expRelation}
\end{equation}
which is exactly the exponent relation one obtains by subtracting
Eq.\ref{kappa_fixed} from Eq.\ref{gy_fixed}, and using the expressions
Eq.\ref{eta_kappa} and \ref{eta_u} for $\eta_\kappa$ and $\eta_u$, all
of which were obtained using renormalization group arguments.

Since the above physical shell argument is very general,
Eqs.\ref{kappaInduced} and \ref{expRelation} hold independent of the
mechanism that generates anomalous elasticity. For the case of the
phantom membrane (for $D>3/2$) Eq.\ref{kappaInduced} reveals that
$\kappa$ is not anomalous because the softening of the shear modulus
$g_y({\bf q})$ by thermal fluctuations precisely compensates the
bending rigidity produced by the finite diameter $R_G$ of the
tubule. Equation \ref{expRelation} then correctly predicts for the
{\it phantom} tubule that $\eta_u=2\nu/z$, which is consistent with
the phantom tubule results $\eta_u=5-2D$, $\nu=(5-2D)/4$, and
$z=1/2$. Furthermore, because the anharmonic elasticity $V_h({\bf q})$
is irrelevant for $d>d_*$,
\begin{equation}
\eta_u=2\nu/z\;,
\label{eta_u_nu}
\end{equation}
is valid, even in a self-avoiding tubule embedded in these high
dimensions.

\begin{figure}[bth]
\centering
\setlength{\unitlength}{1mm}
\begin{picture}(150,70)(0,0)
\put(-20,-70){\begin{picture}(150,70)(0,0)
\includegraphics{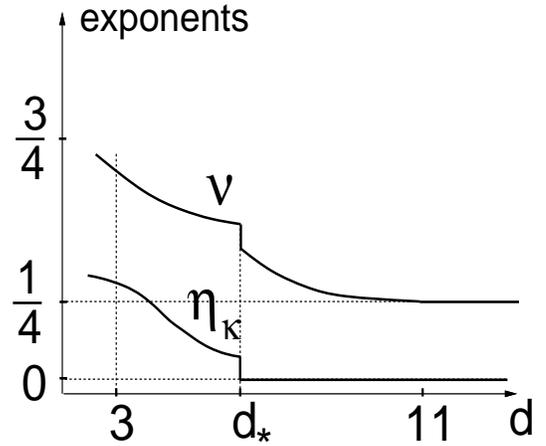}
\end{picture}}
\end{picture}
\caption{Schematic graph of the shape exponent $\nu$ and anomalous
bend exponent $\eta_\kappa$ (for $D=2$). Note the jump discontinuity
as a function of embedding dimension $d$, occurring at $d=d_*\approx
6$.}
\label{exponent_fig}
\end{figure}

We note, finally, that all of the exponents must show a jump
discontinuity at $d_*$, as shown in Fig.\ref{exponent_fig}. Therefore,
unfortunately, an extrapolation from $\epsilon=11-d$--expansion in a
truncated model with linear elasticity\cite{BG} down to the physical
dimension of $d=3$ (which is below $d_*$) gives little information
about the properties of a real tubule.
The computations for a physical tubule must be performed for $d<d_*$,
where both the self-avoidance and the anharmonic nonlinearities are
relevant and must be handled simultaneously. As we discussed above,
for $d<d_*$, the eigenvalue $\lambda_{g_\perp}>0$, leading to the flow
of $g_\perp(l)$ to infinity, which in turn leads to $V_h({\bf q})=
g_y$. Physically this regime of $g_\perp\rightarrow\infty$ corresponds
to freezing out the phonons $u$, i.e. setting $u=0$ in the free energy
$F[\vec{h},u]$ in Eq.\ref{SAtubuleRG}. This is consistent with our
finding that for $d<d_*$, in the effective free energy $F[\vec{h}]$
(with phonons integrated out), Eq.\ref{SAtubuleRG2}, the kernel
$V_h=g_y$. The resulting effective free energy functional for a
physical self-avoiding tubule is 
\begin{eqnarray}
F&=&{1\over2}\int d^{D-1}x_\perp dy
\bigg[\kappa(\partial^2_y{\vec h})^2 + t(\partial^\perp_\alpha \vec h)^2
+ {1\over4}g_y(\partial_y{\vec h})^4\nonumber\\
&+&{v}\int\hspace{-0.1cm}dy\hspace{0.05cm} 
d^{D-1}x_\perp d^{D-1}x'_\perp
\hspace{0.05cm}\delta^{(d-1)}\big({\vec h}({\bf x_\perp},y)-
{\vec h}({\bf x'_\perp},y)\big)\;,\nonumber\\
\label{SAtubuleRGeff}
\end{eqnarray}
Unfortunately, no controlled perturbative study is possible for
$d<d_*$, since one must perturb in $g_y$ around a nontrivial, {\it
strong} coupling fixed described by $v^*=O(1)$ and
$g_y^*=0$. Furthermore, as we will show below, at {\em this} fixed
point there is no upper critical dimension for $g_y$, i.e. anharmonic
nonlinearities are always relevant for $d=3<d_*$, for {\em any}
$D$. This strongly contrasts with the Gaussian fixed point (describing
phantom membranes) at which the anharmonic nonlinearity is only
relevant for $D<D_{uc}=5/2$.

In what follows, we will illustrate how one might attempt to actually
calculate the exponents $\nu$, $z$, $\eta_\kappa$, and $\eta_u$, for
$d<d_*$, and enumerate the (many) technical difficulties that prevent
us from doing so, and conclude with a cautionary list of several
unsuccessful uncontrolled approximations that we have tried.

In principle, all we need to do is calculate the $f_i$
($i=t,v,g,\kappa$) functions in the recursion relation
Eqs.\ref{t_rr}-\ref{v_rr}, which represent the perturbative
(``graphical'') corrections to the associated coupling constants.
Once these $f$-functions are known they give $4$ equations
(Eqs.\ref{t_fixed}-\ref{v_fixed}) that uniquely determining the $4$
unknowns tubule shape exponents, $\nu$, $z$, $v^*$, and $g_y^*$, as
well as the the flow of $g_\perp(l)$, and therefore completely
characterize the long wavelength properties of self-avoiding
anharmonic tubules.

\begin{figure}[bth]
\centering
\setlength{\unitlength}{1mm}
\begin{picture}(150,70)(0,0)
\put(-20,-70){\begin{picture}(150,70)(0,0)
\includegraphics{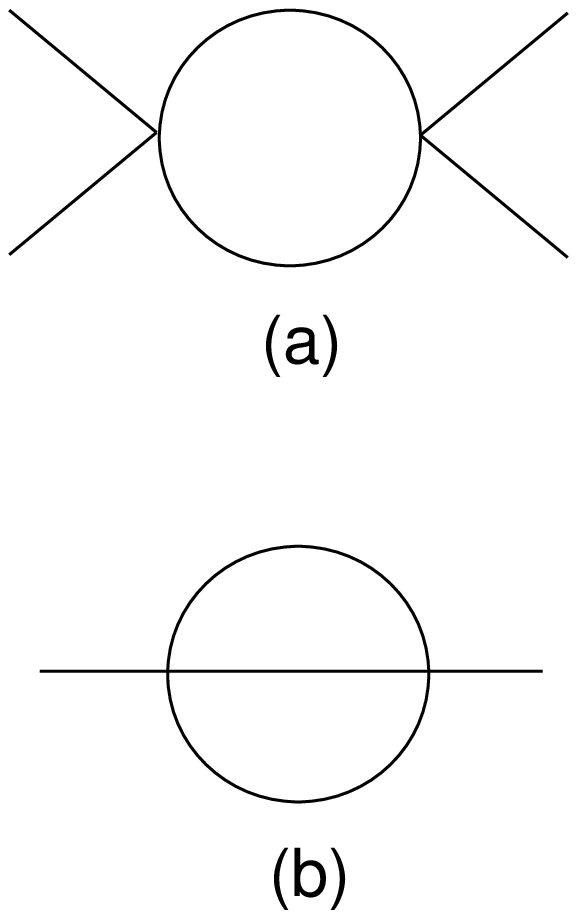}
\end{picture}}
\end{picture}
\caption{Feynman graphs that renormalize: (a) the anharmonic
elasticity $g_y$,and (b) the bending rigidity $\kappa$.}
\label{d_kappa_gy}
\end{figure}

Our goal then is to calculate $f_t(v)$, $f_v(v)$, $f_\kappa(g_y)$, and
$f_g(g_y)$. The functions  $f_g(g_y)$ and $f_\kappa(g_y)$ are
determined by the diagrammatic corrections to $g_y$ and $\kappa$,
with the corresponding Feynman diagrams displayed in
Fig.\ref{d_kappa_gy}. The results to leading order in $g_y$, are
\begin{eqnarray}
f_\kappa(g_y)&=&C_\kappa g_y^2\;,\label{f_kappa2}\\
f_g(g_y)&=&C_g g_y\;,\label{f_gy}
\end{eqnarray}
where $C_\kappa$ and $C_g$ are $d$ and $D$-dependent constants, whose
calculation proves to be the sticking point, as we will describe
below.

Of course, once $d$ is below $d_*$, {\it no matter how close it is to
$d_*$}, the fixed point that controls the elastic properties of the
tubule phase is {\it not} perturbative in $g_y$. That is, we do {\it
not} expect $g_y$ to be $O(d_*-d)$, but, rather, $O(1)$, even for
$d_*-d<<1$. Furthermore, of course, since $d_*\approx 6$, $d_*-d$ is
not small in the physical case $d=3$ anyway. For both of these
reasons, truncating the calculations of $f_\kappa$ and $f_g$ at the
leading order in $g_y$, as we have done in Eqs.\ref{f_kappa2} and
\ref{f_gy}, is an uncontrolled, and far from trustworthy
approximation. However, we know of no other analytical
approach. Furthermore, as we shall see, even this uncontrolled
analytic approach proves intractable: a reliable calculation of the
values of the constants $C_\kappa$ and $C_g$ has eluded us.

To complete the characterization of the fixed point we can proceed in
two ways. The most direct way is to simply perturbatively evaluate the
functions $f_t(v)$, $f_v(v)$. Luckily (for us) this has recently been
done by Bowick and Guitter\cite{BG} in a truncated harmonic tubule
model (previously introduced and studied by us\cite{RT}) near
$d=d_{uc}^{SA}$. Although, for the reasons that we discussed above,
these calculations are not rigorously applicable to a physical tubule
in $d=3<d_*$ (where anharmonic elasticity is certainly important), for
lack of being able to do any better we extrapolate {\em these}
functions, computed near $d=11$,\cite{BG} down to $d=3$
\begin{eqnarray}
f_t(v)&=&C_t v\;,\label{f_t}\\
f_v(v)&=&C_v v\;,\label{f_v}
\end{eqnarray}

Now using Eqs.\ref{f_kappa2}--\ref{f_v} in
Eqs.\ref{t_fixed}--\ref{v_fixed} we obtain four equations for four
unknowns ($z$, $\nu$, $g_y^*$, and $v^*$), expressed in terms
constants $C_\kappa$, $C_g$, $C_t$ and $C_v$, (specialized here to
$D=2$).
\begin{eqnarray}
2\nu+z-1-C_t v^*&=&0\;,
\label{t_fixed2}\\
2\nu-3z+1+C_\kappa {g_y^*}^2&=&0\;,
\label{kappa_fixed2}\\
4\nu-3z+1-C_g g_y^*&=&0\;,
\label{gy_fixed2}\\
2+z-\nu(d-1)-C_v v^*&=&0\;.
\label{v_fixed2}
\end{eqnarray}
where the constants $C_t$ and $C_v$ (computed in the truncated
tubule model near $d=d_{uc}^{SA}$ for $D=2$) are given by\cite{BG}
\begin{eqnarray}
C_t&=&{1\over8\pi^2}\;,\label{Ct}\\
C_v&=&{0.068\over\pi^{5/2}}\;,\label{Cv}
\end{eqnarray}
These equations can be uniquely solved for $\nu$, $z$, $g_y^*$, and
$v^*$. In terms of $C_\kappa$ and $C_g$, in $D=2$ and $d=3$ we obtain
for $\nu$ and $z$
\begin{eqnarray}
\nu&=&{1\over 4 C_\kappa}\;,
\label{nu1}\\
z&=&{1\over 3}+{1\over 6 C_\kappa}+ {C_g\over 6 C_\kappa}\;,
\label{z1}
\end{eqnarray}
from which $\eta_\kappa$ and $\eta_u$ can also be determined using the
solution for $g_y^*$ inside Eqs.\ref{eta_kappa} and \ref{eta_u}
\begin{eqnarray}
\eta_\kappa&=&{3C_g\over 1+C_g+2C_\kappa}\;,
\label{eta_kappa2}\\
\eta_u&=&{3-3C_g\over 1+C_g+2C_\kappa}\;,
\label{eta_u2}
\end{eqnarray}

Another approach to estimating the tubule shape exponents is to rely
on the usual accuracy of the Flory theory (in treating the effects of
self-avoidance), instead of the extrapolation of functions $f_v(v)$
and $f_t(v)$ down from $\epsilon$-expansion. Although it is usually
not stated this way, in the language of renormalization group, Flory
theory amounts to assuming that the graphical corrections to $t$ and
to $v$ are the same, i.e. $f_v(v^*)=f_t(v^*)$. Using this in
Eqs.\ref{t_fixed} and \ref{v_fixed}, we obtain the Flory result for
$\nu$
\begin{eqnarray}
{\nu_{F}}&=&{D+1\over d+1}\;,\label{SAnuF}\\
&=&{3\over4}\;,\;\;\;\mbox{for}\;\;d=3\;,\;\;D=2\;,\label{SAnuFeval}
\end{eqnarray}
consistent with our earlier analysis in subsection \ref{Flory}.  Note
that, if $f_v(v)=f_t(v)$ for {\it all} $v$, this result would be exact
{\it independent} of the jump in the other exponents $z$, $\eta_\kappa$, and
$\eta_u$ {\it at} $d_*$. That is, it would apply even {\it below}
$d_*$, and $\nu$ would not jump, or be in any way non-analytic, at
$d_*$.

Now, of course, we know from the explicit leading order calculation in
Ref.\onlinecite{BG} that $f_v(v)$ does {\it not} $=f_t(v)$
exactly. However, we {\it do} know from that calculation that they are
quite close, at least to leading order, as illustrated by the good
agreement between Flory theory and the extrapolated
$\epsilon$-expansion. {\it If} this persists down to $d=3$, and to
large $v$, and our experience with polymers suggests that it will,
then $\nu$ may be quite accurately predicted by Flory theory, {\it
despite} the complications associated with the onset of anomalous bend
elasticity at $d_*$.

Using the Flory value for $\nu$ (Eq.\ref{SAnuF}) inside
Eqs.\ref{kappa_fixed2} and \ref{gy_fixed2}, together with the
diagrammatic corrections to $\kappa$ and $g_y$ given in
Eqs.\ref{f_kappa2} and \ref{f_gy}, we obtain two equations
(specialized to $D=2$)
\begin{eqnarray}
6/(d+1)-3z+1+C_\kappa {g_y^*}^2&=&0\;,
\label{kappa_fixed3}\\
12/(d+1)-3z+1-C_g g_y^*&=&0\;,
\label{gy_fixed3}
\end{eqnarray}
which gives for $d=3$
\begin{equation}
z={4\over3} + {C_g^2\over 6C_\kappa}-{C_g(C_g^2+6C_\kappa)^{1/2}
\over 6C_\kappa}\;,
\label{z2}
\end{equation}

Now, at least in this uncontrolled approximation of truncated
perturbation theory at one loop order, it seems that we are left with
the straightforward task of calculating the constants $C_\kappa$ and
$C_g$. Alas, things are not so simple, for reasons that are
undoubtedly connected with the fact that $d_*$ is not perturbatively
close to $d_{uc}^{SA}$, which is the only dimension about which one
can do a genuinely controlled approximation\cite{RT,BG}, {\it and} the
much more surprising fact that, even though $\epsilon_I\equiv5/2-D$ is
only $1/2$ (for $D=2$), this $\epsilon_I$-expansion in {\it intrinsic}
dimension, as we will show, is demonstrably {\it extremely}
unreliable, giving {\it qualitatively} different answers, such as a
{\it reduction}, rather than an increase of $\kappa$ due to
fluctuations.

Our unsuccessful (but heroic) attempts to calculate $C_g$ and
$C_\kappa$ were as follows:

(I) 

Calculate them in an $\epsilon_I\equiv5/2-D$-expansion for a {\it
phantom} membrane, then use these same constants $C_g$ and $C_\kappa$
for the real, self-avoiding membrane. This approach obviously makes
many errors, since, by the time we get down to $d_*(5/2)$, the
correlation functions of the true, self-avoiding membrane are already
quite different from those of the phantom membrane, due to the effects
of self-avoidance. Furthermore, these effects are particularly
pronounced for intrinsic dimensions $D=5/2$, since
$d_{uc}^{SA}(5/2)=\infty$, as illustrated in Fig.\ref{d_vs_D}.

Nonetheless, since no other analytical calculation is available (and
we are persistent young lads), we attempted this
$\epsilon_I\equiv5/2-D$-expansion. However, the results made no
physical sense: we found a {\it negative} $\eta_\kappa$, i.e., a {\it
downward} renormalization of $\kappa$. The detailed calculations are
virtually identical to those for the renormalization of $\kappa$ at
the tubule-to-crumpled phase transition, which are described in
Sec.\ref{transition_section}. We note here simply that the origin of
this negative contribution to $\kappa$ is a negative region of the
real-space correlation function $G({\bf x}_\perp,
y)=C(x_\perp/y^2)^{1/4} Y(x_\perp/y^2)$, as given by Eq.\ref{Gii}. The
integrand $x^{5/4} Y^3(x)$ in the $x$-integral of Eq.\ref{C} has a
negative region which, though narrow, actually overwhelms the positive
contribution to $\eta_\kappa$ from the much longer, but smaller, tail,
as we have verified by direct numerical integration\cite{stephany}.

This negative region is purely an artifact of calculating in a {\it
fractional} intrinsic dimension $D=5/2$. In $D=2$ for a phantom
membrane, where there is no relevant anomalous elasticity for
$\vec{h}$, and hence we can calculate $\vec{h}$-$\vec{h}$ correlation
functions exactly, we find the analog of Eq.\ref{Gii} is
\begin{eqnarray}
G(x,y)&=&\int {d q_x d q_y\over(2\pi)^{2}}
{e^{i q_x x + i q_y y}\;q_y^2\over q_x^2 + q_y^4}\;,\label{G2}\\
&=&{1\over 4(\pi|x|)^{1/2}} e^{-y^2/(4|x|)}\;,\label{G2ii}
\end{eqnarray}
which, {\it unlike} the analogous correlation function in $D=5/2$,
Eq.\ref{Gii}, is positive definite. Thus, the anomalous contribution
to $\kappa$ in $D=2$ will also be positive, as we expect on physical
grounds (i.e., the shell theory argument summarized in
Eq.\ref{kappaInduced}), while the $5/2-D$-expansion is {\it
qualitatively} wrong in predicting a negative renormalization of
$\kappa$. Clearly, it cannot be trusted quantitatively either, and is,
in fact, totally useless.

(II)

Direct, uncontrolled RG in $D=2$. Now, we at least get qualitatively
correct upward renormalization of $\kappa$. However, here we have a
different problem, that appears in {\it any} perturbative calculation
away from an upper critical dimension (and is usually ``swept under
the rug''): even though $D=2$ would not, a priori, {\it appear} to be
far below $D=5/2$, it is, in the sense that graphs that {\it only}
diverge logarithmically in $D=5/2$ diverge {\it extremely} strongly in
$D=2$. In particular, following very closely the manipulations that
lead to Eq.\ref{Bii}, we find a contribution to $\kappa$ of the form
\begin{equation}
\delta\kappa=c_1\int_0^{c_2 q_y^{-1}} y d y\int_0^\infty d
x{e^{-3/(4x)}\over x^{3/2}}\;,\label{delta_kappa2}
\end{equation}
where $c_1$ is a well-determined constant that we could calculate, and
$c_2$ is an {\it arbitrary} constant which depends on precisely how
the infrared divergence of the above integral is cutoff with
$q_y$. This arbitrary constant is the problem: {\it if} the integral
Eq.\ref{delta_kappa2} {\it had} diverged logarithmically, the precise
value of the constant $c_2$ would be unimportant (it would just lead
to a finite additive constant). But, since the integral in
Eq.\ref{delta_kappa2} diverges so strongly (like $(c_2/q_y)^2$) in
$D=2$, it is {\it extremely} sensitive to the precise value of $c_2$,
which we have {\it no clue} as how to choose. Thus, we have {\it no}
ability to predict $\eta_\kappa$ at all by this approach.

This strong divergence indicates that in this sense $D=2$ is quite far
from $D=5/2$, and {\it any} kind of perturbative approach, even to
simply calculating one loop constants like $C_g$ and $C_\kappa$, is
doomed.

\subsection{Gaussian variational theory of self-avoiding tubules}
\label{GaussianVariational}

Here we study the effects of self-avoidance within the tubule phase
using the Gaussian variational method, which was previously applied to
the study of self-avoidance in crumpled isotropic
membranes\cite{Goulian91,Ledoussal92} and in
polymers\cite{Edwards}. It is important to emphasize that both Flory
theory and the Gaussian variational method are uncontrolled
approximations in that there is no way to systematically estimate and
reduce the error.

We begin with the effective Hamiltonian that describes the long
wavelength behavior of the tubule for $d<d_*$.
\begin{eqnarray}
H&=&{1\over2}\int d^{D-1}x_\perp dy
\bigg[\kappa(\partial^2_y{\vec h})^2 + 
t(\partial^\perp_\alpha \vec h)^2
+ {1\over4}g_y(\partial_y{\vec h})^4\bigg]\nonumber\\
&+&{v}\int\hspace{-0.1cm}dy\hspace{0.05cm} d^{D-1}x_\perp d^{D-1}x'_\perp
\hspace{0.05cm}\delta^{(d-1)}\big({\vec h}({\bf x_\perp},y)-
{\vec h}({\bf x'_\perp},y)\big)\;,\nonumber\\
\label{SAtubule}
\end{eqnarray}
where, in contrast to other sections we use the notation $H$ to
distinguish long-wavelength effective Hamiltonian (the free energy
{\it functional}) from the actual free energy $F$.  Computation of
correlation functions in the presence of the self-avoiding
nonlinearity cannot be done exactly. However, we can replace the
Hamiltonian $H$, Eq.\ref{SAtubule} by a variational Hamiltonian $H_v$,
quadratic in the fields ${\vec h}({\bf x_\perp},y)$, which allows
exact calculations of any correlation function. Following the standard
variational procedure, we then pick the ``best'' form of this
variational Hamiltonian, where by ``best'' we mean that it minimizes
an upper bound on the true free energy $F$:\cite{Feynman}
\begin{equation}
F\leq \tilde{F}\equiv\langle H - H_v\rangle_v + F_v\;.
\label{convex}
\end{equation}

We take our variation ansatz Hamiltonian to be
\begin{equation}
H_v={1\over2}\int{d k_y d^{D-1}{\bf k_\perp}\over(2\pi)^D}
G_v({\bf k_\perp},k_y)|{\vec h}({\bf k_\perp},k_y)|^2\;,
\end{equation}
where $G_v({\bf k_\perp},k_y)$ is the variational kernel to be
optimized over. Note that because of anisotropy intrinsic to the
tubule, $G_v({\bf k_\perp},k_y)$ is {\it not} rotationally invariant
as it is for the analogous analysis of the crumpled phase.

We now compute the right-hand-side of Eq.\ref{convex} and minimize it over
$G_v({\bf k_\perp},k_y)$.
\begin{eqnarray}
&&\hspace{-0.5cm}\langle H-H_v\rangle_v={A\over 2}
\int_k\bigg(\kappa k_y^4+t k_\perp^2 - G_v({\bf k_\perp},k_y)\bigg)
\langle|{\vec h}({\bf k_\perp},k_y) |^2\rangle_v\nonumber\\
&+&{g_y\over8}\int_{\bf x}\langle(\partial_y{\vec h})^4\rangle_v
+{v}\int_{\bf x}\hspace{0.05cm}
\big\langle\delta^{(d-1)}\big({\vec h}({\bf x_\perp},y)-
{\vec h}({\bf x'_\perp},y)\big)\big\rangle_v\;,\nonumber\\
\label{HH}
\end{eqnarray}
where $A=L_y L_\perp^{D-1}$ is the ``area'' of the membrane and we
defined $\int_{\bf x}\equiv\int\hspace{-0.1cm}dy
\hspace{0.05cm}d^{D-1}x_\perp d^{D-1}x'_\perp$, and $\int_k\equiv\int
d k_y d^{D-1} {\bf k_\perp}/(2\pi)^D$.

The above averages are easily evaluated with $\langle|{\vec h}({\bf
k_\perp},k_y) |^2\rangle_v=(d-1)/G_v({\bf k})$ and
$\langle\delta\rangle_v\equiv \langle\delta^{(d-1)}\big({\vec h}({\bf
x_\perp},y)- {\vec h}({\bf x'_\perp},y)\big)\rangle_v$ given by
\begin{eqnarray}
\langle\delta\rangle_v
&=&\big\langle\int{d^{d-1}{\vec q}\over(2\pi)^{d-1}}
e^{i{\vec q}\cdot\big({\vec h}({\bf x_\perp},y)-
{\vec h}({\bf x'_\perp},y)\big)}\big\rangle_v\;,\nonumber\\
&=&\int{d^{d-1}{\vec q}\over(2\pi)^{d-1}}
e^{-q^2 K(|{\bf x_\perp}-{\bf x'_\perp}|)}\;,\nonumber\\
&=&{1\over(2\pi)^{d-1}}
\left({\pi\over K(|{\bf x_\perp}-{\bf x'_\perp}|)}\right)^{(d-1)/2}\;,
\label{delta}
\end{eqnarray}
where,
\begin{eqnarray}
K(|{\bf x_\perp}|)&=&\frac{1}{2(d-1)}
\langle|{\vec h}({\bf x_\perp},0)
-{\vec h}({\bf 0_\perp},0)|^2\rangle_v\;,\nonumber\\
&=&\int_k{\big(1-\cos({\bf k_\perp}\cdot{\bf x_\perp})\big)\over
G_v({\bf k})}\;,
\label{K}
\end{eqnarray}
and we have used in Eq.\ref{delta} the Fourier representation of the
$d-1$-dimensional delta-function.

Putting all this together we obtain for the right-hand-side of
Eq.\ref{convex}
\begin{eqnarray}
\hspace{-.25in}
{\tilde{F}\over(d-1)A/2}&=&\int_k\left({\kappa k_y^4 + t k_\perp^2\over
G_v({\bf k})} - 1\right)+ {g_y(d+1)\over4(d-1)}\left(\int_k {k_y^2\over
G_v({\bf k})}\right)^2\;\nonumber\\
&+&{4v\over(2\pi)^{d-1}(d-1)}
\int d^{D-1}x_\perp\left({\pi\over K(x_\perp)}\right)^{(d-1)/2}\;\nonumber\\
&+&\int_k\log\left(G_v({\bf k})\right)\;,
\label{Ftilde}
\end{eqnarray}
which when minimized with respect to $G_v({\bf k})$, 
$\delta\tilde F/\delta G_v({\bf k})=0$ gives an integral equation 
\begin{equation}
\hspace{-3mm}G_v(\vec k)=\kappa k_y^4 + t k_\perp^2 -
{2 v\over(4\pi)^{(d-1)/2}}\hspace{-1mm}\int\hspace{-1mm}{d^{D-1}x_\perp
\left(1-\cos({\bf k_\perp}\cdot{\bf x_\perp})\right)\over
K(x_\perp)^{(d+1)/2}}\;.
\label{geqn}
\end{equation}
The only effect of the anharmonic elasticity term $g_y$ is to generate
an upward renormalization of the effective tension along the $y$-axis
\begin{equation}
\delta t_y={g_y(d+1)\over2(d-1)}\int_k {k_y^2\over G_v({\bf k})}\;.
\label{delta_ty}
\end{equation}
Since we must choose the {\it renormalized} tension along the extended
tubule axis ($y$) to be exactly zero in order to treat the free
tubule, all of the anharmonic elastic effects disappear in this
Gaussian variational approximation. That is, to correctly model a
tubule with free boundaries, we should have started with an elastic
Hamiltonian with a bare, negative tension piece that exactly cancelled
thermally generated positive contribution in Eq.\ref{delta_ty}.

The simultaneous integral equations Eq.\ref{K} and Eq.\ref{geqn}
determine $G_v({\bf k})$ and $K(x_\perp)$. At long length scales they
are solved by $K(x_\perp)\sim x_\perp^{2\nu}$, where from the
definition Eq.\ref{K}, we see that $K(x_\perp=L_\perp)$ is
proportional to the square of the radius of gyration or the tubule
thickness that we are after, and hence the $\nu$ that solves these
coupled non-linear integral equations will be the Gaussian variational
prediction for the radius of gyration exponent as well. We substitute
this scaling ansatz into Eq.\ref{geqn} for $G_v(\vec k)$, and find
that while for $d>d_{uc}^{SA}$ the self-avoidance is irrelevant and
$\nu=(5-2D)/4$ (as found in Sec.\ref{Flory}), for $d<d_{uc}^{SA}$
these integral equations can only be solved if the $t k_\perp^2$ term
in Eq.\ref{geqn} is exactly cancelled by a part coming from the
integral in the last term and the resulting propagator takes the form
\begin{equation}
G_v({\bf k})=\kappa k_y^4 + \tilde{v} k_\perp^{(d+1)\nu-D+1}\;,
\label{grenorm}
\end{equation}
where $\tilde v\propto v$ is an effective self-avoiding interaction
parameter. Substituting this form into Eq.\ref{K} for $G_v({\bf k})$,
and requiring self-consistency with our original ansatz
$K(x_\perp)\sim x_\perp^{2\nu}$ gives
\begin{equation}
x_\perp^{2\nu}\propto
\int{d^{D-1}q_\perp d q_y\;
\big(1-\cos({\bf k_\perp}\cdot{\bf x_\perp})\big)
\over \kappa k_y^4 + \tilde{v} k_\perp^{(d+1)\nu-D+1}}\;.
\label{sc1}
\end{equation}
Making the change of variables ${\bf q}_\perp\equiv\tilde{\bf
q}_\perp/|{\bf x}_\perp|$ and $q_y\equiv\tilde q_y/|{\bf
x}_\perp|^\alpha$, with $\alpha=(\nu (d+1)-D+1)/4$ reveals that the
right hand side of Eq.\ref{sc1} is proportional to $x_\perp^\gamma$
with
\begin{eqnarray}
\gamma&=&1-D+3\alpha\;,\nonumber\\
&=&{7(1-D)+3\nu(d+1)\over4}\;.
\label{sc2}
\end{eqnarray}

To satisfy the self-consistent condition Eq.\ref{sc1}, $\gamma$ must
be equal to $2\nu$. The resulting simple linear equation for $\nu$ has
a solution (for $d<d_{uc}^{SA}(D)$):
\begin{equation}
\nu={7D-7\over3d-5}\;,
\label{nu_Dsolution}
\end{equation}
which for the physical case of $D=2$ gives
\begin{eqnarray}
\nu&=&{7\over 3d-5}\;\;,\;\;\;\mbox{for}\;\;\;d<11\;,\\
&=&{1\over 4}\;\;,\;\;\;\mbox{for}\;\;\;d\geq11\;
\label{nuGaussian}
\end{eqnarray}

We observe that $\nu(d=4)=1$, and therefore (according to the Gaussian
variational approximation) the tubule is no longer crumpled along the
$\perp$-direction. This suggests that the tubule phase is unstable to
the flat phase in embedding dimensions $d<4$ (which unfortunately
includes the physical case of $d=3$). However, as discussed in the
Introduction, the Gaussian variational method is an uncontrolled
approximation. It probably does give the correct {\it trends} of,
e.g., exponents with dimensionality $d$. However, the variational
approach is very close, in spirit and technically, to the large $d$
expansion methods, and therefore intrinsically unable to obtain the
small $d$ dependence correctly. It is therefore difficult to place any
faith in the actual values of exponents, particularly when the value
of $\nu$ at small $d$ actually determines whether the tubule phase
survives or not.

We believe that this Gaussian variation theory is incorrect in
predicting that the tubule phase does not exist in the presence of
self-avoidance in $d=3$, and reiterate our earlier observation that
both Flory theory\cite{RT} and the $\epsilon=11-d$-expansion\cite{BG}
predict that the tubule phase survives self-avoidance. Since both
these latter approaches agree quite closely with each other, and
since, furthermore, the $\epsilon$-expansion is the only controlled
approximation, we are far more inclined to trust them than the
uncontrolled Gaussian approximation, which agrees with neither.

The final determination of whether or not the tubule phase survives
self-avoidance will, or course, rest upon simulations and experiments,
both of which we hope our analytic work stimulates.

\section{Fluctuation Effects at Crumpled-to-Tubule and Tubule-to-Flat 
Transitions}
\label{transition_section}

The transition from the crumpled-to-flat phase in isotropic membranes
has been previously studied\cite{PKN} and is predicted to be driven
first order by fluctuations for embedding dimensions $d<d_c=219$. As
can be seen from Fig.\ref{phase_diagram1} this direct transition is
very special for anisotropic membranes. It is easy to see that any
path finely tuned to pass through the tetracritical point will undergo
a direct crumpled-to-flat transition identical to that of isotropic
membranes, discussed in Ref.~\onlinecite{PKN}.

Here we focus on the new transitions crumpled-to-tubule and
tubule-to-flat, which are generic for membranes with {\it any} amount
of anisotropy. As we discussed at the end of Sec.\ref{mft}, there are
two possible mean field phase diagram topologies depending on the
values of microscopic elastic moduli of the membrane. However, for the
crumpled-to-tubule transition there is no difference. In this section
we first study the crumpled-to-tubule transition for a phantom
membrane using a detailed renormalization group analyses. We then
study both the crumpled-to-tubule and tubule-to-flat transition using
scaling theory, incorporating the effects of both the anharmonic
elasticity and self-avoidance. We postpone the more technically
challenging renormalization group analysis of the phantom
tubule-to-flat transition\cite{JR} and renormalization group analysis
of crumpled-to-tubule and tubule-to-flat transitions for self-avoiding
membranes\cite{RTfuture} for future publications.

\subsection{Renormalization group analysis of crumpled-to-tubule
transition}
\label{crumpled-to-tubuleRG}
We start out with the general free energy defined in Eq.\ref{Fc}, for
now ignoring the self-avoiding interaction. Without loss of generality
we will study the transition from the crumpled to the y-tubule phase.
As discussed above, in mean-field theory, this transition occurs when
$t_y\rightarrow 0$ from above, while $t_\perp$ remains finite and
positive. Hence, simple power-counting on the quadratic part of the
free energy leads to anisotropic scaling at the transition with
$q_\perp\propto q_y^2$. Therefore, the only relevant terms quadratic
in $\vec r$ near the transition are: the bending rigidity along the
$y$-direction ($\kappa_y\left(\partial_y^2\vec{r}\right)^2$), and the
surface tension terms along the $y$ and $\perp$-directions
$t_y\left(\partial_y\vec{r}\right)^2$ and
$t_\perp\left(\partial_\alpha^\perp\vec{r}\right)^2$, respectively.
The corresponding noninteracting propagator at the transition is
\begin{equation}
\langle r_i({\bf q}) r_j(-{\bf q})\rangle=
{\delta_{ij}\over t_\perp q_\perp^2 + t_y q_y^2 + \kappa_y q_y^4}
\equiv C({\bf q})\delta_{ij}\;,
\label{propagator}
\end{equation}

The anisotropic scaling dictated by this noninteracting propagator at
the transition ($t_y=0$) leads to significant simplification of the
interaction term in the free energy. Keeping only the dominant
nonlinearity we obtain
\begin{eqnarray}
F[{\vec r}({\bf x})]&=& {1\over2}\int d^{D-1}x_\perp dy
\bigg[\kappa_y\left(\partial_y^2\vec{r}\right)^2
+ t_\perp\left(\partial^\perp_\alpha\vec{r}\right)^2\nonumber\\
&+& t_y\left(\partial_y\vec{r}\right)^2 +
{u_{y y}\over2}\left(\partial_y\vec{r}\cdot\partial_y\vec{r}\right)^2\bigg]\;.
\label{Fcii}
\end{eqnarray}
The critical properties of the crumpled-to-tubule transition can be
obtained by applying scaling theory and the renormalization group to
this free energy exactly as we did earlier in treating fluctuations in
the tubule phase itself. In this case, ``lengths'' means intrinsic
coordinates ${\bf x}=({\bf x}_\perp, y)$, and the ``fields'' are the
extrinsic positions $\vec{r}({\bf x})$. Because of the strong {\em
scaling} anisotropy of the quadratic pieces of the free energy, we
rescale ${\bf x}_\perp$ and $y$ anisotropically:
\begin{eqnarray}
{\bf x}_\perp&=&{\bf x}_\perp'e^{l}\;,\\
y&=&y' e^{z l}\;,
\end{eqnarray}
and rescale the ``fields'' according to 
\begin{equation}
{\vec r}({\bf x})=e^{\chi l}{\vec r}'({\bf x}')\;.
\end{equation}
Under this transformation
\begin{eqnarray}
\kappa_y(l)&=&\kappa_y e^{(D-1-3z+2\chi)l}\;,\label{kappa_scale}\\
t_\perp(l)&=&t_\perp e^{(D-3+z+2\chi)l}\;,\label{t_scale}
\end{eqnarray}
Requiring that both $\kappa_y$ and $t_\perp$ remain fixed under this
rescaling (zeroth order RG transformation) fixes the ``anisotropy''
exponent $z$ and the ``roughness'' exponent $\chi$ (which is the
analog of $\nu$ for the tubule phase):
\begin{eqnarray}
z&=&{1\over2}\;,\label{z_quad}\\
\chi&=&(5/2-D)/2\;.\label{chi_quad}
\end{eqnarray}
Although this choice keeps the quadratic (in $\vec{r}$) part of $F$
Eq.\ref{Fcii}) unchanged, it {\em does} change the quartic piece:
\begin{eqnarray}
u_{yy}(l)&=& u_{yy}e^{(D-1-3z+4\chi)l}\;,\\
&=&u_{yy}e^{(5/2-D)l}\;,
\label{uyy_scale}\;
\end{eqnarray}
where in the second equality we have used the results Eqs.\ref{z_quad}
and \ref{chi_quad} for $z$ and $\chi$. We see that for $D<5/2$,
$u_{yy}$ grows upon rescaling. Physically, this means that its effects
become more important at longer length scales. At sufficiently long
length scales, it completely invalidates the harmonic elastic theory
and the naive perturbation theory in the nonlinearity $u_{yy}$ around
it, even for arbitrarily small coupling $u_{yy}$. Simple additional
anisotropic rescaling of ${\bf x}_\perp=\alpha {\bf x}'_\perp$ and
$y=\beta y'$, with $\beta=(t_\perp/\kappa_y)^{1/2}\alpha^2$, which
rescales $\kappa_y$ and $t_\perp$ to $1$, reveals that the effective
coupling constant of the nonlinearity is $u_{yy}/\kappa_y$. This,
together with Eq.\ref{uyy_scale}, predicts that the characteristic
length scale $L_\perp^{nl}$ beyond which the {\em dimensionless}
coupling constant becomes of order $1$ and the harmonic elastic theory
and perturbation theory (around it) break down is
\begin{equation}
L_\perp^{nl}=\left(\kappa_y\over u_{yy}\right)^{1/(5/2-D)}\;.
\label{Lnl}
\end{equation}

To analyze the new behavior that prevails on even {\em longer} length
scales requires a full-blown renormalization group analysis. 

Such an analysis\cite{Wilson} will lead to corrections to the simple
rescaling of $\kappa_y$, $t_\perp$, and $t_y$, due to the
non-linearities (in this case $u_{yy}$, as discussed above). These
corrections can be absorbed into ``anomalous'' exponents
$\eta_\kappa$, $\eta_t$, and $\delta\theta$, defined by the large
renormalization group ``time'' ($l\rightarrow\infty$) limits of
$\kappa_y(l)$, $t_\perp(l)$, and $t_y(l)$, respectively:
\begin{eqnarray}
\kappa_y(l)&=& \kappa_y e^{(D-1-3z+z\eta_\kappa +2\chi) l}\;,\\
\label{kappa}
t_\perp(l)&=& t_\perp e^{(D-3+z+\eta_t+2\chi) l}\;,\\
\label{tperp}
t_y(l)&=& t_y e^{(D-1-z-\delta\theta+2\chi) l}\equiv t_y e^{\lambda_t l}\;,\\
\label{ty}
\nonumber
\end{eqnarray}
The exponent $\lambda_t$ defined above is the thermal eigenvalue of
the reduced temperature (surface tension along y-direction) which is
an inverse of the correlation length exponent along the
$\perp$-direction (see below). Requiring that $\kappa_y$ and $t_\perp$
remain invariant under the renormalization group transformation
determines the values the anisotropy exponent $z$ and the field
rescaling exponent $\chi$,
\begin{eqnarray}
z&=&{2-\eta_t\over4-\eta_\kappa}\;,\\
\label{z_transition}
\chi&=& {10-4D+\eta_\kappa(D-3+\eta_t)-3\eta_t\over8-2\eta_\kappa}\;,
\label{chi}
\nonumber
\end{eqnarray}
which, as quoted above in Eqs.\ref{z_quad} and \ref{chi_quad}, reduce
to $z=1/2$ and $\chi=(5/2-D)/2$, for $\eta_\kappa=\eta_t=0$, as is
valid at zero order in perturbation theory in $u_{yy}$.

Once the values of $\eta_t$, $\eta_\kappa$ and $\chi$ at the critical
point are determined, the renormalization group gives a relation
between correlation functions at or near criticality (small $t_y$) and
at small wavectors (functions that are difficult to compute, because
direct perturbation theory is divergent) to the same correlation
functions away from criticality and at large wavevectors (functions
that can be accurately computed using perturbation theory). For
example the behavior of the correlation lengths near the transition
can be deduced in this way:
\begin{eqnarray}
\xi_\perp(t_y)&=& e^{l}\xi_\perp(t_y e^{\lambda_t l})\;,\\
\label{xiperp}
\xi_y(t_y)&=& e^{z l}\xi_y(t_y e^{\lambda_t l})\;,\\
\label{xiy}
\nonumber
\end{eqnarray}
where in the above we assumed that a critical fixed point exists and
all other coupling constants have well-defined values at the fixed
point. Using the above equations for $t_y e^{\lambda_t l}\approx 1$,
we obtain,
\begin{eqnarray}
\xi_\perp(t_y)&\approx& a\; t_y^{-\nu_\perp}\;,\\
\label{xiperpii}
\xi_y(t_y)&\approx& a\; t_y^{-\nu_y}\;, 
\label{xiyii}
\nonumber
\end{eqnarray}
where $a\approx\xi(1)$ is the microscopic cutoff and,
\begin{eqnarray}
\nu_\perp&=&{1\over\lambda_t}
={4-\eta_\kappa\over 2(2-\eta_t-2\delta\theta)-
\eta_\kappa(2-\eta_t-\delta\theta)}\;,
\label{nuperp}\\
&&\nonumber\\ 
\nu_y&=&{z\nu_\perp}\;.
\label{nuy}
\end{eqnarray}

We now compute the anomalous exponents to lowest non-zero order in
$\epsilon$, where $\epsilon=5/2-D$. As usual in the
$\epsilon$-expansion, the order at which a given graphical correction
enters the perturbation theory is equal to the number of loops in the
associated Feynman graph. We split the field ${\vec r}({\bf x})$ into
short and long wavevector parts ${\vec r}({\bf x})={\vec r}_<({\bf
x})+{\vec r}_>({\bf x})$ and integrate over the fast fields ${\vec
r}_>({\bf x})$. Diagrammatically this leads to one-loop corrections to
$u_{yy}$ and $t_y$. There are no corrections to $\kappa_y$ to first
order in $\epsilon$, i.e., $\eta_\kappa=O(\epsilon^2)$. Furthermore,
since the interaction $u_{yy}$ always carries a factor of $q_y$ with
every field $\vec r$, the $t_\perp$ tension remains unrenormalized,
and $\eta_t=0$ to all orders, implying $z=1/2 + O(\epsilon^2)$ and
$\chi=(5/2-D)/2 + O(\epsilon^2)$.
\begin{figure}[bth]
\centering
\setlength{\unitlength}{1mm}
\begin{picture}(150,100)(0,0)
\put(-20,-40){\begin{picture}(150,100)(0,0)
\includegraphics{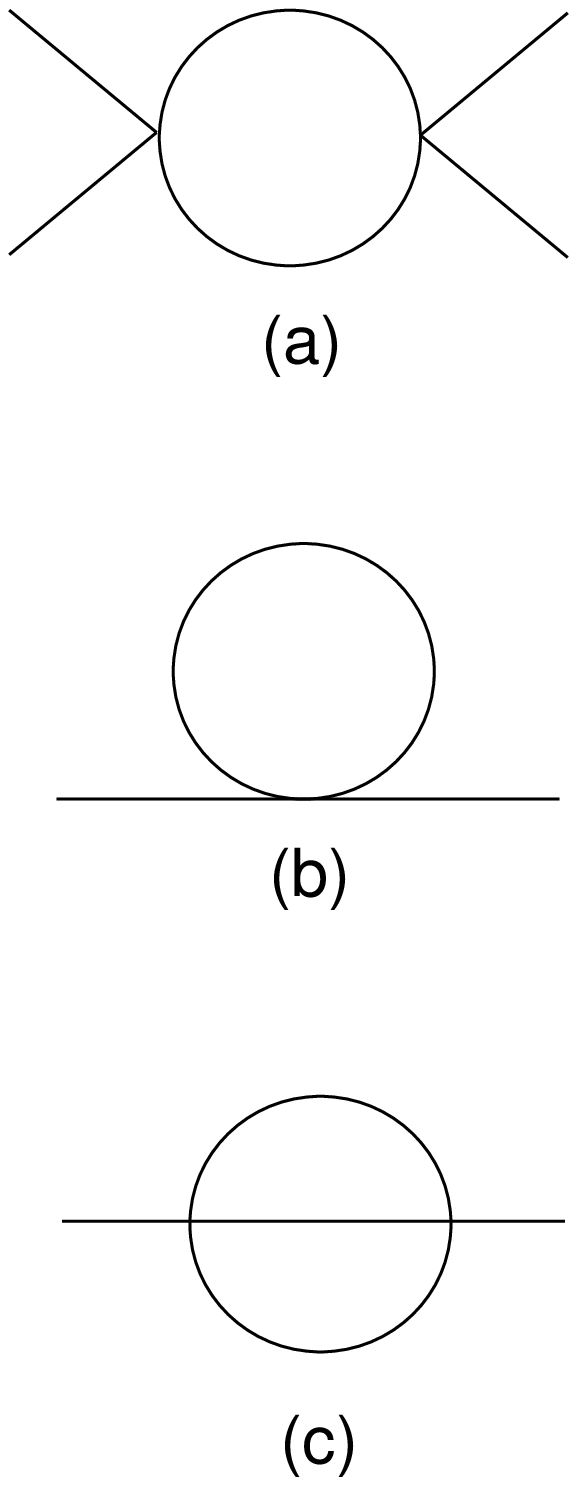}
\end{picture}}
\end{picture}
\caption{Feynman graphs that renormalize: (a) the nonlinearity
$u_{yy}$, (b) the tension $t_y$ and (c) the bending rigidity
$\kappa_y$.}
\label{transitionRGloops}
\end{figure}

The first two diagrams in Fig.\ref{transitionRGloops}, followed by the
rescaling introduced above (necessary to restore the original uv
cutoff), lead to the one-loop recursion relations for
$u=(K_{3/2}/\sqrt{2}) u_{yy}/(\kappa_y^{5/4} t_\perp^{3/4})$ and
$t_y$, respectively,
\begin{eqnarray}
{\partial u\over\partial l}&=& \epsilon u - (d+8) u^2\;,\\
\label{uRecursion}
{\partial t_y\over\partial l}&=& (1-(d+2) u)t_y\;,
\label{tRecursion}
\end{eqnarray}
where $\epsilon=5/2-D$ and $K_{3/2}$ is the surface of area of a
$3/2$-dimensional sphere divided by $(2\pi)^{3/2}$. As usual, in the
above, we also redefined $t_y$ to be the reduced temperature, measured
from its true value at the transition (which in mean-field theory
starts out at $0$, but is shifted to a negative value by
fluctuations). Note that, in contrast to the familiar $\epsilon=4-D$-
expansion for critical phenomena, for which $\epsilon=1$ in the
physical case $D=3$, here we have $\epsilon=1/2$ in the physical case
$D=2$. Hence, our one-loop expansion should be quantitatively more
accurate by a factor of $\epsilon^{-2}=4$, than the $4-D$-expansion at
the same order. Thus, we expect our one-loop values for $\nu_\perp$
and $\nu_y$ to be accurate to $\pm 0.02$.

Examining Eq.\ref{uRecursion}, we observe that for $D<D_{uc}=5/2$,
(i.e., for $\epsilon>0$) the Gaussian fixed point is unstable and the
critical properties of the crumpled-to-tubule transition are
characterized by a nontrivial fixed point with a fixed point value
$u_*$ of $u$ given by:
\begin{equation}
u_*={\epsilon\over d+8}\;.
\label{u_f}
\end{equation}
Note that, in contrast to the treatment of crumpled-to-flat transition
in isotropic membranes\cite{PKN}, where the critical point was only
stable for an unphysically large value of the embedding dimension
$d>219$, the critical point characterizing the crumpled-to-tubule
transition found here is stable for all $d$.

Equation \ref{tRecursion} can be easily integrated once the fixed
point value $u_*$, Eq.\ref{u_f}, is inserted for $u$; comparison with
the general Eq.\ref{ty} then gives $\lambda_t$:
\begin{equation}
\lambda_t=1-\left({d+2}\over{d+8}\right)\epsilon\;,
\label{lambda_t}
\end{equation}
which, upon using Eqs.\ref{nuperp} and 
\ref{nuy}, gives for a physical
membrane ($D=2$, $d=3$)
\begin{eqnarray}
\nu_\perp&\approx& 1.227\;,\\
\label{nuperp2D}
\nu_y&\approx& 0.614\;.
\label{nuy2D}
\end{eqnarray}

The $\eta_\kappa$ exponent to $O(\epsilon^2)$ is determined by the
diagram in Fig.\ref{transitionRGloops}{c}. Evaluating this diagram in
real space and then Fourier transforming, we find that this
contributes to the free energy
\begin{equation}
\delta F = 
- 16 u^2 (d+2)\int_q q_y^2 |{\vec r}({\bf q})|^2 \Gamma({\bf q})\;,
\label{deltaF}
\end{equation}
where
\begin{equation}
\Gamma({\bf q})\equiv\int d^{3/2}{\bf x_\perp} dy\; 
e^{i{\bf q}\cdot{\bf x}}\; G^3({\bf x})\;,
\label{Gamma}
\end{equation}
with, in turn,
\begin{equation} 
G({\bf x_\perp},y)=\int {d^{3/2}{\bf q_\perp} d q_y\over(2\pi)^{5/2}}
{e^{i{\bf q_\perp}\cdot{\bf x_\perp}}\;
e^{i q_y y}\;q_y^2\over q_\perp^2 + q_y^4}\;,
\label{G}
\end{equation}
where we have rescaled lengths so that $\kappa_y=t_\perp=1$.

After a contour integral over $q_y$, and an angular integral
$\int_0^\pi d\theta (\sin\theta)^{(D-2)}e^{i{\bf q_\perp}\cdot{\bf
x_\perp}}$, we obtain,
\begin{equation} 
G({\bf x_\perp},y)=2^{-7/4}\pi^{-3/4} 
y^{-2}\left({x_\perp\over y^2}\right)^{1/4} Y\left({x_\perp\over
y^2}\right)\;,
\label{Gii}
\end{equation}
where we have defined
\begin{equation} 
Y(x)\equiv\int_0^\infty du\; u^{1/4} J_{-1/4}(x u)\;
e^{-\sqrt{u/2}}\cos(\sqrt{u/2}+\pi/4)\;.
\label{Y}
\end{equation}

Now going back to Eq.\ref{deltaF} we observe that the $\bf q=0$ piece
of $\Gamma({\bf q})$ contributes to the $q_y^2 |{\vec r}({\bf q})|^2$
part of $F$, additively renormalizing $t_y$ which corresponds to the
usual inconsequential $T_c$ (critical tension) shift. The order
$q_y^2$ piece of $\Gamma({\bf q})$ renormalizes $\kappa_y$. We define
\begin{equation}
\Gamma({\bf q})=\Gamma(0)-{1\over 2} q_y^2 B(q_y)\;,
\label{Gammaii}
\end{equation}
where
\begin{equation}
B(q_y)\equiv\int_{\Lambda^{-1}<|y|<q_y^{-1}} dy\; d^{3/2}x_\perp\; y^2 
G^3({\bf x_\perp}, y)\;.
\label{B}
\end{equation}
Note that the infra-red cutoff on the integral over $y$ is
$q_y^{-1}$. This integral diverges logarithmically as $q_y\rightarrow
0$. We can identify the coefficient of the logarithm with
$\eta_\kappa$ in the expression $\kappa_y(q_y)\propto
q_y^{-\eta_\kappa}$.

To extract this logarithmic divergence, we make a change of variables
in the integral $|{\bf x_\perp}|\equiv x y^2$ and find
\begin{equation}
B(q_y)\equiv-{1\over2^{13/4}\pi^{3/2}\Gamma(3/4)}\int_{\Lambda^{-1}}^{q_y^{-1}}
{dy\over y}\int_0^\infty dx\;x^{5/4}\; Y^3(x)\;,
\label{Bii}
\end{equation}
where we have used the fact that the surface area of a
$3/2$-dimensional sphere is $2\pi^{3/2}/\Gamma(3/4)$, and taken into
account the factor of $2$ coming from the fact that the original
integral over $y$ extends over both $y>0$ and $y<0$.

Putting all of the above together and evaluating the coefficient of
the $\log(q_y)$ at the fixed point value of $g_y^*$ from
Eq.\ref{u_f}, we obtain
\begin{equation}
\eta_\kappa={C(2)(d+2)\over 8(d+8)^2}\epsilon^2\;,
\label{eta}
\end{equation}
where
\begin{equation}
C(2)\equiv 2^{23/4}\Gamma(3/4)\int_0^\infty dx\;x^{5/4}\; Y^3(x)\;,
\label{C}
\end{equation}
The value of $C(2)$ has been calculated numerically\cite{TT} to be
$C(2)\approx -1.166\pm 0.001$. Using this value, $\epsilon=1/2$, and
$d=3$ in Eq.\ref{eta}, we find that $\eta_\kappa$ is very small,
\begin{equation}
\eta_\kappa(D=2,d=3)\approx-0.0015\;.
\label{etaii}
\end{equation}

As noted earlier in our discussion of the tubule phase itself, we do
not trust this negative value of $\eta_\kappa$, but, rather, believe
it to be an artifact of the peculiar negative regime that appears in
the correlation function $G({\bf x}_\perp,y)$ in $D=5/2$. We expect
$\eta_\kappa$ to be positive, but still quite small, at the {\it
phantom} tubule-to-crumpled transition.

Given the smallness of $\eta_\kappa$ and $\epsilon$, and the vanishing
of $\eta_t$, the exponents computed here to first order in $\epsilon$
are expected to be very accurate.

\subsection{Scaling theory of crumpled-to-tubule and tubule-to-flat
transitions}
\label{scaling_theory}
We will now incorporate the effects of self-avoidance on these
transitions. We have not yet done the full renormalization group
analysis of this problem (which must include {\em both} the elastic
and self-avoiding interaction nonlinearities)\cite{RTfuture}, and
limit ourselves here to discussing scaling theory and the Flory
approximation.

Near the crumpled-to-tubule transition, for square membranes of
internal size $L$, we make the following general scaling ansatz for
the extensions $R_y$ and $R_G$ of the membrane along and orthogonal to
the tubule axis, respectively:
\begin{eqnarray}
R_{G, y}&=& L^{\nu_{ct}^{G,y}} f_{G,y} (t_y L^{\phi})\;,\nonumber\\
&\propto&
\left\{ \begin{array}{lr}
t_y^{\gamma_{+}^{G,y}} L^{\nu_c}, & t_y > 0,  L >> \xi_{ct} \\
L^{\nu_{ct}^{G,y}} , & L << \xi_{ct} \\
|t_y|^{\gamma_{-}^{G,y}} L^{\nu_{t}^{G,y}}, & t_y < 0, L >> \xi_{ct}
\label{Rtrans}
\end{array} \right.
\end{eqnarray}
where subscripts $t$, $c$ and $ct$ refer to tubule, crumpled and
tubule-to-crumpled transition, respectively, and $\xi_{ct}\propto
|t_y|^{-1/\phi}$ is a correlation length for the crumpled-to-tubule
transition, $t_y=(T-T_{ct})/T_{ct}$, $T_{ct}$ is the
crumpled-to-tubule transition temperature, and $t_y>0$ corresponds to
the crumpled phase.

Note that we have built into the scaling laws the fact that both $R_y$
and $R_G$ scale like $L^{\nu_c}$ in the crumpled phase, with $\nu_c$
the radius of gyration exponent for the crumpled phase (which, as
noted earlier, is the same for anisotropic and isotropic
membranes). Due to the extended nature of the tubule phase,
$\nu_t^y=1$, of course. The anisotropy in manifested in the crumpled
phase only through the different temperature dependences of $R_G$ and
$R_y$. The former of these vanishes as $t_y\rightarrow 0^+$ (since the
radius of gyration in the tubule phase is much less than that in the
crumpled phase, since $\nu_t < \nu_c$), which implies $\gamma_+^G >
0$, while the latter diverges as $t_y\rightarrow 0^+$, since the
tubule ultimately extends in that direction, which implies $\gamma_+^y
<0$.

Note also that our expression Eq.\ref{Rtrans}, and, in particular, the
fact that $R_G\neq R_y$ {\it even} above the crumpled-to-tubule
transition (i.e., {\it in} the {\it crumpled} phase), implies a
spontaneous breaking of rotational invariance even in the crumpled
phase! This seemingly bizarre (but correct) result is actually not all
that unfamiliar: polymers, which are always crumpled, nonetheless
assume, on average, non-spherical shapes\cite{Aronovitz}, as can be
seen, for example, by looking at the ratio of the average maximum and
minimum eigenvalues of the moment of inertia tensor. Our result
Eq.\ref{Rtrans} for $t_y>0$ is only a little more surprising, since it
predicts an aspect ratio $R_y/R_G$ that actually {\it diverges} as
$T\rightarrow T^+_{ct}$, and membrane begins to extend into a tubule
configuration.

The exponents $\gamma_{+/-}^{G,y}$ defined in above equation obey the
scaling laws 
\begin{eqnarray}
\gamma_{+}^{G,y}&=&{\nu_c-\nu_{ct}^{G,y}\over\phi}\;,\label{scaling_above}\\
\gamma_{-}^{G,y}&=&{\nu_t^{G,y}-\nu_{ct}^{G,y}\over\phi}\;.
\label{scaling_below}
\end{eqnarray}
As always, these scaling laws follow from requiring that the
generalized scaling form matches on to known results in the
appropriate limits.

From Flory theory, we can derive the values of the critical exponents
in Eq.\ref{Rtrans}, as we have already derived the exponents $\nu_t$
and $\nu_c$ characterizing the tubule and crumpled phases, simply by
being more careful about temperature dependent factors in that
derivation.  Again, we start by estimating the total self-avoidance
energy Eq.\ref{Fc} in the tubule phase (i.e., $t_y < 0$) as
$E_{SA}\approx V\rho^2$. Now, however, we very carefully write the
volume $V$ in the embedding space occupied by the tubule as $V\approx
R_G^{d-1} R_y$. Writing 
\begin{equation}
R_y = \zeta_y L_y\;,\label{R_y}
\end{equation}
as we did earlier in our discussion of mean field theory in the {\it
absence} of self-avoidance, and using $\rho=M/V$ for the embedding
space density of the tubule, and again using the fact that the tubule
mass $M \approx L_\perp^{D-1} L_y$, we see that
\begin{equation}
E_{SA}\approx v {L_y L_\perp^{2(D-1)}\over \zeta_y
R_G^{d-1}}\;.\label{Esacrit}
\end{equation}

Using this estimate of the self-avoidance energy in Eq.\ref{Fc}, and
estimating the other terms in that expression by scaling, we obtain
the full Flory theory for the tubule phase, with all temperature
dependent effects taken (admittedly crudely) into account:
\begin{equation}
\hspace{-.3cm}E_{FL}=\bigg(t_y \zeta_y^2 + u_{yy} \zeta_y^4 +
t_\perp\left({R_G\over L_\perp}\right)^2\bigg) L_\perp^{D-1} L_y +
v {L_y L_\perp^{2(D-1)}\over \zeta_y
R_G^{d-1}}
\;.
\label{Ecritub}
\end{equation}
Minimizing this over $R_G$, we obtain
\begin{equation}
R_G\approx
L_\perp^{\nu_t} \bigg({ v \over t_\perp \zeta_y}\bigg)^{1/(d + 1)}\;,
\label{RGcrit}
\end{equation}
where, as we found earlier, $\nu_t= {D+1 \over d+1}$, but now we have
the singular temperature dependence of $R_G$ near the
crumpled-to-tubule transition explicit through the presence of the
$\zeta_y$ term.  Inserting this expression for $R_G$ into
Eq.\ref{Ecritub}, we find
\begin{equation}
\hspace{-0.2cm}E_{FL}=\left(t_y \zeta_y^2 + 
u_{yy} \zeta_y^4 + t_\perp^{d-1\over d+1}
\left(v\over\zeta_y\right)^{2\over d+1}L_{\perp}^{-{2(d-D)\over
d+1}}\right) L_\perp^{D-1} L_y\;.
\label{Ecritub2}
\end{equation}

The exponents defined by Eq.\ref{Rtrans} can now be obtained by
minimizing $E_{FL}$ in Eq.\ref{Ecritub2} with respect to $\zeta_y$,
which amounts to balancing two of the three terms in $E_{FL}$, which
two, depending on whether one is interested in the crumpled phase
($t_y>0$), the tubule phase ($t_y<0$), or the transition between them
($t_y=0$).

In the crumpled phase $t_y>0$, as a result, the order parameter
$\zeta_y$ vanishes in the thermodynamic limit, allowing us to neglect
the quartic $\zeta_y^4$ term relative to the quadratic $\zeta_y^2$
one. Balancing the remaining two terms
\begin{equation}
t_y\zeta_y^2\approx t_\perp^{d-1\over d+1}
\left(v\over\zeta_y\right)^{2\over d+1}L_{\perp}^{-{2(d-D)\over
d+1}}\;,
\end{equation}
we obtain
\begin{equation}
\zeta_y\approx \left({v^2 t_\perp^{d-1}\over
t_y^{d+1}}\right)^{1\over2(d+2)} L_\perp^{-{d-D\over d+2}}\;,
\label{zeta_y2}
\end{equation}
Using this expression for $\zeta_y$ inside Eq.\ref{R_y} for $R_y$
gives, for a square membrane ($L_y=L_\perp=L$)
\begin{equation}
R_y\approx \left(v^2 t_\perp^{d-1}\right)^{1\over2(d+2)} 
t_y^{-{d+1\over2(d+2)}} L_\perp^{{D+2\over d+2}}\;,
\end{equation}
which, after comparing with the general form for $R_y$,
Eq.\ref{Rtrans}, gives 
\begin{eqnarray}
\nu_c&=&{D+2\over d+2}\;,\label{nu_c}\\
\gamma_+^y&=&-{d+1\over 2(d+2)}\;,\label{gamma_+^y}
\end{eqnarray}
equation\ \ref{nu_c} being a well-known Flory result for the radius of
gyration exponent $\nu_c$ for a $D$-dimensional manifold, embedded in
$d$ dimensions,\cite{KN_sa,KKN,AL_sa} and $\gamma_+^y$ new and special
to anisotropic membranes. Furthermore, inserting $\zeta_y$,
Eq.\ref{zeta_y2} inside Eq.\ref{RGcrit} for $R_G$, we obtain
\begin{equation}
R_G\approx \left(v^{d+3\over(d+2)(d+1)} 
t_\perp^{-{d+5\over2(d+2)(d+1)}}\right)
t_y^{{1\over2(d+2)}} L_\perp^{{D+2\over d+2}}\;,
\end{equation}
which, not surprisingly gives the same expression for $\nu_c$ as in
Eq.\ref{nu_c}, and predicts
\begin{equation}
\gamma_+^G={1\over2(d+2)}\;.\label{gamma_+^G}
\end{equation}
$\gamma_+^y\neq\gamma_+^G$ supports our earlier claim that even the
crumpled phase spontaneously breaks rotational invariance in the
embedding space.  It does so gently by having the identical growth
(for square membranes) of $R_G$ and $R_y$ with $L$, but exhibiting
anisotropy via the prefactors, with the ratio $R_y/R_G$ diverging as
the crumpled-to-tubule transition is approached.

The tubule phase is characterized by $t_y<0$ and a finite order
parameter $\zeta_y>0$. Therefore in this phase, the term proportional
to $t_\perp^{d-1/(d+1)}$ in $E_{FL}$, Eq.\ref{Ecritub2}, clearly becomes
negligible relative to the first two terms when $L_\perp
\rightarrow\infty$.  Therefore, we can neglect that term for a
sufficiently large membrane (i.e., a membrane larger than the critical
correlation length $\xi_{cr}$). Minimizing the remaining first two
terms in $E_{FL}$ therefore gives $\zeta_y \propto \sqrt{|t_y|}$,
(independent of $L_\perp$) as in mean-field theory in the absence of
self-avoidance.  Inserting this inside $R_y$, Eq.\ref{R_y} and
comparing with the general scaling form for $R_y$, implies for a
square membrane
\begin{eqnarray}
\nu_t^y&=&1\;,\label{nu_t^y}\\
\gamma_{-}^y&=&{1\over2}\;.\label{gamma_-^y}
\end{eqnarray}
Using this in the earlier expression Eq.\ref{RGcrit} for $R_G$, we
obtain the last line of Eq.\ref{Rtrans}, with
\begin{eqnarray}
\nu_t^G&=&{D+1\over d+1}\;,\label{nu_t^G}\\
\gamma_-^G &=& - {1 \over 2 ( d + 1 ) }\;.\label{gamma_-^G}\\
\label{gcritfl}
\end{eqnarray}

Finally, right at the crumpled-to-tubule transition, $t_y = 0$ and we
must balance the last two terms in $E_{FL}$, Eq.\ref{Ecritub2}.
Minimizing $E_{FL}$ over $\zeta_y$, we find at the transition
\begin{equation}
\zeta_y \propto L_\perp^{- {(d - D) \over 3 + 2d} }
\label{zetaycrit}
\end{equation}
which, when inserted in Eq.\ref{R_y} for $R_y$ implies for a square
membrane that
\begin{equation}
R_y \propto L^{ D + d +3 \over 3 + 2d }
\label{Rycrit}
\end{equation}
right at the transition. This leads to
\begin{equation}
\nu_{ct}^y = { D + d + 3 \over 2d + 3}
\label{nu_ct^y}
\end{equation}
for a square membrane. Using the result Eq.\ref{zetaycrit} for
$\zeta_y$ in Eq.\ref{RGcrit} for $R_G$ gives, right at the transition,
\begin{equation}
R_G\propto L_\perp^{\nu_{ct}^G}\;,
\label{RGcrit2}
\end{equation}
with
\begin{eqnarray}
\nu_{ct}^G&=&\nu_t+{1\over d+1}\left({d-D\over3+2d}\right)\;,\\
&=& { 2D + 3 \over 2d + 3}\;,\label{nu_ct^G}
\end{eqnarray}
The scaling relations Eqs.\ref{scaling_above} and \ref{scaling_below},
quoted above, then give
\begin{equation}
\phi = {2 ( d - D) \over 2d + 3}\;,\label{phi1}
\end{equation}
and are reassuringly consistent with our independent calculations of
exponents $\gamma_{+,-}^{G,y}$, $\nu_c$, $\nu_t^{G,y}$, and
$\nu_{ct}^{G,y}$, given in
Eqs.\ref{gamma_+^y},\ref{gamma_+^G},\ref{gamma_-^y},\ref{gamma_-^G},
\ref{nu_c},\ref{nu_t^y},\ref{nu_t^G},\ref{nu_ct^y}, and \ref{nu_ct^G},
above.  For the physical case of a two dimensional membrane embedded
in a three dimensional space, ($D=2,d=3$)
\begin{mathletters}
\begin{eqnarray}
\nu_c&=&4/5\;,\label{nu_c2}\\
\nu_{ct}^G&=&7/9\;,\label{nu_ctG}\\
\nu_{ct}^y&=&8/9\;,\label{nu_cty}\\
\nu_t&=&3/4\;,\label{nu_t}\\
\gamma_+^G&=&1/10\;,\label{gamma+G}\\
\gamma_+^y&=&-2/5\;,\label{gamma+y}\\
\gamma_-^G&=&-1/8\;,\label{gamma-G}\\
\gamma_-^y&=&1/2\;.\label{gamma-y}\\
\phi&=&2/9\;,\label{phi}
\end{eqnarray}
\end{mathletters}

Note that the signs of the $\gamma_{+/-}^{G,y}$ imply that $R_G$
shrinks as the crumpled-to-tubule transition is approached from above,
and grows as it is approached from below, while $R_y$ does the
opposite. Note also that the crumpled-to-tubule transition is quite
rounded by finite size effects, even for large membranes, because of
the small value of the crossover exponent $\phi$, which leads to a
large correlation length $\xi_{ct}(t_y)$. Taking an example of a
$L=10\mu m$ membrane with lattice constant $a=10$\AA, we find that the
crumpled-to-tubule transition is rounded at a reduced temperature
$t_y\approx(L/a)^{-\phi}\approx 0.13$, while our hypothetical
simulation of a $10^4$ particle net experiences rounding at
$t_y\approx 0.36$. Thus, the transition may not appear sharp
experimentally or in simulations, even though it is, in principle, in
the thermodynamic limit.

The singular parts of other thermodynamic variables obey scaling laws
similar to that for $R_{G,y}$, Eq.\ref{Rtrans}. For example the
singular part of the specific heat per particle $C_v$, i.e., a second
derivative of the intensive free energy with respect to temperature,
is given by
\begin{equation}
C_v\sim{1\over L^D}{\partial^2\over\partial t_y^2}\left({1\over2}t_y
R_y^2 L^{D-2}\right)\;,\label{Cv1}
\end{equation}
which, using Eq.\ref{Rtrans} leads to the scaling form for $C_v$
\begin{eqnarray}
C_v&=&L^{\beta} g(t_y L^{\phi})\;,\nonumber\\
&\propto&
\left\{ \begin{array}{lr}
t_y^{-\alpha_+} L^{\beta-\alpha_+\phi}, & t_y > 0,  L >> \xi_{ct} \\
L^{\beta} , & L << \xi_{ct} \\
|t_y|^{-\alpha_-} L^{\beta-\alpha_-\phi}, & t_y < 0, L >> \xi_{ct}
\label{heat}
\end{array} \right.
\end{eqnarray}
where, 
\begin{equation}
g(x)\approx {d^2\over d x^2}\left[ f_y^2(x)\right]\;.\label{gx}
\end{equation}
Using the exponents characterizing $R_y$ derived above, we obtain:
\begin{mathletters}
\begin{eqnarray}
\beta&=&2\nu^y_{ct}-2+\phi\;,\label{beta1}\\
&=&0\;,\;\;\;\mbox{Flory theory}\label{beta2}
\end{eqnarray}
\end{mathletters}
\begin{mathletters}
\begin{eqnarray}
\alpha_+&=&-2\gamma^y_{+}+1\;,\label{alpha_+1}\\
&=&{2d+3\over d+2}\;,\;\;\;\mbox{Flory theory}\label{alpha_+2}\\
&=&{9\over5}\;,\;\;\;\mbox{Flory theory},\;\; d=3\label{alpha_+3}
\end{eqnarray}
\end{mathletters}
\begin{mathletters}
\begin{eqnarray}
\alpha_-&=&-2\gamma^y_{-}+1\;,\label{alpha_-1}\\
&=&{0}\;,\;\;\;\mbox{Flory theory}\label{alpha_-2}
\end{eqnarray}
\end{mathletters}

This leads to the unusual feature that outside the critical regime
(i.e. for $L >> \xi_{ct}$), the singular part of the specific heat
above the crumpled-to-tubule transition vanishes in the thermodynamic
limit like $L^{-\alpha_+\phi}\sim L^{-2(d-D)/(d+2)}\sim L^{-2/5}$; in
the last expression we have used the Flory estimates of the exponents,
evaluated in $D=2$ and $d=3$. Only within the critical regime does the
singular part of the specific heat per particle becomes nonvanishing
as $L\rightarrow\infty$. Similar results were first found for the
direct crumpled-to-flat transition by Paczuski et al.\cite{PKN}.

We now turn to the tubule-to-flat (tf) transition. On both sides of
{\it this} transition, $R_y=L_y\times O(1)$. Therefore only the other
two radii of gyration $R_x$ and $R_z$ exhibit critical behavior, which
can be summarized by the scaling law:
\begin{eqnarray}
R_{x,z}&=&L^{\nu_{tf}^{x,z}} f_{x,z} (t_\perp L^{\phi_{tf}})
\;,\nonumber\\
&\propto&
\left\{ \begin{array}{lr}
t_\perp^{\gamma_{+}^{x,z}} L^{\nu_t}, & t_\perp > 0,  L >> \xi_{tf} \\
L^{\nu_{tf}^{x,z}} , & L << \xi_{tf} \\
|t_\perp|^{\gamma_{-}^{x,z}} L^{\nu_f^{x,z}}, & t_\perp < 0, L >> \xi_{tf}
\end{array} \right.
\label{Rtf}
\end{eqnarray}
where $t_\perp=(T-T_{tf})/T_{tf}$, $t_\perp>0$ is assumed to
correspond to the tubule phase,
$\xi_{tf}\propto|t_\perp|^{-1/\phi_{tf}}$ is the correlation length
for this transition, and the exponents obey the scaling relations
\begin{mathletters}
\begin{eqnarray}
\nu_f^z&=&\zeta\approx0.59\;,\label{nu_f^z}\\
\nu_f^x&=&1\;,\label{nu_f^x}\\
\gamma_{+}^{x,z}&=&{\nu_t-\nu_{tf}^{x,z}\over\phi_{tf}}\;,\label{gamma+tf1}\\
\gamma_{-}^{x,z}&=&{\nu_f^{x,z}-\nu_{tf}^{x,z}\over\phi_{tf}}\;.\label{gamma-tf1}
\end{eqnarray}
\end{mathletters}
In the above we have taken the $x$-direction to be the new (in
addition to $y$) extended direction in the flat phase (which is why
$\nu_f^x=1$), and $\zeta$ is the roughness exponent\cite{LR} of the
flat phase (quoted for the physical case $D=2$ and $d=3$), giving the
transverse height fluctuations of the $d-2$ components of the
displacement perpendicular to the flat membrane.

To calculate these exponents, we can use Flory theory in the tubule
phase, and at the transition, while in the flat phase, where as
discussed above, self-avoidance is irrelevant, we simply match onto
the scaling theory\cite{LR} of the flat phase. Doing so, we find that
Flory theory predicts {\it identical} behavior for $R_x$ and $R_z$ in
the tubule phase and at the transition:
\begin{mathletters}
\begin{eqnarray}
\nu_{tf}^{x}&=&\nu_{tf}^z={D+3\over d+3}\;,\label{nu_xz}\\
&=&{5\over6}\;,\;\;\;\mbox{for}\;\;\; D=2\;,d=3\;,\label{nu_xz_eval}\\
\gamma_+^{x}&=&\gamma_+^{z}=-{1\over d+1}\;,\label{gamma+xz}\\
&=&-{1\over4}\;,\;\;\;\mbox{for}\;\;\; D=2\;,d=3\;.\label{gamma+xz_eval}
\end{eqnarray}
\end{mathletters}
We believe that the identical temperature ($t_\perp$) and scaling
(with $L$) behavior of $R_x$ and $R_z$ as the tubule-to-flat
transition is approached from the tubule side (Eqs.\ref{nu_xz} and
\ref{gamma+xz}) is an artifact of Flory theory and that in fact
$R_x>>R_z$ throughout this region, with the ratio $R_x/R_z$ actually
diverging as the transition is approached from above. That is, we
expect that in reality $\nu_{tf}^x>\nu_{tf}^z$ and $\gamma_+^x
<\gamma_+^z$.
 
In addition, Flory theory predicts 
\begin{eqnarray}
\phi_{tf}&=&{2(d-D)\over d+3}\;,\label{phi_tf}\\
&=&{1\over3}\;,\;\;\;\mbox{for}\;\;\; D=2\;,d=3\;.\label{phi_tf_eval}
\end{eqnarray}

In the flat phase, $\gamma_-^x$ follows from simply minimizing the
mean field free energy {\it without} self-avoidance (since
self-avoidance is irrelevant in the flat phase), giving 
\begin{equation}
\gamma_-^{x}={1\over2}\;,\label{gamma-x}
\end{equation}
while matching $R_z=L^\zeta|t_\perp|^{\gamma_-^z}$ onto the critical
prediction $R_z\propto L^{\nu_{tf}^z}$ at the correlation length
$L=\xi=|t_\perp|^{-1/\phi_{tf}}$ gives
\begin{eqnarray}
\gamma_-^{z}&=&{\zeta-\nu_{tf}^{z}\over\phi_{tf}}\;,\label{gamma-z}\\
&\approx&-0.73\;,\label{gamma-z_eval}
\end{eqnarray}
where the first equality is an exact scaling law, while the second,
approximate one uses Flory theory for $\phi_{tf}$ and $\nu_{tf}^x$, and
the SCSA calculation\cite{LR} of $\zeta$ for the flat phase, all
evaluated in the physical case $D=2$ and $d=3$.

As the tubule-to-flat transition is approached from below (the flat
phase side) $R_x$ shrinks as $R_x\approx|t_\perp|^{1/2}L$ and $R_z$
increases as $R_z\approx|t_\perp|^{-0.73}L^{0.59}$ with vanishing
$|t_\perp|$. Approaching this transition from above (the tubule phase
side) $R_x$ and $R_z$ both extend as
$R_{x,z}\approx|t_\perp|^{-1/4}L^{3/4}$ with vanishing $t_\perp$ to
the $L^{5/6}$ scaling at the tubule-to-flat critical point.

The singular part of the specific heat again obeys a scaling law:
\begin{eqnarray}
C_v&=&L^{\beta_{tf}} g_{tf}(t_\perp L^{\phi_{tf}})\;,\nonumber\\
&\propto&
\left\{ \begin{array}{lr}
t_\perp^{-\alpha_+^{tf}} L^{\beta_{tf}-\alpha_+^{tf}\phi_{tf}}, & t_\perp > 0,  L >> \xi_{tf} \\
L^{\beta_{tf}} , & L << \xi_{tf} \\
|t_\perp|^{-\alpha_-^{tf}} L^{\beta_{tf}-\alpha_-^{tf}\phi_{tf}}, & t_y < 0, L >> \xi_{tf}
\end{array} \right.
\label{cf}
\end{eqnarray}

where, in Flory theory,
\begin{mathletters}
\begin{eqnarray}
\alpha_+^{tf}&=&{3\over2}\;,\label{alpha+tf}\\
\alpha_-^{tf}&=&0\;,\label{alpha-tf}\\
\beta_{tf}=2\nu_{tf}+\phi_{tf}-2&=&0\;,\label{beta_tf}
\end{eqnarray}
\end{mathletters}

Thus, again, the singular part of the specific heat vanishes (now like
$L^{-1/2}$) in the thermodynamic limit above (i.e., on the tubule
side) of the transition, while it is $O(1)$ and smooth as a function
of temperature in both the critical regime and in the flat phase.

\section{Summary and Conclusions}
\label{summary_section}

In summary, we have studied the effects of intrinsic anisotropy in
polymerized membrane. We found that this seemingly innocuous
generalization leads to a wealth of new phenomena, most remarkable of
which is that {\it any} amount of anisotropy leads to a new, tubule
phase which intervenes between the previously predicted flat and
crumpled phases in anisotropic membranes (See
Fig.\ref{phase_diagram1}). We have presented a detailed theory of the
anisotropic membrane focusing on the new tubule phase.  Considering
thermal fluctuations in the tubule phase we have shown that the {\em
phantom} tubule phase exhibits anomalous elasticity, and calculated
the elasticity and size exponents {\em exactly}, as summarized in
Eqs.\ref{z_phantom},\ref{etau_phantom},\ref{nu_phantom},\ref{zeta}.
We then considered the physically more relevant case of a {\em
self-avoiding} tubule, finding that self-avoiding interaction is
important for physical dimensionalities. Establishing relations
between the exponent characterizing the diameter of the tubule and the
exponents describing anomalous elasticity and transverse undulations,
we calculated the tubule diameter, size of the undulations and the
anomalous elasticity within the Flory and
$\epsilon=d_{uc}-d$--expansion theories. We have also studied
self-avoidance within a Gaussian variational approximation, which
unfortunately but, we believe incorrectly predicts that self-avoiding
interaction destroys the tubule phase (as it does the crumpled phase)
for $d<4$.  We studied the crumpled-to-tubule transition in mean-field
theory and with the $\epsilon=4-D$-expansion. Finally we developed a
scaling theory of the crumpled-to-tubule and tubule-to-flat
transitions.

Our {\em exact} predictions for the phantom tubules
Eqs.\ref{z_phantom},\ref{etau_phantom},\ref{nu_phantom},\ref{zeta}
have been {\em quantitatively} verified in the recent simulations by
the authors of Ref.~\onlinecite{BFT}.

The possibility of the existence of a new tubule phase intermediate
between the fully disordered crumpled phase and fully ordered flat
phase is exciting from both basic physics and potential applications
points of view. Recently much attention has focussed on utilizing
self-assembled microstructures for encapsulations for various
applications, most notably controlled and slow drug
delivery.\cite{Rudolf} The structural stability of polymerized
membranes is superior to their liquid membrane analogs. The
theoretical discovery of the tubule phase significantly expands the
number of possibilities, and also offers the potential tunability (by,
e.g., adjusting the strength of self-avoidance) of the tubule diameter
and therefore the amount of encapsulation and rate of delivery.

The realization of the tubule phase in polymerized membranes carries
even more significance if the claims that the fully crumpled phase in
polymerized membranes does not exist are in fact correct, since in
this case the tubule phase is the only disordered phase of a
polymerized membrane.

With the recent focus on self-assembly, it may be possible in the near
future to freeze in intrinsic anisotropy by polymerizing tilted phase
of liquid membranes or cross-linking polymers.  Further numerical
simulations which include self-avoidance offer another avenue to
investigate our predictions. We hope that our work stimulates further
theory, simulations and experiments in this area.

\acknowledgements
 
\vskip .2in 

We both thank The Institute for Theoretical Physics at the UCSB, and
the organizers of the Biomembranes Workshop held there, where this
work was initiated, for our good tans, their hospitality and financial
support under NSF Grant No. PHY94-07194. We acknowledge partial
support from the Colorado Center for Chaos and Complexity through the
Summer 1997 Workshop on Nucleation and Critical Phenomena in Complex
Nonlinear Systems, during which this work was completed. Leo
Radzihovsky acknowledges support by the NSF CAREER award, through
Grant DMR--9625111, and partial support by the A.P. Sloan
Foundation. John Toner acknowledges financial support by the NSF
through Grants DMR-9625111 and DMR-9634596.

%

\end{multicols} 

\begin{references}
\bibitem{NotGeneric} Actually, in most experimental realizations, the
polymerization is random and leads to an isotropic disordered
membrane, making the isotropic case quite general.
\bibitem{CommentSA} The failure to find the crumpled phase in computer
simulations for self-avoiding membranes might be because
self-avoidance interaction shifts the bare $\kappa$ to high values,
such that in the almost universally employed numerical ball and spring
models this always put the isotropic membrane on the flat side of the
crumpling transition.
\bibitem{magnets} This is analogous to the familiar $1/n$ expansion
for critical phenomena, in which one expands about the number of spin
components $n\rightarrow\infty$ limit.
\bibitem{noteSC}In doing this power counting, one must assume that the
$q = 0$ value of the integral in Eq.\ref{SCgy} is precisely cancelled
by the bare $g_y$. This assumption is the only way to solve
Eq.\ref{SCgy} for $D < D_{uc}$.
\end{references}
\end{document}